\documentclass[rmp,twocolumn,aps,amsmath,amssymb]{revtex4-1}
\usepackage{graphicx}
\usepackage{dcolumn,xcolor,ulem}
\usepackage{amsmath} 
\usepackage{amssymb}
\usepackage{amsfonts}
\usepackage{bm}
\usepackage{scalerel}
\usepackage[makeroom]{cancel}
\usepackage{array}
\usepackage{overpic}
\usepackage{pifont}
\usepackage{wasysym} 

\makeatletter
\providecommand{\leftsquigarrow}{%
  \mathrel{\mathpalette\reflect@squig\relax}%
}
\newcommand{\reflect@squig}[2]{%
  \reflectbox{$\m@th#1\rightsquigarrow$}%
}
\makeatother

\usepackage{hyperref}

\newcommand{\myG}{\raisebox{\depth}{\rotatebox{180}{G}} } 

    \newcolumntype{P}[1]{>{\centering\arraybackslash}p{#1}}
    \newcolumntype{M}[1]{>{\centering\arraybackslash}m{#1}}
\usepackage{enumitem}
\newlist{abbrv}{itemize}{1}
\setlist[abbrv,1]{label=,labelwidth=1in,align=parleft,itemsep=0.1\baselineskip,leftmargin=!}

\usepackage[latin1]{inputenc}

\newcommand{\Maf}[1]{\textcolor{black}{#1}}

\begin{document}



\title{Dynamic duos: the building blocks of dimensional mechanics}

\author{M.A.~Fardin}
\altaffiliation[Corresponding author ]{}
\email{marc-antoine.fardin@ijm.fr}
\affiliation{Universit\'{e} de Paris, CNRS, Institut Jacques Monod, F-75013 Paris, France}
\affiliation{The Academy of Bradylogists}
\author{M.~Hautefeuille}
\affiliation{Institut de Biologie Paris Seine, Sorbonne Universit\'{e}, 7 quai Saint Bernard, 75005 Paris, France}
\author{V.~Sharma}
\affiliation{Department  of Chemical  Engineering, University of Illinois at Chicago, Chicago, Illinois 60608, United States}
\affiliation{The Academy of Bradylogists}
\date{\today}

\begin{abstract}
Mechanics studies the relationships between space, time, and matter. These relationships can be expressed in terms of the dimensions of length $\mathcal{L}$, time $\mathcal{T}$, and mass $\mathcal{M}$. Each dimension broadens the scope of mechanics. Historically, mechanics emerged from geometry, which considers quantities like lengths or areas, with dimensions of the form $\mathcal{L}^x$. With the Renaissance quantities combining space and time were considered, like speeds, accelerations or later diffusivities, all of the form $\mathcal{L}^x\mathcal{T}^y$. Eventually, mechanics reached its full potential by including ``mass-carrying'' quantities such as mass, force, momentum, energy, action, power, viscosity, etc. These standard mechanical quantities have dimensions of the form $\mathcal{M}\mathcal{L}^x\mathcal{T}^y$, where $x$ and $y$ are integers. In this contribution, we show that thanks to this dimensional structure these mass-carrying quantities can be readily arranged into a table such that $x$ and $y$ increase along the row and column respectively. Ratios of quantities in the same rows provide characteristic lengths, and in the same columns characteristic times, encompassing a great variety of physical phenomena from atomic to astronomical scales. Most generally, we show that picking duos of mechanical quantities that are neither on the same row nor column of the table yields dynamics, where one mechanical quantity is understood as impelling motion, while the other is impeding it. The force and the mass are the prototypes of impelling and impeding factors, but many other duos are possible. We present examples from the physical and biological realms, including planetary motion, sedimentation, explosions, fluid flows, turbulence, diffusion, cell mechanics, capillary and gravity waves, and spreading, pinching, and coalescence of drops and bubbles. This review provides a novel synthesis revealing the power of scaling or dimensional analysis, to understand processes governed by the interplay of two mechanical quantities. This elementary decomposition of space, time and motion into pairs of mechanical factors is the foundation of ``dimensional mechanics'', a method that this review wishes to promote and advance. Pairs are the fundamental building blocks, but they are only a starting point. Beyond this simple world of mechanical duos, we envision a richer universe that beckons with an interplay of three, four, or more quantities, yielding multiple characteristic lengths, times, and kinematics. The review is complemented by online video lectures, which initiate a discussion on the elaborate interplay of two or more mechanical quantities. 
\end{abstract}

\maketitle

\tableofcontents

\section{Introduction}
Mechanics is the bedrock of physics and is influential to all sciences. Mechanics has had such a far reaching impact on our understanding of the natural world that it is hard to contain it under a single definition. In the 19th century, it was an effort to integrate new disciplines like thermodynamics and electromagnetism under its fold that led Fourier~\cite{Fourier1822}, Gauss and Weber~\cite{Assis2002}, Maxwell and Kelvin~\cite{Maxwell1873,Mitchell2017}, and their contemporaries to one of the most commonly accepted definition of mechanics~\cite{Maxwell1873,Macagno1971}. Mechanics deals with the relationships between space, time and matter, usually quantified by the dimensions of length $\mathcal{L}$, time $\mathcal{T}$, and mass~$\mathcal{M}$. 

Mechanics in a general sense includes \textit{geometric} quantities, with dimensions of the form $\mathcal{L}^x$ (lengths, areas, etc.). More broadly, mechanics also includes \textit{kinematic} quantities, with dimensions $\mathcal{L}^x \mathcal{T}^y$ (speed, acceleration, diffusivity, etc.). These kinematic quantities describe motion, but without any reference to the ``causes'' of these motions. It is the quest for these causes that led to the definition of the mass, and all its offsprings: force, density, momentum, energy, action, power, stiffness, pressure, viscosity, etc. The bestiary of mechanics includes many creatures, but they are cast from the same mold.  All these mechanical quantities have dimensions of the form $\mathcal{M} \mathcal{L}^x \mathcal{T}^y$, they are ``mass-carrying quantities''. As we will see in this review, this shared structure allows a representation of the mechanical quantities in a plane with coordinates $x$ and $y$, the exponents of the space and time dimensions. Moreover, since $x$ and $y$ are usually integers, the \textit{standard mechanical quantities} can be arranged into a table, which is a  great guide for researchers and teachers, and the perfect cheat sheet for students. We have spent the last three years toying with this enigmatic map of the mechanical quantities. We put this table together in order to provide a Rosetta stone to help translate knowledge across the boundaries of the many sub-fields of science. We invite readers to contribute to this table, and to suggest additions or modifications. 

Our investigations led us to a reformulation of the dimensional approach to mechanics, which we are sharing in series of lectures on a \href{https://www.youtube.com/@naturesnumbers}{Youtube channel} that we created for this purpose (youtube.com/@naturesnumbers). These lectures explain in detail how to use this table to identify the ``causes'' of a wide range of motions, to transform a kinematic description into a dynamical understanding. These videos serve as ``supplementary material'' to this review, which focuses on the decomposition of geometric or kinematic quantities into ratios of mechanical quantities. From a dimensional perspective, this kind of decomposition is elementary, but it has far reaching consequences on the understanding of the relationship between mechanics and motion, and it provides a systematic way to approach the ``causes'' of motion. 

The basis for this dimensional approach to mechanics dates back to Archimedes. To measure a volume $\Omega$, Archimedes proposed to express it as a ratio, $\Omega=m/\rho$, between the mass $m$ of the object and its density $\rho$. This old formula seems so elementary today that we do not realize the great leap that it encompasses: a geometric quantity (the volume) is given from a ratio of two mechanical quantities (the mass and the density). Dimensionally, the logic is flawless: $[\Omega]=\mathcal{L}^3=[m/\rho]=\cancel{\mathcal{M}}/\cancel{\mathcal{M}} \mathcal{L}^{-3}$, where the brackets return the dimensions of their content. The extra dimension of mass is a sort of ``dummy'' dimension, disappearing from the final result, a very useful intermediary in the computation.

Almost two thousand years after Archimedes, Newton pushed this logic even further. What Newton sought to compute was not a geometric quantity, but a kinematic one, an acceleration $a$, but he used the same principle. He expressed the acceleration as the ratio between two mechanical quantities: $a=F/m$. Again, the dimensions match: $[a]=\mathcal{L} \mathcal{T}^{-2}=[F/m]=\cancel{\mathcal{M}} \mathcal{L} \mathcal{T}^{-2}/\cancel{\mathcal{M}}$. The example is so classical that it may not seem too impressive today. 

Fast forward almost three centuries to the 1940s and consider this other example, often found in textbooks on dimensional analysis~\cite{Barenblatt2003}. The Second World War is raging and the British physicist G.I. Taylor is trying to compute the dynamics of an explosion blast. Experiments suggest that the radius of the explosion follows a `power law' of time, $d(t)\simeq Kt^\alpha$. To understand the value of the kinematic prefactor $K$, Taylor uses the old trick again. In this context, Taylor identifies the energy $E$ of the bomb and the density $\rho$ of the ambient air as the relevant mechanical parameters. Then, the dimensions of the mechanical ratio provide an answer: $[E/\rho]=\cancel{\mathcal{M}} \mathcal{L}^2 \mathcal{T}^{-2}/\cancel{\mathcal{M}} \mathcal{L}^{-3}=\mathcal{L}^5 \mathcal{T}^{-2}$. Taylor concludes that $K\simeq (E/\rho)^\frac{1}{5}$, that is $d(t)\simeq (E/\rho)^\frac{1}{5} t^\frac{2}{5}$~\cite{Taylor1950,Taylor1950b}. Not so trivial anymore!

From Archimedes to Newton and Taylor the procedure remains the same. A geometric or more broadly kinematic quantity (without any mass dimension) can always be expressed as some ratio of mechanical quantities. The only thing that varies from one example to another is the pair of mechanical quantities that are involved in the decomposition. In any case, the dimension of mass comes to the rescue, providing a way to understand sizes, durations and motions of all sorts, from a ratio, or ``balance'', or ``struggle'', between ``competing'' mechanical factors. 

Even if we restrict ourselves to the standard mechanical quantities in Table~\ref{masscary}, there are hundreds of possible pairs, and quite a few with a rich history. The purpose of this review is to discuss a few of these pairs. Each pair tells a different story, synthesizes different ``physics'', and retraces the steps of those who sought to explore this mechanical landscape. 

Pairs of mechanical quantities are the building blocks of the relationship between mechanics and kinematics, but they are only a starting point. If motion can be understood from the interplay of mechanical quantities, what can we expect from the interaction of three, four or even more quantities? We asked ourselves these very questions three years ago and we have been working on answering them since, our \href{https://www.youtube.com/@naturesnumbers}{lecture series} providing a diary of this journey. With this review, we solely focus on the interplay of pairs of mechanical quantities, but we will return later with more on the impact of additional players. 

To illustrate the scope of a dimensional analysis of mechanics we will use examples from a wide spectrum of fields. This diversity constrains us to limit our citations to a few papers, which can be used as gates toward larger bodies of literature. Our background in fluid dynamics, soft matter, and biophysics, has biased us toward references from these fields. For instance, we are indebted to several reviews and textbooks on spreading, pinching and coalescence, including~\citet{Dussan1979}, \citet{PGG1985}, \citet{Leger1992}, \citet{Middleman1995}, \citet{Oron1997},  \citet{McKinley2005}, \citet{Starov2007}, \citet{Kalliadasis2007}, \citet{Craster2009}, \citet{Bonn2009}, \citet{Popescu2012}, \citet{PGG2013}, \citet{Snoeijer2013}, \citet{Lu2016}, \citet{Bico2018}, \citet{Andreotti2020}, and \citet{Lohse2015,Lohse2020}. However, we have tried as much as possible to diversify our references to include a literature more familiar to biologists and engineers. In particular, for explosions we relied on~\citet{Bethe1958}, \citet{Glasstone1977}, \citet{Sedov1993}, \citet{Krehl2008}, \citet{Westine2012}, \citet{Kinney2013} and \citet{Sachdev2016}. For biological systems we relied on~\citet{Thompson1917}, \citet{Mitchison1996}, \citet{Alt1997}, \citet{Sheetz2001}, \citet{Roberts2002}, \citet{Lecuit2007}, \citet{LeClainche2008}, \citet{Pollard2009}, \citet{Phillips2012}, \citet{Marchetti2013} and \citet{Schwarz2013}. We have also benefited from seminal texts on dimensional analysis, including  \citet{Fourier1822}, \citet{Maxwell1873}, \citet{Buckingham1914}, \citet{Rayleigh1915}, \citet{Bridgman1922}, \citet{Barenblatt1996,Barenblatt2003}, and \citet{Santiago2019}. 

In this review, terms first appearing between `single quotes' are technical terms from the literature. A search of this term on the Web will generally lead to its definition. Terms appearing \textit{in italics} are those we first define here, or which substantially deviate from traditional usage. Terms appearing between ``double quotes'' are actual quotes, or colloquialisms. The sign `$\equiv$' symbolizes a definition, where the left-hand side is a shorthand notation for the right-hand side. The sign `$\simeq$' means that the two sides of the equation are expected to be of the same `order of magnitude' (other authors may use $\propto$ or $\sim$). The sign  `$\sim$' will be used to state an incomplete scaling relation, as in $d\sim t^\alpha$, where ``incomplete'' means that the left and right-hand sides do not have the same dimensions. The sign `$=$' refers to a standard equality, which is presumably exact. 

Links to the \href{https://www.youtube.com/@naturesnumbers}{video lectures} are given at the beginning of each associated section.

\section{The mechanical quantities\label{sec1}}
\href{https://youtu.be/XDQ21ngtpkI?si=a1PeLW6ICFV8bP5i}{Mechanics 1: Mechanical Quantities}\\

\begin{table*}
\centering
\includegraphics[width=17cm,clip]{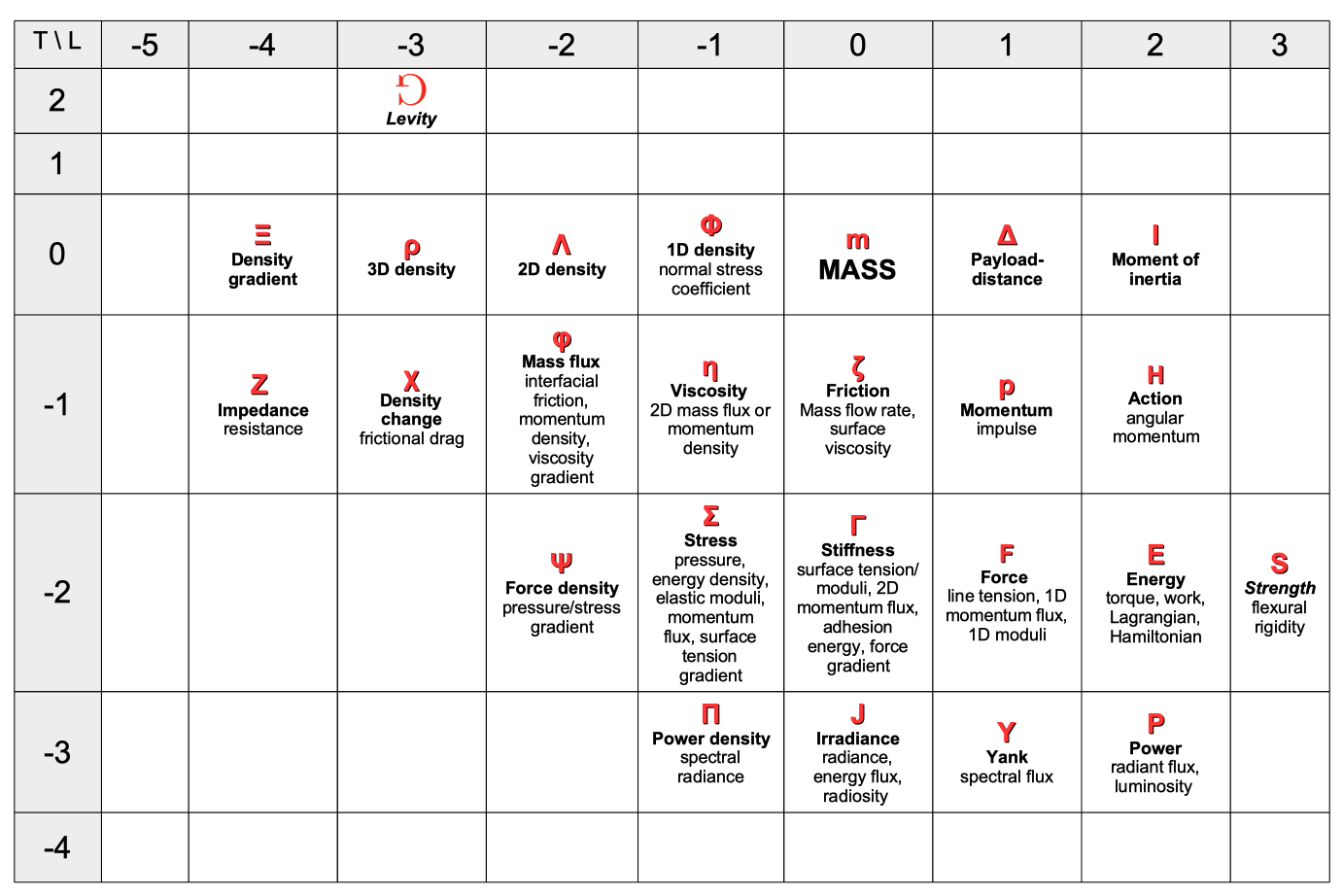}
\caption{Table of standard mechanical quantities, with dimensions $\mathcal{M}\mathcal{L}^x\mathcal{T}^y$. Values of the exponents $x$ and $y$ refer to the columns and rows. A given quantity associated with a choice of couple $(x,y)$ usually have several names. The symbol and the name in bold are the one we chose when generically referring to that quantity in this review. \Maf{The dimensions of a mechanical quantity dictate its symbol. We systematically use the same symbols for all the mechanical quantities that share the same dimensions. For instance, we do not use ``$P$'' for a pressure, ``$G$'' for an elastic modulus and ``$\sigma$'' for a shear stress, we just use $\Sigma$ for all quantities with dimensions $\mathcal{M}\mathcal{L}^{-1}\mathcal{T}^{-2}$}. We have found useful to define the \textit{levity} and \textit{strength}, which are respectively discussed in sections~\ref{Klev} and~\ref{Bohrr}. Please feel free to complete this table in any way you see fit.  
\label{masscary}}
\end{table*} 
Mechanics includes geometry, kinematics and everything beyond, if it can be expressed with the addition of the dimension of mass. So the most generous definition of the term ``mechanical quantity'' could encompass any quantity with dimensions of the form $\mathcal{M}^z\mathcal{L}^x\mathcal{T}^y$, where $x$, $y$ and $z$ could a priori be real numbers. However, this is not how we will use this term in this review. We will call \textit{mechanical quantities} those with dimensions of the form $\mathcal{M}^1\mathcal{L}^x\mathcal{T}^y$. We will use the term \textit{kinematic quantities} to describe quantities with dimensions of the form $\mathcal{M}^0\mathcal{L}^x \mathcal{T}^y$, with $x\neq0$ and $y\neq0$. And we will use the adjectives \textit{geometric/spatial} and \textit{chronometric/temporal}  to respectively describe quantities with dimensions of the form $\mathcal{M}^0 \mathcal{L}^x \mathcal{T}^0$ and $\mathcal{M}^0 \mathcal{L}^0 \mathcal{T}^y$  ($x\neq0$ and $y\neq0$). \Maf{What about quantities like $\mathcal{M}^z \mathcal{L}^x \mathcal{T}^y$ ($z\neq0$)?  We will disregard them on account of the fact that they can be reduced to mechanical quantities by factorization: $(\mathcal{M}^1\mathcal{L}^\frac{x}{z}\mathcal{T}^\frac{y}{z})^z$. }

Examples of geometric quantities include the well-known length ($\mathcal{L}^1$), area ($\mathcal{L}^2$), and volume ($\mathcal{L}^3$), but also more technical quantities like the `wavenumber' ($\mathcal{L}^{-1}$). Chronometric quantities include the duration or period ($\mathcal{T}^1$), or the frequency ($\mathcal{T}^{-1}$). The three most well-known examples of kinematic quantities are the speed or velocity ($\mathcal{L}^1\mathcal{T}^{-1}$), the acceleration ($\mathcal{L}^1\mathcal{T}^{-2}$), and the diffusivity ($\mathcal{L}^2\mathcal{T}^{-1}$). Progressive time derivatives of the position beyond acceleration lead to the so-called `jerk', `snap', `crackle' and `pop', but these colorful terms are seldom used. More broadly, as we will see later, there are many more possible kinematic quantities, although they are less known and rarely have names. 

Now, what about mechanical quantities ($\mathcal{M} \mathcal{L}^x \mathcal{T}^y$)? In a colloquial sense, the mechanical quantities are the ``forces'' that are in turn pushing or pulling, driving or resisting, impelling or impairing, all the processes at play behind space, time and their combination: motion. We have seen a few examples of these mechanical quantities in the introduction: the mass ($x=0$,$y=0$), the density ($x=-3$,$y=0$), the force ($x=1$,$y=-2$), and the energy ($x=2$,$y=-2$). Each quantity is specified by its coordinates ($x$,$y$), so mechanical quantities can be represented as points on a plane. As we said, a priori, the coordinates $x$ and $y$ could take any value, but the small integers are of notable importance. Indeed the well-known geometric, kinematic and mechanical quantities have integer exponents. Investigating the reasons for this preference for integers is a fascinating task, but it goes beyond the scope of this review. In this review, we will only take note of this fact, and we will use it to our advantage. Because if the coordinates are small integers, we can represent the standard mechanical quantities they correspond to in a table. Thus, \textit{standard mechanical quantities} refer to mechanical quantities where the exponents $x$ and $y$ are small integers, but since these are the only mechanical quantities we will be dealing with here, we will drop the adjective ``standard''.  

In Table~\ref{masscary}, we tabulated the mechanical quantities we could find in the literature, highlighting the fact that they may bear different names depending on the context. Surprisingly such table does not seem to have been drawn before, although as we will see it provides a great way to understand the mechanical underpinning of space-time. The table is organized around the mass ($x=y=0$), with columns set by the exponent $x$, and rows by the exponent $y$. You can think of each mechanical quantity as being located in the dimensional space with two coordinates ($x$,$y$). We will use the symbol $Q(x,y)$ to designate the mechanical quantity with dimensions $\mathcal{M}\mathcal{L}^x \mathcal{T}^y$, and the symbol $K(x,y)$ to designate the kinematic quantity with dimensions $\mathcal{L}^x\mathcal{T}^y$, or simply $Q$ and $K$ when the exponents are implicit. 

Table~\ref{masscary} is a map of the explorations of mechanics in the past centuries, but this mechanical universe is still mostly uncharted territory. Quantities on this table were discovered step by step. Just a few centuries ago, the table would have been mostly empty. Beyond the mass, the density, the force and the momentum, contemporaries of Newton had very little to play with. Newton himself formalized the concept of `viscosity', while his rival Hooke was quantifying the concept of `stiffness'. It is the painstaking recording of natural phenomena that progressively enlarged the mechanical cartography. And this exploration is still ongoing. There are blank spots to fill. We took the liberty to name two quantities we felt deserved their place, but for which we could not find names in the literature: the \textit{levity} ($\mathcal{M}\mathcal{L}^{-3}\mathcal{T}^2$) and the \textit{strength} ($\mathcal{M}\mathcal{L}^{3}\mathcal{T}^{-2}$). A famous example of levity is the inverse of the gravitational constant $G$. Almost equally famous examples of strengths are $\hbar c$, and $k_c e^2$, where $\hbar$, $k_c$, $c$ and $e$ are respectively the Planck and Coulomb constants, the speed of light, and the elementary charge. These two expressions respectively give the ``strength'' of the nuclear and electromagnetic interactions. We will return to these important examples later in the review. 

Each mechanical quantity can a priori be independent from the others. However, as we shall see, mechanical quantities are revealed by their interactions with one another. The most elementary form of such interaction is between pairs of mechanical quantities. Ratios of different mechanical quantities can produce space, time and motion. 

Ratios of quantities in the same row produce purely spatial results:
\begin{equation}
\Big(\frac{Q(x,y)}{Q(x-n,y)}\Big)^\frac{1}{n}= K(1,0)= \ell
\label{lengthscaling}
\end{equation}
We will discuss examples of such lengths in section~\ref{lengths}. The symbol $\ell$ will generally be used to refer to any kind of length, size or distance, when this length is constant. We will rather use the symbol $d$ when referring to a variable length. When multiple lengths are present we may occasionally use alternate symbols for lengths, like $h$ for heights, or $r$ for radii. 

Ratios of quantities in the same column produce purely temporal results:
\begin{equation}
\Big(\frac{Q(x,y)}{Q(x,y-n)}\Big)^\frac{1}{n}=K(0,1)= \tau
\end{equation}
We will discuss examples of such times in section~\ref{times}. The symbol $\tau$ is used to refer to any kind of constant time, duration or period. We will rather use $t$ to refer to a variable time.

Ratios of quantities on a diagonal of slope -1 produce speeds:
\begin{equation}
\Big(\frac{Q(x,y)}{Q(x-n,y+n)}\Big)^\frac{1}{n} = K(1,-1)= u
\end{equation}
We will discuss examples of such speeds in section~\ref{speeds}. \Maf{We will use the symbol $u$ to designate any constant speed, and $v$ to refer to a variable speed.} 

Ratios of quantities on a diagonal of slope -2 produce accelerations:
\begin{equation}
\Big(\frac{Q(x,y)}{Q(x-n,y+2n)}\Big)^\frac{1}{n} = K(1,-2)=g
\end{equation}
We will discuss examples of such accelerations in section~\ref{sprexa}. We will use the symbol $g$ to designate any constant acceleration, and $a$ for a variable acceleration.  

These ratios giving rise to lengths, times, speeds, and accelerations are the most well-known, but we shall see that others are of interest. Note also that the relationship between two mechanical quantities is sometimes encoded in the very names of these quantities. In particular, the quantity $Q(x-2,y-1)$ can be thought of as the flux of $Q(x,y)$. For instance, a stress can be thought of as a flux of momentum. The quantity $Q(x-3,y)$ can be thought of as the density of the quantity $Q(x,y)$. For instance, the stress can be thought of as a density of energy. However, except in a few instances were traditions obliged us (as with the mass flux, or the force-density), we tried to use names that did not explicitly refer to a parent quantity. All quantities on the table are related to another, but every quantity exist\Maf{s} in its own right. 

As illustrated by the different names of the mechanical quantities shown in Table~\ref{masscary}, the overall spatial dimension of a particular context can lead to a confusing usage of the same words. For instance, in a 2D setting, one may call the stiffness $Q(0,-2)$ an elasticity (this is very common in the study of cells and tissues~\cite{Marchetti2013,Schwarz2013}). Our choice here will be to use the 3D naming conventions written in bold in Table~\ref{masscary}. Note that the existence of many names for a quantity with the same dimensions is correlated to the existence of many units as well. For instance, Table~\ref{masscary} will make it quite obvious that a stiffness is sometimes expressed in N/m, where a Newton is a unit of force, or in J/m$^2$, where a Joule is a unit of energy. Although less conventional, it could very well be given in poise.m/s, where a poise is a unit of viscosity. We have found that the table of mechanical quantities can make it easier to juggle with all these overlapping names and units.

\section{The mechanics of space or time}
Before we address the relationship between mechanics ($\mathcal{M}\mathcal{L}^x \mathcal{T}^y$) and kinematics ($\mathcal{L}^x\mathcal{T}^y$), that is between mass-carrying quantities and motion, we should first discuss how space or time can separately be understood mechanically. Any constant length or duration can be decomposed into a pair of mechanical factors. At the very least, these factors are interpreted as providing a way to compute the values of lengths or durations, but they can also be regarded as the ``origin'', or ``reason'', or ``cause'' behind these lengths or durations. 

\subsection{Simple lengths\label{lengths}}
\href{https://youtu.be/QiZ--8tkMdo?si=2IERNoMuT8Jq1A6w}{Mechanics 2: Simple Lengths}\\

\begin{figure}
\centering
\includegraphics[width=8.5cm,clip]{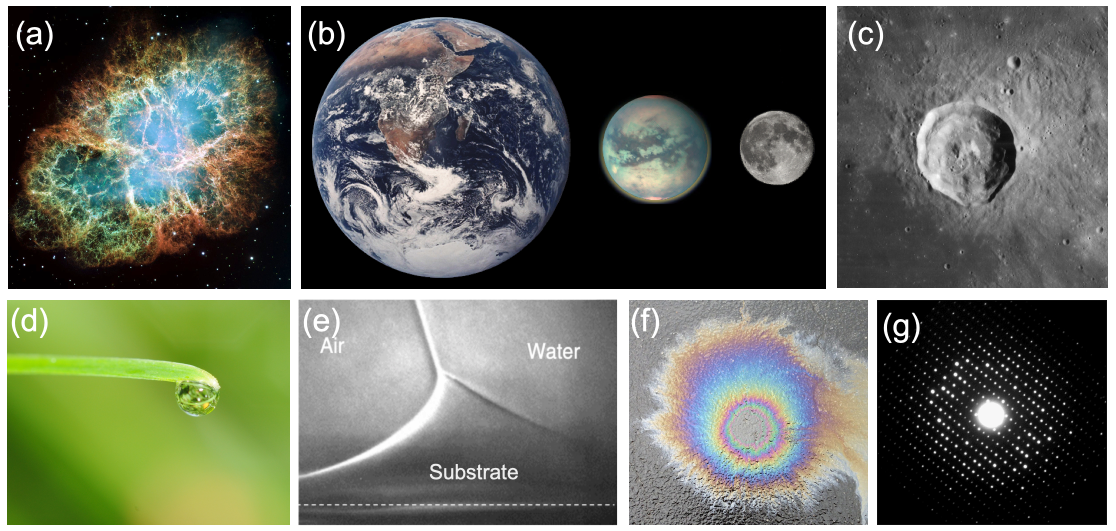}
\caption{Examples of simple lengths. (a) The Crab nebula, a supernova remnant, which will grow for millennia up to a size $\ell \simeq (E/\Sigma)^{\frac{1}{3}}$ (here $\ell\simeq10^{17}$m). \Maf{Image credit: NASA (public domain)}. (b) Earth, Titan and the Moon, three astronomical bodies with sizes of the form $\ell \simeq  \Sigma/\Psi$ (here $\ell\simeq 6371$, 2575, 1737 km). \Maf{Image credit: NASA (public domain)}. (c) Manilius crater on the Moon, an example of length given by $\ell \simeq (E/\Psi)^{\frac{1}{4}}$ (here $\ell \simeq 19$~km). \Maf{Image credit: NASA (public domain)}. (d) Illustration of the capillary length on a pendant drop before its fall \Maf{(public domain)}, with size $\ell\simeq (\Gamma/\Psi)^\frac{1}{2}$ (here $\ell \simeq 3$~mm). (e) An elastic substrate is deformed by capillarity at the contact line, over a distance $\ell\simeq \Gamma/\Sigma$ (here $\ell \simeq 10~\mu$m)\Maf{~\cite{Bico2018}}. (f) A thin film of oil on water produce\Maf{s} an interference pattern testifying of its small thickness expressed as $(E/\Gamma)^\frac{1}{2}$ \Maf{(public domain)}. (g) Beyond the optical resolution, the diffraction pattern of an atomic crystal obtained by electron crystallography reveals the typical size of the atoms \Maf{(public domain)}. This size if of the form $\ell\simeq S/E$ or $\ell\simeq H/p$ (here $\ell\simeq 0.1$~nm).     
\label{fig5}}
\end{figure} 
As mentioned in the introduction, Archimedes showed us the way when it comes to relating geometry and mechanics, when he expressed a volume as the ratio of a mass and a density $\Omega=m/\rho$. The story has been told a thousand times, it is the original ``Eureka!'' moment. \href{https://math.nyu.edu/~crorres/Archimedes/Crown/bilancetta.html}{Galileo's insight on this ancient story} shows the importance that it had on the mechanical Renaissance~\cite{Mottana2017}. 

If the mass and density of an object are known then its volume is known. For a simple volume, like that of a cube, we could simply compute the volume from a knowledge of the length of the side $\ell$, as $\Omega=\ell^3$. Conversely, for a given volume we can always compute the length of the side of a cube with the same volume, as $\ell=\Omega^\frac{1}{3}$, which is a kind of ``average size'' of the object. Using the notations introduced in the previous section (Eq.~\ref{lengthscaling}), we can write: 
\begin{equation}
\Big(\frac{Q(0,0)}{Q(0-3,0)}\Big)^\frac{1}{3}= \Big(\frac{m}{\rho}\Big)^\frac{1}{3}= K(1,0)= \ell
\end{equation}
We will use the symbol $\ell$ to denote any length, when no confusion is possible. Once we shall start dealing with multiple such lengths simultaneously, we will introduce more specific notations. In particular, the length built from the quantities $Q_1$ and $Q_2$ shall be called $\ell_{\scaleto{Q_1Q_2}{5pt}}$. So in the example from Archimedes, the average size is $\ell_{\scaleto{m\rho}{5pt}}$, and the volume is $\ell_{\scaleto{m\rho}{5pt}}^3$. Note that the order of the indices does not matter, so  $\ell_{\scaleto{m\rho}{5pt}}=\ell_{\scaleto{\rho m}{5pt}}$. We will come back to this important point in section~\ref{II}. 

Objects can have all sorts of shapes and a different height, width and length. When a measurement of the ``size'' of this object is performed, this measure may not exactly coincide with the size  $\ell_{\scaleto{Q_1Q_2}{5pt}}\equiv (Q_1(x,y)/Q_2(x-n,y))^\frac{1}{n}$. For instance, if the object is spherical its volume will be $\Omega=(4\pi/3) r^3$, where $r$ is the radius, so $\ell_{\scaleto{m\rho}{5pt}}\equiv (4\pi/3)^\frac{1}{3} r \simeq 1.6 r$. If we call the radius $r$ the ``size'', then this size is only approximately given by the ratio of mass and density, $r\simeq (m/\rho)^\frac{1}{3}$. Because tracking the fine effects of shape is often challenging, and because we do not seek precision but generality, we will often rely on this approximate equality sign `$\simeq$'. 

In all generality, lengths could be built by combining an arbitrary number of quantities (geometric, kinematic, mechanical and even beyond) such that the overall dimension is a length. However, in this review we will focus on cases where the decomposition only involves two mechanical quantities. We will call these lengths \textit{simple lengths}, and as we will see they have been useful in a very wide range of situations. Even if we restrict ourselves to the standard mechanical quantities tabulated in Table~\ref{masscary} there are over sixty pairs that can produce lengths. We will only discuss a few, enough to illustrate the generality of this mechanical approach to space: 
\begin{align*}
   & (E/\Sigma)^{\frac{1}{3}} \quad &\text{energy \& stress} \\
    & (E/\Psi)^{\frac{1}{4}} \quad &\text{energy \& force-density}   \\
    & \Sigma/\Psi \quad &\text{stress \& force-density}\\
    & (\Gamma/\Psi)^{\frac{1}{2}} \quad &\text{stiffness \& force-density}  \\
     & \Gamma/\Sigma &\quad \text{stiffness \& stress} \\
     & (E/\Gamma)^{\frac{1}{2}} &\text{energy \& stiffness}\\
     & S/E &\text{strength \& energy}\\
     & H/p &\text{action \& momentum}\\
     & (H/\eta)^\frac{1}{3} &\text{action \& viscosity}\\
     & F/\Gamma \quad &\text{force \& stiffness}\\
     & \zeta/\eta \quad &\text{friction \& viscosity}\\
     & (F/\Sigma)^\frac{1}{2} \quad &\text{force \& stress}
\end{align*}
Each of these characteristic lengths have several famous examples, and some are presented in detail below. We invite the reader to add to this list. 

\subsubsection{Energy and stress: explosions and ideal gases}
\begin{equation}
\ell_{\scaleto{E\Sigma}{4pt}} \equiv \Big(\frac{E}{\Sigma}\Big)^{\frac{1}{3}} \label{ESigma}
\end{equation}
The length $\ell_{\scaleto{E\Sigma}{4pt}}$ applies in particular in the context of explosions. In the introduction we mentioned the scaling derived by Taylor for the dynamics of the radius of an explosion, $d(t)\simeq (E/\rho)^\frac{1}{5} t^\frac{2}{5}$. We will return to this scaling later in the review, it concerns a type of motion  connected to a ratio of energy and density. Evidently, this motion cannot continue indefinitely and eventually a `final blast radius' is reached~\cite{Hopkinson1915,Cranz1926,Sachs1944,Glasstone1977,Westine2012,Kinney2013,Wei2021}. This radius gives the extent of the zone where most damages occur. For the nuclear test studied by Taylor (Trinity), the energy is that of the bomb, around $E\simeq 10^{14}$~J, and the stress is the bulk modulus of the air, which is not far from the atmospheric pressure $\Sigma\simeq 10^{5}$~Pa. Overall, this gives: $\ell \simeq (10^{14}/10^5)^{\frac{1}{3}}\simeq 1$km. \Maf{As a comparison, bombs of 1 Mt, 10 Mt and 100 Mt of TNT (where 1 ton of TNT is equal to 4.184 gigajoules) correspond to blasts radii of 3 km, 8 km, and 16 km respectively~\cite{Glasstone1977}}. The length $\ell_{\scaleto{E\Sigma}{4pt}}$ also applies for blast cavities of underground explosions, where $\Sigma$ is the elastic modulus of the ground materials~\cite{Fokin2000}. 

As illustrated in Fig.~\ref{fig5}a, the same formula can be applied all the way up to supernovae explosions~\cite{Reynolds2008,Asvarov2014}, which release energies on the order of $10^{44}$~J, and can extend their blast to a distance of at least $10^{19}$~m (i.e. over 300 parsecs). This gives a pressure of the interstellar medium of $E/\ell^3 \simeq 10^{44}/10^{19\times 3}\simeq 10^{-13}$ Pa, which is the right order of magnitude~\cite{Asvarov2014}. In regions of interstellar space with even smaller pressures, the supernova remnants can extend even further. 

The length $\ell_{\scaleto{E\Sigma}{4pt}}$ can also be used in situations far from explosions. For instance, in microscopic physics influenced by thermal effects, $E$ can be the thermal energy $E=k_B \Theta$, where $k_B$ is Boltzmann constant and $\Theta$ is the temperature. This Boltzmann constant conveniently allows to translate a temperature into an energy, incorporating thermodynamics into the realm of mechanics. In this context, the equation $\ell\simeq (E/\Sigma)^\frac{1}{3}$ is better known as $\Sigma \ell^3 \simeq k_B \Theta$, which is called the `ideal gas law', and is usually written as `$PV=nk_B T$', where $V/n\simeq \ell^3$ is the average volume associated to each microscopic constituent of concentration, \Maf{$n$ is the number of these constituents, and $P$, $V$ and $T$ are the symbols usually used for pressure, volume and temperature (note that we shall not use these notations in this review)}. This denomination is a bit misleading since this formula is not restricted to ideal gases but can be useful to connect (thermal) energy, pressure, and microscales, for a wide variety of materials. \Maf{For instance, in 1905, Einstein's theory of Brownian motion was developed with the postulate that the osmotic pressure due to dilute suspension of particles was analogous to that of molecular solutes~\cite{Einstein1905}.} In some cases the stress $\Sigma$ can then be interpreted as an elastic modulus. For instance, in polymer physics, $\ell_{\scaleto{E\Sigma}{4pt}}$ gives the typical `blob size'. Assuming room temperature ($E\simeq \Maf{4}~10^{-21}$~J) and a modulus around $\Sigma\simeq 100$~Pa, which is typical for soft gels, then $\ell \simeq (E/\Sigma)^\frac{1}{3}\simeq \Maf{30}$~nm, a scale characteristic of the biological frontier of physics~\cite{Phillips2006}. Clearly, the same formula can underpin very different interpretations. Similar formulas have also been used to explain the size of cells, where the thermal energy is multiplied by the effective number of proteins in the cell~\cite{Guo2017,Xie2018,Adar2020}.

\subsubsection{Energy and force-density: craters and Brownian particles\label{EPsisec}}
\begin{equation}
\ell_{\scaleto{E\Psi}{4pt}} \equiv \Big(\frac{E}{\Psi}\Big)^{\frac{1}{4}}  \label{EPsi}
\end{equation}
The length given in this equation describes a situation where energy is balanced by force-density. The force-density is usually the weight per unit volume, that is $\Psi=\rho g_0$, where $g_0$ is the standard acceleration of gravity. This length scale is for instance relevant to the size of the crater of an explosion~\cite{Housen1983,Holsapple1993}, if the density $\rho$ is taken at the value of the ground materials. Indeed, this type of simple length is used to study all sorts of craters from explosions or impacts, including those of asteroids on the moon, as illustrated in Fig.~\ref{fig5}c~\cite{Katsuragi2016}. \Maf{Studies of cratering in granular media also found the same length scale, when the impactors have moderate speeds~\cite{Uehara2003,Takita2013}. The same scaling can also be applied for the size of cavity created by explosions at the surface of liquids~\cite{Benusiglio2014}. }

The length $\ell_{\scaleto{E\Psi}{4pt}}$ can also be used to describe the average `height of a Brownian particle' in sedimentation or centrifugation. There, $E$ is the thermal energy and the force-density is $\Psi\simeq \rho g$, with $g=g_0$ for sedimentation and $g= r \omega^2$ for centrifugation (where $r$ is the distance from the axis of rotation at rate $\omega$)~\cite{Sharma2009}. For instance, at room temperature $E\simeq \Maf{4}~10^{-21}$~J, and if $\rho\simeq 10^3$~kg/m$^3$, then $(E/\rho g_0)^\frac{1}{4}\simeq \Maf{0.8}~\mu$m. Particles below this size are `Brownian', \Maf{and they remain suspended or dispersed, with a number distribution at any height set by the sedimentation-diffusion equilibrium~\cite{Perrin1926}.}

Note that for objects embedded in a fluid, the force-density will generally be built from `buoyancy',  i.e. from the difference in density with the surrounding medium $\Psi\simeq \mid\rho_f -\rho\mid g$. In particular, in the case of Brownian particles, the density difference between dispersed particle and outside medium determines the length scale that can be identified as the upper limit to the size of  Brownian particles. The argument explains why metal nanoparticles are Brownian only below 100 nm, whereas polymer microbeads can be over a micron. Centrifugation can provide a much larger value of effective g, and therefore leads to sedimentation and separation~\cite{Sharma2009}. 

An example of characteristic size of the form $\ell_{\scaleto{E\Psi}{4pt}}$ for which the force-density is not the weight density occurs for drop impact. When the viscosity of the drop is negligible in contrast to its inertia and surface-tension, \Maf{some studies have found that} the maximum drop radius after impact is given by $(E/\Psi)^\frac{1}{4}$, where $E\simeq \rho r^3 u^2$ is the kinetic energy of the impacting drop of radius $r$, speed $u$ and density $\rho$. The force-density $\Psi\simeq \Gamma/r^2$ originates from capillarity~\cite{Clanet2004}. \Maf{We will see in section~\ref{EGamsec} that other studies suggest a different scaling.} 

\subsubsection{Stress and force-density: the hydrostatic equilibrium}
 \begin{equation}
\ell_{\scaleto{\Sigma\Psi}{4pt}} \equiv \frac{\Sigma}{\Psi} \label{SigmaPsi}
\end{equation}
In this length, the force-density is here again often the weight per unit volume, giving $\ell \simeq \Sigma/\rho g_0$. The simplest examples consider that the stress is the isotropic pressure. Such formula is particularly useful in the context of the formation of astronomical objects like planets or stars~\cite{Choudhuri2010}. In these cases, the size of a ball of matter is understood as a compromise between the compression by gravity and the resistance of an internal pressure. The size is then that of the planet or star. In this context, Eq.~\ref{SigmaPsi} is sometimes referred to as `the hydrostatic equilibrium' and written as $\Sigma\simeq \ell \Psi$, where the internal pressure equilibrates the weight per unit volume~\cite{Kippenhahn1990}, as in the context of barometers where pressure was first defined~\cite{Frontali2013}. As an example, in the case of Earth, $\rho g_0\simeq 5~10^4$~N/m$^3$, and $\ell \simeq 6~10^6$~m, giving $\Sigma\simeq 10^{11}$~Pa. Such pressure typically corresponds to the elastic modulus of metallic or amorphous solids constituting the Earth ($\simeq 100$GPa). Through this scaling, the various sizes of astronomical objects shown in Fig.~\ref{fig5}b are directly related to their densities and elastic moduli~\cite{Choudhuri2010}. For more information, we refer the reader to a pedagogical presentation on scaling approaches to the size of stars and planets, which some of us have recently published~\cite{Fardin2022s}. 

\subsubsection{Stiffness and force-density: the capillary length}
\begin{equation}
\ell_{\scaleto{\Gamma\Psi}{4pt}} \equiv  \Big(\frac{\Gamma}{\Psi}\Big)^{\frac{1}{2}} \label{GammaPsi} 
\end{equation}
Usually $\Psi\simeq\rho g_0$ and the stiffness $\Gamma$ is typically understood as a surface energy, also called `surface-tension'. In the context of the wetting of fluids, such length is called the `capillary length'~\cite{PGG2013}. This capillary length sets the scale where surface energy and gravity are of comparable influence. For instance, for the water-air interface, the typical surface-tension is $\Gamma\simeq 7~10^{-2}$~N/m, the density is $\rho\simeq 10^{3}$~kg/m$^3$, and $g_0\simeq 9.8$~m/s$^2$, such that the capillary length is around 3~mm. On the moon, where $g\simeq 1.6$~m/s$^2$, the capillary length of water is more than twice bigger.  

When a drop hangs from a leaf as in Fig.~\ref{fig5}d, it may grow in size only up to the capillary length, after which it will fall. Generally, gravitation flattens drops of size larger than $\ell_{\scaleto{\Gamma\Psi}{4pt}}$, \Maf{forming puddles}, while capillarity keeps smaller drops spherical. The capillary length also influences the shape of a meniscus near an immersed or floating object~\cite{PGG2013}. Understanding the interplay of gravity and capillarity can actually be used to determine surface-tension, using pendant drop analysis or using capillary rise~\cite{PGG2013}.

 
\subsubsection{Stiffness and stress: the elasto-capillary and elasto-adhesive lengths}
\begin{equation}
\ell_{\scaleto{\Gamma\Sigma}{4pt}} \equiv  \frac{\Gamma}{\Sigma} \label{GammaSigma} 
\end{equation}
This length provides a balance between stiffness (i.e. surface energy) and stress (i.e. volume energy). The \Maf{relevant surface energy can be dominated by the contact with a solid substrate (in which case it is sometimes called `adhesion energy'), or with a fluid medium (in which case it is usually referred to as `surface-tension')}. Since $\Gamma$ is in the numerator, it is the ``driving'' term. Larger values of $\Gamma$ lead to larger lengths. One particular instance of this kind of length scale is when the stiffness is understood as an adhesion energy, and when the stress is elastic. The elasticity generates a recoil that is balanced by adhesion. In this context, $\ell_{\scaleto{\Gamma\Sigma}{4pt}}$ may be called the `elasto-adhesive' length~\cite{Creton2016}.  

When $\Gamma$ comes from surface-tension, the length $\ell_{\scaleto{\Gamma\Sigma}{4pt}}$ is often called the `elasto-capillary length'~\cite{Bico2018}. This length is relevant for the spreading of drops on soft substrates, associating surface energy and elasticity~\cite{Andreotti2020}. The elasticity can be that of the spreading object or of its environment. For instance,  as illustrated in Fig.~\ref{fig5}e,  $\ell_{\scaleto{\Gamma\Sigma}{4pt}}$ applies to the height of the wetting ridge near the solid-liquid-air triple line~\cite{Jerison2011}, where the surface-tension $\Gamma$ acts perpendicularly to a substrate of elastic modulus $\Sigma$. Experimentally, a drop of glycerol ($\Gamma\simeq 63$~\Maf{mN/m}) on a soft silicone gel ($\Sigma\simeq 2.4$~kPa) produces a ridge of about 12~$\mu$m~\cite{Coux2020}. Other orders of magnitude can be obtained, for instance a length of 30~nm was found for tricresyl phosphate ($\Gamma\simeq 28.5$~\Maf{mN/m}) on a silicone elastomer ($\Sigma\simeq 0.6$~MPa)~\cite{Carre1996}.  

\subsubsection{Energy and stiffness: Scheludko-Vrij length\label{EGamsec}}
\begin{equation}
\ell_{\scaleto{E\Gamma}{4pt}} \equiv \Big(\frac{E}{\Gamma}\Big)^\frac{1}{2} \label{EGam}
\end{equation}
This length is particularly relevant for thin films, where it can be called the `Scheludko-Vrij length'~\cite{Sheludko1967,Nikolov2014}. One example considers that the energy $E$ comes from van der Waals interactions and is called the `Hamaker constant'~\cite{PGG2013,Israelachvili2015}. For typical fluids this length is around a few angstroms. In this context, one usually defines a `disjoining pressure' $\Sigma\simeq E/\ell^3$, where $\ell$ is the film thickness. Disjoining pressure and the Hamaker constant play an important role in the climbing and spreading of thin films~\cite{Leger1992,Popescu2012,Starov2019}, and in setting the nano-topography of foam films~\cite{Zhang2018}. 

Note that the linear stability analysis of both freestanding and supported ultra-thin films results in a prediction of a spinodal-like instability into thick-thin regions, with a typical size $\ell_{\scaleto{\Gamma\Psi}{4pt}}$ given in Eq.~\ref{GammaPsi}, where the force-density is defined as the gradient of disjoining pressure, i.e. $\Psi\simeq E/\ell_{\scaleto{E\Gamma}{4pt}}^4$. This length scale has been observed experimentally in spinodal dewetting and spinodal stratification~\citep{Kalliadasis2007,Yilixiati2019}.

\Maf{As mentioned in section~\ref{EPsisec}, in the context of drop impact, when the viscosity of the drop is negligible, some studies have found that the maximum drop radius after impact is a simple length of the form $\ell_{\scaleto{E\Psi}{4pt}}$, where $E$ is the kinetic energy of the drop. In contrast, other studies suggest that the maximum radius may be of the form given by $\ell_{\scaleto{E\Gamma}{4pt}}$~\cite{Bennett1993,Eggers2010,Laan2014}}. 

\subsubsection{Strength and energy: Bohr radius and Bjerrum length\label{Bohrr}}
\begin{equation}
\ell_{\scaleto{SE}{4pt}}  \equiv \frac{S}{E}\label{SE}
\end{equation}
This length has very deep roots since it can be used to express the size of the atom. This length also gives us the opportunity to say a few words about the mechanical quantity $S=Q(3,-2)$. To the best of our knowledge this quantity does not have a standard name in the literature. In the context of the deformations of elastic beams, it is sometimes called the `flexural rigidity'~\cite{Landau1986}. We call $S$ the \textit{strength} because it is often used to compare the relative strength of fundamental forces~\cite{Dirac1937}. Newton's force of gravity \Maf{between two masses $m_1$ and $m_2$ separated by a distance $r$} can be expressed as $Gm_1 m_2/r^2$, whereas Coulomb's force \Maf{between two charges $q_1$ and $q_2$} can be expressed as $k_C q_1 q_2/r^2$, where $G$ and $k_C=1/4\pi \epsilon_0$ are respectively the gravitational and Coulomb constants (with $\epsilon_0$ the vacuum permittivity). Both $Gm_1 m_2$ and $k_C q_1 q_2$ have the dimensions of a strength. Most notably, if the charges $q_1$ and $q_2$ are elementary $S_0=k_C q_1 q_2=k_C e^2\simeq 2~10^{-28}$~kg.m$^3$.s$^{-2}$. In the microscopic realm one says that electromagnetism has a greater strength than gravity because $S_0\gg Gm^2$, where $m$ is for instance the mass of a proton. 

The length $\ell_{\scaleto{SE}{4pt}}$ can then be used to express the size of an atom, using $S_0$ and the Hartree energy $E=m_e u^2$, where $m_e$ is the mass of the electron and where $u=\alpha_0 c$ is the semi-classical speed of the electron ($\alpha_0\simeq 1/137$ is the `fine structure constant' we shall discuss later, $c$ is the speed of light). Under these assumptions Eq.~\ref{SE} gives \Maf{$\ell=e^2/(4\pi \epsilon_0 m_e u^2)$, which is the first historical expression of the `Bohr radius'~\cite{Griffiths2018} (an alternate way of writing this radius will be given in the next sub-section).} 

In plasma and electrolytes, the strength $S_0$ also appears in the definition of the Bjerrum and Debye lengths~\cite{Israelachvili2015}. The `Bjerrum length' follows Eq.~\ref{SE} with $S=S_0/\epsilon_r$, which takes into account the dimensionless relative dielectric constant $\epsilon_r$, and $E=k_B \Theta$. \Maf{This length provides the scale at which the electrostatic and thermal effects are of comparable influence. It arises in the context of electrolytes and colloidal dispersions~\cite{Israelachvili2015,Muthukumar2023}}. For water at room temperature, $\epsilon_r\simeq 80$ and the Bjerrum length is around 0.7~nm. The different assumptions leading from $S/E$ to the Bohr or Bjerrum lengths are summarized here: \Maf{
\begin{equation}
 \ell =\frac{S}{E} \&
\left\{
  \begin{array}{lr}
    S=S_0~\&~E=m_e u^2  &\Rightarrow \ell = \frac{e^2}{4\pi\epsilon_0 m_e u^2}  \\
    S=\frac{S_0}{\epsilon_r}~\&~E= k_B \Theta &\Rightarrow \ell=\frac{e^2}{4\pi\epsilon_0 \epsilon_r k_B \Theta}
  \end{array}
\right.
\end{equation}
}
\Maf{In the context of plasmas and electrolytes the `Debye length' also involves a strength $S_0$. The Debye length characterizes the screening of electrostatic interactions between two charges in the presence of other charges. In colloidal dispersions, as the Debye length depends on ion concentration, it is a property of the solution, unlike the Bjerrum length which depends on the solvent and its dielectric constant~\cite{Muthukumar2023}}. In this context one can define an effective stiffness as a \textit{density of strength}, $\Gamma=S_0/r^3$, which is then combined with the thermal energy using Eq.~\ref{EGam}. Here, the distance $r$ is the mean distance between electrons and $1/r^3$ is the electron number density, so $\Gamma$ can be understood as a charge density expressed in units of mass, length and time. The Debye length can vary widely, from atomic scale in the solar core to thousands of kilometers in the intergalactic medium. In electrolyte media, encountered in soft matter and within cells, the strength is built from the number density of ions, whereas in semiconductors, the number density of dopants makes the relevant contribution~\cite{Robinson2012,Clemmow2018}. 

\subsubsection{Action and momentum: Bohr radius and de Broglie wavelength}
\begin{equation}
\ell_{\scaleto{Hp}{4pt}}  \equiv \frac{H}{p}\label{Hp}
\end{equation}
We have seen in the preceding sub-section that the very size of the atom can be expressed as a ratio of two mechanical quantities. More precisely, the Bohr radius can be expressed as a ratio between the electromagnetic strength and the kinetic energy of the electron, $\ell\simeq S_0/E$. The kinetic energy can be written in terms of the mass $m_e$ of the electron and its speed $u=\alpha_0 c$, so $\ell\simeq \Maf{S_0}/(m_e u \alpha_0 c)\simeq S_0/(p\alpha_0 c)$, where $p=m_e u$ is the momentum of the electron. Historically, the dimensionless `fine structure constant' $\alpha_0$ was understood precisely in this fashion, as the ratio between the speed of the electron and the speed of light~\cite{Kragh2003}. However, $\alpha_0$ quickly showed up in other situations, and in particular in the comparison between the electromagnetic strength $S_0$ and the nuclear strength $\hbar c$, since $\alpha_0 = S_0/\hbar c$. Using this formula we can rewrite the Bohr radius as $\ell\simeq \hbar/p$. Expressed in this way the Bohr radius is understood as the `de Broglie wavelength of the electron'. Generally, when $H$ is the quantum of action $\hbar$, Eq.~\ref{Hp} encompasses one of the central concept of quantum mechanics, the relationship between waves and particles~\cite{deBroglie1925}. 

We have seen that the Bohr radius can be expressed by two different pairs of mechanical quantities, $S$ and $E$ or $H$ and $p$. The existence of multiple mechanical decompositions is not at all special to this case. Any length can always be decomposed into a ratio of two mechanical quantities, but this decomposition is not unique, and this plurality encourages a diversity of mechanical models. The pair chosen in a particular situation depends on the greater context where the length is found, and on historical circumstances. A more complete investigation of this ``plurality'' would require more than two mechanical quantities and is therefore out of the scope of this review. We will say a few more words about this in the conclusion. 

\subsubsection{Action and viscosity: Viscosity as a density of action}
\begin{equation}
\ell_{\scaleto{H\eta}{4pt}}  \equiv \Big(\frac{H}{\eta}\Big)^\frac{1}{3} \label{Heta}
\end{equation}
As we said in section~\ref{sec1}, a mechanical quantity of the form $Q(x-3,y)$ can always be thought as a 3D density of the quantity $Q(x,y)$. This is famously true for the density itself, which is a mass-density and the template for all the others. This is also true for the stress, which can be understood as a density of energy, as we saw with Eq.~\ref{ESigma}. It is also true in less traditional cases, as with the viscosity $\eta\equiv Q(2-3,-1)$, which can be thought of as a density of action $H\equiv Q(2,-1)$. Take water as an example. Water has a viscosity around $\eta\simeq 10^{-3}$~Pa.s (where 1~Pa.s$\equiv$1~kg.m$^{-1}$.s$^{-1}$). A water molecule has a radius around $\ell\simeq 1$~\r{A}. So if we multiply the viscosity by the typical volume of a water molecule we get an action: $H\simeq \eta \ell^3 \simeq  10^{-33}$~J.s. This value is not far from the quantum of action, $\hbar\simeq 10^{-34}$~J.s, so the viscosity of water almost corresponds to one quantum of action per molecule. \Maf{Although this link between viscosity and action is here only sketched by rough orders of magnitude, it can be formalized more rigorously, as exemplified in a recent paper by~\citet{Trachenko2020}. }

\subsubsection{Old lengths under new light}
\begin{align}
\ell_{\scaleto{F\Gamma}{4pt}}&  \equiv \frac{F}{\Gamma} \label{FGamma}\\
\ell_{\scaleto{\zeta\eta}{4pt}}&  \equiv \frac{\zeta}{\eta} \label{zetaeta}\\
\ell_{\scaleto{F\Sigma}{4pt}}&  \equiv \Big(\frac{F}{\Sigma}\Big)^\frac{1}{2} \label{FSig}
\end{align}
These last examples provide ratios that are well known but often represented differently. 

In Eq.~\ref{FGamma}, $\Gamma$ can be interpreted as the stiffness of a material behaving as a spring, then Eq.~\ref{FGamma} is just Hooke's law, $F= \Gamma\ell_{\scaleto{F\Gamma}{4pt}}$, where $F$ and $\ell_{\scaleto{F\Gamma}{4pt}}$ are usually understood as variable. In the context of spreading drops or cells, this ratio can state a balance between a driving force $F$ and a surface-tension or stiffness $\Gamma$. For cell spreading, the length $\ell_{\scaleto{F\Gamma}{4pt}}$ can be used to characterize the portion of the cell behind the edge, which is rich in a very dynamic polymer called `actin'~\cite{Mitchison1996,Roberts2002,Pollard2009}. The polymerization of actin can be associated with a `protrusion force' $F$, which is balanced by a surface energy $\Gamma$, with contributions form the plasma membrane, the cell stiffness, and the adhesion with the substrate~\cite{Cuvelier2007,Fardin2010}. 

In Eq.~\ref{zetaeta}, the length is $\ell_{\scaleto{\zeta\eta}{4pt}}\Maf{\equiv} \zeta/\eta$, that is a ratio between a `friction' or `mobility' and a viscosity. This equation is more often seen in the form $\zeta=\eta\ell_{\scaleto{\zeta\eta}{4pt}}$, in the context of Stokes drag~\cite{Stokes1850,Landau1959}, where it gives the effective friction $\zeta$ on an object of size $\ell_{\scaleto{\zeta\eta}{4pt}}$ moving slowly in a fluid of viscosity $\eta$. Indeed, for high viscosity and low speed, the friction force is proportional to speed $u$, and given by $F\simeq \zeta u\simeq \eta \ell_{\scaleto{\zeta\eta}{4pt}} u$. This connection is the basis for Brownian motion in the Stokes Einstein relation~\cite{Landau1959}, and thus lies at the heart of colloidal physics and chemistry~\cite{Russel1991}. 

In Eq.~\ref{FSig}, a length is defined as the ratio between a force and a stress. This formula is more often seen as $\Sigma\simeq F/\ell^2$, which defines a stress from the force $F$ on the area $\ell^2$. From this perspective, the stress is typically understood as intensive, whereas the force is extensive but normalized by the area. When $F$ and $\Sigma$ are independent constants, $\ell_{\scaleto{F\Sigma}{4pt}}$ is a simple length in its own right. This is for instance the case in the physics of nematic and polar materials, which includes a large class of living systems~\cite{Marchetti2013,Schwarz2013}. In this context, $\ell_{\scaleto{F\Sigma}{4pt}}$ is sometimes called the `nematic length', where $F$ is understood as the `Frank constant', which represents a 1D elasticity associated with differences in alignment, and where $\Sigma$ represents the energy per unit volume associated with the alignment of the nematic components~\cite{PGG1993}. This length scale gives the typical extent of orientational boundary layers~\cite{Marchetti2013}. Another important simple length in the study of active matter is the crossover from `wet' to `dry' active particles, which can be written as $\ell_{\scaleto{\eta\chi}{4pt}}$ where $\eta$ is the viscosity of the embedding fluid, and $\chi$ is a bit understood as a so-called `frictional drag'~\cite{Marchetti2013}.

\subsection{Simple times\label{times}}
\href{https://youtu.be/KNeNx9mi2ao?si=CSSXh1kCe7_DcNKv}{Mechanics 3: Simple Times}\\

We have seen that ratios of mechanical quantities can produce length scales that show up in a wide variety of situations. In these examples a length emerges out of a kind of ``balance'' between conflicting ``forces'', where the term ``force'' is here used quite generously to encompass any mechanical quantity ($\mathcal{M}\mathcal{L}^x\mathcal{T}^y$). Similarly, pairs of mechanical quantities can be used to understand time, durations and periods, leading to what we can call \textit{simple times}. We will use the symbol $\tau$ when no ambiguity is possible, and $\tau_{\scaleto{Q_1Q_2}{5pt}}$ when specificity is required. Over thirty such simple times can be built from the standard quantities of Table~\ref{masscary}. We list here the ones we shall discuss in this section: 
\begin{align*}
  & (m/\Gamma)^\frac{1}{2} \quad &\text{mass \& stiffness} \\
   & E/P \quad &\text{energy \& power} \\
   & H/E \quad &\text{action \& energy} \\
    & \eta/\Sigma \quad &\text{viscosity \& stress}\\
    & \Phi/\eta \quad &\text{normal stress coefficient \& viscosity} \\
    & \zeta/\Gamma \quad &\text{friction \& stiffness} \\
     & \rho/\chi &\quad \text{density \& density variation}
\end{align*}

\subsubsection{Mass and stiffness: Hooke-Rayleigh time}
\begin{equation}
\tau_{m\Gamma} \equiv \Big(\frac{m}{\Gamma}\Big)^\frac{1}{2} \label{mGamma} 
\end{equation}
This time is the archetypal example of a simple time. When $\Gamma$ is interpreted as the stiffness of a spring from which is attached a mass $m$, Eq.~\ref{mGamma} is the familiar expression of the period of oscillation. The standard formula found in textbooks usually used the symbol `$k$' instead of $\Gamma$, and includes a prefactor of $2\pi$, so $\tau_{m\Gamma}$ is more precisely the inverse of the `angular frequency'. 

 In the context of the dynamics of droplets, the mass is usually given by $m\simeq \rho r^3$, where $\rho$ is the density of the fluid and $r$ is the radius of the droplet. In this context one speaks of the `Rayleigh time'~\cite{Rayleigh1879}, which applies to the oscillation frequency of drops, as well as to the contact time of rebounding drops~\cite{Richard2002}. Despite very different rebound profiles depending on the impact speed, the contact time remains the same and is set by $\tau_{m\Gamma}$. The timescale also appears in capillarity-driven flows of `inviscid fluids' (i.e. negligible viscosity)~\cite{Rayleigh1879,Middleman1995,Eggers1997,McKinley2005,Fardin2022}.

\subsubsection{Energy and power: Energy consumption and Ritter-Kelvin-Helmholtz time}
\begin{equation}
 \tau_{EP} \equiv \frac{E}{P} \label{EP}
\end{equation}
The power relates to a transfer or conversion of energy over time, and so the dimension of $E/P$ is naturally $\mathcal{T}$. Common units of energy like the kilowatt-hour reflect this proximity, with 1~kWh$=3.6~10^6$~J.s$^{-1}$.s, so simply $3.6~10^6$~J. For a given energy $E$, the time scale in Eq.~\ref{EP} gives the time range to be expected when the energy consumption rate is the power $P$. This time scale can be used for a wide variety of purposes, to estimate how long you can drive on a full tank, as well as the life expectancy of the Sun. 

A typical small car will have something like 70 horsepower, so $P\simeq 53~10^3$~W. The energy comes from the fuel. Assuming a gas tank of 35 liters of standard fuel, with $8.9$~kWh/liter, yields $E\simeq 10^9$~J per gas tank. Then, $\tau_{EP}\simeq 5$~hours. This is roughly how long this car can drive without refueling. 

The principle behind the formula in Eq.~\ref{EP} remains the same for all kinds of fuel and all kinds of systems consuming this fuel. In particular, this formula can also be used to obtain an estimate of the lifetime of a star like the Sun. In this case, the power is well estimated by the solar luminosity, and $P\simeq 3.8~10^{26}$~W~\cite{Kippenhahn1990}. If the fuel of the Sun was standard gasoline as in the car, then the lifetime of the Sun would only be around 3000 years, according to Eq.~\ref{EP}. This is obviously not the case. 

So what is the fuel of the Sun? The quest to answer this question spanned from the mid 19th to the mid 20th century and involved some of the greatest minds of this time. The story is told beautifully in a paper by~\citet{Shaviv2008}. An important step in the quest was to consider the energy to be due to the self-gravitation of the Sun, so $E\simeq Gm^2/\ell$, where $m$ and $\ell$ are respectively the mass and size of the Sun. In this scenario, the power of the Sun, that is its luminosity is due to the gravitational potential energy. This time scale is sometimes called the `thermal timescale', or the `Kelvin-Helmholtz timescale', to honor Kelvin and Helmholtz contributions to this field of research. However, as noted by\Maf{~\citet{Shaviv2008}}, August Ritter was the first to derive this formula. This timescale plays an important role in astronomy, in particular to set the timescale of the collapse of protostars, however, it fails to estimate the age of the Sun and similar stars. Indeed, using $\ell\simeq 7~10^8$~m and $m\simeq 2~ 10^{30}$~kg, we get $\tau_{EP}\simeq 30$~million years. The inadequacy of this figure with the geological records led to intense debate, and the controversy was only resolved at the beginning of the 20th century, when it was realized that the fuel of the Sun is nuclear~\cite{Shaviv2008}. By considering \Maf{the} conversion of hydrogen into helium, it was estimated that the energy of the Sun is around $E\simeq 6~10^{-4} mc^2$, giving $\tau_{EP}\simeq 10$~billion years, which is the currently accepted order of magnitude, and is sometimes called the `nuclear time scale'~\cite{Kippenhahn1990}. 

\subsubsection{Action and energy: Planck relation}
\begin{equation}
 \tau_{HE} \equiv \frac{H}{E} \label{HE}
\end{equation}
Staying on the same column of Table~\ref{masscary} than in the previous example, we have the pair combining action and energy. The part of physics where a constant action is most dramatically felt is quantum mechanics, where the action is the Planck constant $\hbar$. Using this value, we can rearrange Eq.~\ref{HE} to express the energy from the Planck constant and the inverse of the time, which is usually written as a frequency, $E=\hbar \omega$. This equation started the whole quantum revolution, it is the Planck relation, which gives the energy of a photon of frequency $\omega$, or the frequency from the energy. This is the formula behind Einstein's Nobel prize on the photoelectric effect~\cite{Einstein1905}. With this relationship, Einstein calculated the frequency of a photon required to eject an electron from a metallic target. For instance, if the target is made of Zinc, the binding energy of an electron is around $E\simeq 9$~electronvolts, so $E\simeq 10^{-18}$~J. Thus, according to Eq.~\ref{HE} the frequency of light above which electrons can be extracted is around $10^{16}$~Hz, corresponding to ultraviolet light. 

In the special case where the energy is the thermal energy ($E=k_B\Theta$), the time $\tau=\hbar/k_B\Theta$ is called the `Debye time'~\cite{Kittel2005}. \Maf{The inverse of this time scale provides an important thermal cut-off in the propagation of waves in crystal lattices.} At room temperature, the Debye time is around twenty femtoseconds. Note that the term `Debye time' can also be used in a slightly different way~\cite{Bazant2004}. The two formulas could be reconciled by using the relationship between viscosity and action, as given in Eq.~\ref{Heta}. 

\subsubsection{Viscosity and stress: rheological time}
\begin{equation}
   \tau_{\eta\Sigma} \equiv \frac{\eta}{\Sigma} \label{etaSigma}
\end{equation}
This ratio most notoriously apply to Newton's relation, $\Sigma\simeq\eta\dot\gamma$, where $\dot\gamma$ is the deformation rate. The time scale is then $\dot\gamma^{-1}$. In general, the deformation rate is not a constant. However, in complex fluids there are often remarkable values of $\dot\gamma$. For instance, many materials display rather elastic properties on short time scales, and are viscous on longer time scales~\cite{Larson1999}. These materials are usually called `visco-elastic', and the threshold between short and long time scales is the `relaxation time' $\tau$. In `Maxwell's model', which is the simplest model of visco-elastic fluid, the elasticity of the material $\Sigma$, the viscosity, and the relaxation time are connected by the equation $\tau\simeq \eta/\Sigma$~\cite{Bird1987,Larson1999}. The greater the viscosity the longer the time, and the greater the elasticity the shorter the time. The relaxation time scale of visco-elastic fluids can range from milliseconds to decades~\cite{Bird1987,Larson1999}. At any rate, in simple visco-elastic fluids the time $\tau$ is a constant of the material and it can be used to understand the transition between different flow regimes~\cite{Larson1999,McKinley2005,Fardin2014}. 

In more complex visco-elastic fluids beyond Maxwell's model, there can be more than one relaxation time~\cite{Doi1988,Larson1999}. Polymer solutions typically have a spectrum of relaxation times. In addition, some materials may behave as Maxwell fluids under small deformations, but display flow-induced changes in their structure at higher deformations. For instance, wormlike micelles~\cite{Larson1999} solutions have a viscosity $\eta_1$ at low deformation rates, and above a threshold $\dot\gamma_1$ a different flow-induced ``phase'' of viscosity $\eta_2$ is generated and coexists at constant stress $\Sigma_p$ with the original one until a second threshold $\dot\gamma_2$. Above this threshold the whole material has viscosity $\eta_2$. This phenomenon is usually called `shear-banding'~\cite{Divoux2016}. Both $\dot\gamma_1^{-1}\simeq\eta_1/\Sigma_p$ and $\dot\gamma_2^{-1}\simeq\eta_2/\Sigma_p$ are time scales of the form $\eta/\Sigma$. 
\Maf{More broadly, remarkable values of viscosity or stress can occur in a large class of `complex' or `non-Newtonian' fluids, including shear-thinning, shear-thickening and yield-stress fluids. In turn, these quantities provide multiple ways to construct time scales of the form given in Eq.~\ref{etaSigma}~\cite{Larson1999}. }

\subsubsection{Viscosity and normal stresses: Weissenberg time}
\begin{equation}
\tau_{\Phi\eta} \equiv \frac{\Phi}{\eta} \label{Phieta} 
\end{equation}
In addition to the visco-elastic time scales they are often associated with, non-Newtonian materials can also display quite remarkable `normal stress effects'~\cite{Bird1987,Larson1999}.  In `Newtonian fluids', shear stresses are of the form $\Sigma_s \simeq \eta\dot\gamma$, where $\dot\gamma\simeq u/\ell$ is a velocity gradient over the distance $\ell$. In contrast, normal stresses come from inertia and are usually of the form $\Sigma_n\simeq\rho u^2\simeq\rho \ell^2 \dot\gamma^2$. This stress is sometimes called the `dynamic pressure', or the `Ram pressure' in astrophysics~\cite{Clarke2007}. For Newtonian or non-Newtonian fluids, normal stresses can be expressed as $\Sigma_n\simeq\Phi \dot\gamma^2$, such that $[\Phi]=\mathcal{M}\mathcal{L}^{-1}$, the same dimensions as a 1D mass-density. In the Newtonian case, $\Phi\simeq\rho \ell^2$, but in non-Newtonian fluids, including magnetic fluids relevant to astrophysics, the normal stress coefficient $\Phi$ can be completely disconnected from $\rho$ and inertia in general~\cite{Ogilvie2003}. Whereas the positive value of $\Phi$ for Newtonian fluids tend to generate centrifugal forces pushing a rotating fluid outward, for non-Newtonian fluids $\Phi$ can be negative and push the material inward in the so-called `rod-climbing' or `Weissenberg effect'~\cite{Bird1987,Larson1999}. This is but one among many examples of non-Newtonian normal stress effects. 

For non-Newtonian fluids, the normal stress coefficient $\Phi$ is a material property as important as the viscosity, and disconnected from the density. It is a mechanical quantity of its own, from which a time scale $\tau_{\Phi\eta}$ can be constructed, as in Eq.~\ref{Phieta}. In simple visco-elastic models like Maxwell's model, this time scale is identical to the relaxation time $\tau_{\eta\Sigma}$. Indeed, in Maxwell's model, one has $\Phi\simeq\eta \tau_{\eta\Sigma}$~\cite{Larson1999}. This identity is not true in general. Currently, the differences between the two non-Newtonian time scales are most often investigated in the context of flows with an extensional component, where the dual effects of normal stresses and relaxation time are factored into the differences between shear and extensional rheology~\cite{McKinley2005,Dinic2020}. 

\subsubsection{Friction and stiffness: damping time}
\begin{equation}
 \tau_{\zeta\Gamma}\equiv \frac{\zeta}{\Gamma}\label{zetaGamma} 
\end{equation}
One way to understand this time scale is as the 2D equivalent of $\tau_{\eta\Sigma}$. For fluid films, the details of the dynamics of the height can usually be neglected when the horizontal extent is much larger than the thickness. Under such `lubrication approximation'~\cite{Oron1997,Hamrock2004,Leal2007}, $ \tau_{\zeta\Gamma}$ can be understood as the characteristic time separating the short time dynamics driven by the stiffness, and the long time scales dominated by friction. In the most elementary expression of this time scale, $\Gamma$ is the stiffness of a spring, and $\zeta$ is the damping coefficient. If the spring is initially compressed, it first snaps back fast until a cross-over time $\zeta/\Gamma$, after which it relaxes more slowly. A time scale of this nature is for instance seen for the dewetting time of islands of cells on unwelcoming substrates~\cite{Perez2019}. In this situation, a monolayer of cells progressively retracts into a 3D aggregate. In this context, the friction is $\zeta\simeq \eta h$, where $h$ is the cell height, and the stiffness is the `tension' over the portion of the monolayer close to the edge, $\Gamma\simeq \Sigma \ell$, where $\Sigma$ is the `traction stress' exerted by the cells on the substrate, and the width $\ell$ near the edge is given by the nematic length discussed with Eq.~\ref{FSig}. 

\subsubsection{Density and its variation: proliferation time}
\begin{equation}
\tau_{\rho\chi} \equiv \frac{\rho}{\chi} \label{rhodotrho} 
\end{equation}
Because the standard name we have chosen for the mechanical quantity $\chi\equiv Q(-3,-1)$ is \textit{density change}, the fact that this ratio is a time scale seems trivial. It is the time scale over which the density changes. This time scale is particularly useful in dynamics due to the proliferation of objects with a characteristic mass $m$ and a `number density' $n$ (dimension $\mathcal{L}^{-3}$ or $\mathcal{L}^{-2}$ in 2D). If the mass is constant, then $\tau_{\rho\chi}\Maf{\equiv}\rho/\chi\simeq n/\dot{n}$, where $\dot{n}\simeq \partial n/\partial t$. Thus, the time scale reflects the rate of change of the number of objects. For instance in tissues of cells the time scale $\tau_{\rho\chi}$ is connected to the characteristic time separating two cell divisions. This time scale is relevant for the spreading of tissues~\cite{Puliafito2012}, as well as for some organisms like ants~\cite{Mlot2011}.

\section{The mechanics of motion\label{simplespreadings}}
\begin{figure*}
\centering
\includegraphics[width=15cm,clip]{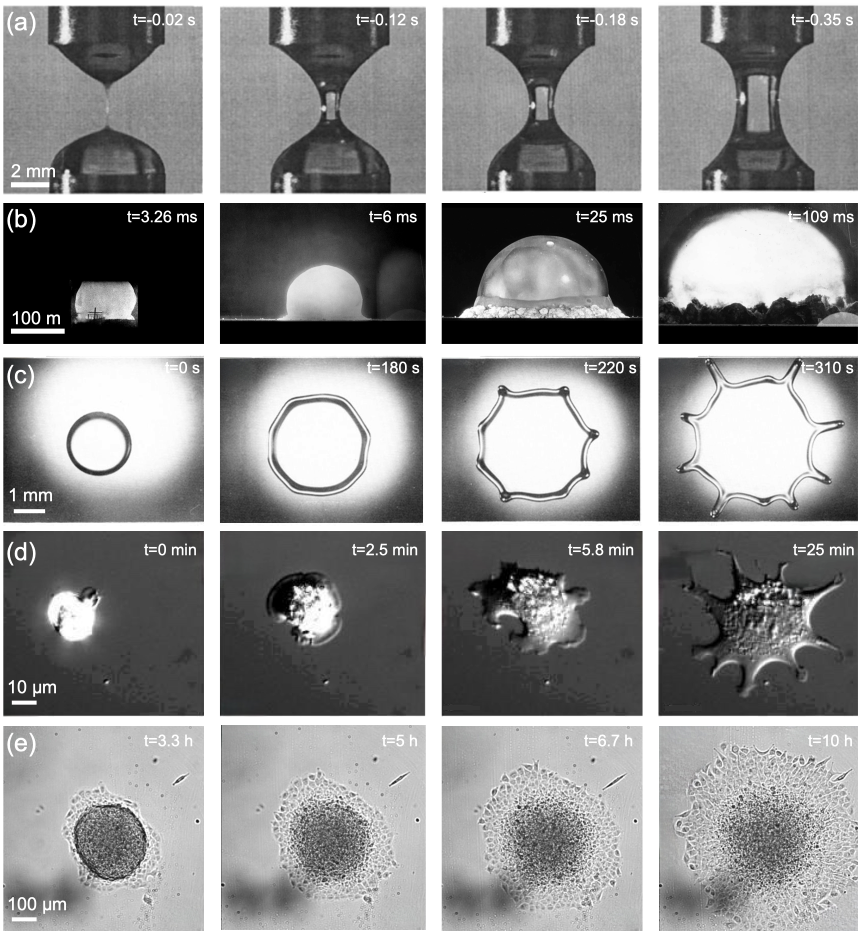}
\caption{Examples of dynamics associated with scalings of the form $d\sim t^\alpha$, from the pinching of viscous liquid threads, to atomic explosions and the motions of living cells. (a) Side view of the pinching dynamics of a viscous glycerol filament~\cite{McKinley2000}, governed by $d\simeq (\Gamma/\eta) t$ (Eq.~\ref{Gammaeta}), where the time $t$ is the duration before pinch-off, so in this example the ``actual time'' runs from right to left. The length $d$ is the radius of the filament, and $\Gamma$ and $\eta$ are the surface-tension and viscosity of the fluid. (b) Side view of the first atomic explosion (Trinity test, 1945), discussed in the introduction~\cite{Taylor1950}. The dynamics of the blast radius $d$ follow $d\simeq (E/\rho)^\frac{1}{5} t^\frac{2}{5}$, where $E$ is the yield of the explosion and $\rho$ is the density of the surrounding air. (c) Top view of the spreading of a silicon oil droplet that is being spin coated~\cite{Melo1989}. The dynamics of the contact radius $d$ follows $d\simeq (E/\varphi)^\frac{1}{4} t^\frac{1}{4}$ (Eq.~\ref{Eflux}), where $E$ is the centrifugal energy and $\varphi\simeq \eta/\ell$ is an \Maf{interfacial friction} built from the viscosity and size of the drop. (d) Top view of the spreading of a cell onto a rigid substrate covered by extracellular matrix~\cite{Fardin2010}. The dynamics of the contact radius $d$ follows $d\simeq (F/\eta)^\frac{1}{2} t^\frac{1}{2}$ (Eq.~\ref{Feta}), where $F$ is the `protrusion force' and $\eta$ is the viscosity of the cell. (e) Top view of the spreading of an aggregate of cells onto a rigid substrate covered by extracellular matrix~\cite{Douezan2011}. The dynamics of the contact radius $d$ follows $d\simeq (E/\eta)^\frac{1}{3} t^\frac{1}{3}$ (Eq.~\ref{Evis}), where $E\simeq \Gamma\ell^2$ is the adhesion energy, and $\eta$ is the viscosity of the aggregate. 
\label{illu}}
\end{figure*} 
In the previous section we investigated the mechanics of space or time, taken separately. Here, we will see how mechanical quantities can be used to rationalize motion, so we will address the connection between mechanics and kinematics, that is between mass-carrying quantities and space-time. 

We will present a few instructive examples of dynamic scalings from the literature, in particular those illustrated in Fig.~\ref{illu}. 

\subsection{General formula}
In the previous section we only considered pairs of mechanical quantities on the same lines or on the same columns of Table~\ref{masscary}. Pairs on the same line yield simple lengths, and pairs on the same column yield simple times. We now consider any arbitrary pair of mechanical quantities, $Q_1(x_1,y_1)$ and $Q_2(x_2,y_2)$. In this general case, the dimensions of the mechanical ratio combine space and time, and the dimension of mass naturally disappears: 
 \begin{equation}
 \Big[\frac{Q_1}{Q_2}\Big] = \mathcal{L}^{x_1-x_2} \mathcal{T}^{y_1-y_2}
 \label{General}
 \end{equation}
This general formula includes simple lengths in the case where $y_1=y_2$ (same line), and simple times when $x_1=x_2$ (same column). Also included are all sorts of fully kinematic results, where $x_1\neq x_2$ and $y_1\neq y_2$. As we will see now, these kinematic cases provide a deep connection between mechanical quantities and motion.

\subsection{Simple speeds\label{speeds}}
\href{https://youtu.be/VvMKwZSNEX4?si=l6aLuzYAhzjqjxfv}{Mechanics 4: Simple Speeds}\\

In the same way that pairs of mechanical quantities can combine to give simple lengths or times, they can also produce speeds. Indeed, for pairs of mechanical quantities satisfying $x_1-x_2=y_2-y_1$, Eq.~\ref{General} implies 
\begin{equation}
\Big[\frac{Q_1}{Q_2}\Big] = \mathcal{L}^{x_1-x_2} \mathcal{T}^{-(x_1-x_2)}= (\mathcal{L} \mathcal{T}^{-1})^{x_1-x_2}=[u]^{x_1-x_2}
\end{equation}
Graphically, the constraint on the exponents, $x_1-x_2=y_2-y_1$, means that the two quantities $Q_1$ and $Q_2$ are on the same diagonal of slope -1 in Table~\ref{masscary}. So in this case, the ratio of such pair of mechanical quantities produces a speed or powers of a speed. Taking the appropriate root we can systematically express the result as a \textit{simple speed}, since:
\begin{equation}
\Big[\frac{Q_1}{Q_2}\Big]^\frac{1}{x_1-x_2} = \mathcal{L}\mathcal{T}^{-1}
\end{equation}
We shall discuss five important examples: 
\begin{align*}
   & (E/m)^{\frac{1}{2}} \quad &\text{energy \& mass} \\
   & (\Sigma/\rho)^{\frac{1}{2}} \quad &\text{stress \& density} \\
   & \Gamma/\eta \quad &\text{stiffness \& viscosity}\\
   & F/\zeta &\quad \text{force \& friction}\\
   & S/H &\quad \text{strength \& action} 
\end{align*}
These speeds give us a preview of the relationship between mechanics and motion, in the special case where this motion is `uniform', i.e. at constant speed. 

\subsubsection{Energy and mass: kinetic energy and projectiles\label{proj}}
\begin{equation}
 u_{Em} \equiv  \Big(\frac{E}{m} \Big)^{\frac{1}{2}} \label{Em} 
\end{equation}
The combination of energy and mass produces a speed, which underlies the concept of kinetic energy, and which can be used to derive the speed of projectiles of known mass and energy. The standard unit of energy is the Joule, which is defined as 1~kg.m$^2$.s$^{-2}$. This definition connects the energy to the mass, $[E]=[m] [u]^2$, where $u$ is some speed. The most famous example of this formula is the most famous formula: $E=mc^2$, the mass-energy equivalence. Another more ancient example of this connection between energy, mass, and speed is the kinetic energy, $E=\frac{1}{2} m u^2$. In this context, the speed $u$ is much smaller than the speed of light. Usually, this formula is used to compute the energy from a known mass $m$ and speed $u$. However, the formula can be rearranged to express the speed from the mass and energy, as in Eq.~\ref{Em}. The mass can be that of a projectile, like a canon ball, or a bullet, and the energy is that delivered by the gun. 

The speed $u_{Em}$ can be used to rationalize the speed of all sorts of projectiles, bullets racing in a straight line, but also debris flying in all directions, \Maf{as in the case of explosions--small, large, or even astronomical. This kind of speed can for instance be used to describe the early stage of supernova explosions.} Some types of supernovae (type Ia) the mass and energy are known with some confidence. The mass is that of the `progenitor', i.e. the exploding star, with a mass around that of our Sun, so $m\simeq 2~10^{30}$~kg, and the energy is around $E\simeq 2~10^{44}$~J. In this context, the early speed of the leading edge of the supernova remnant can be estimated from Eq.~\ref{Em}, reaching a daunting 10,000~km per second! 

\subsubsection{Stress and density: sound speed\label{soundspeed}}
\begin{equation}
 u_{\Sigma\rho} \equiv  \Big(\frac{\Sigma}{\rho} \Big)^{\frac{1}{2}} \label{Sigmarho} 
\end{equation}
This example is probably one of the most well-known. The speed $u_{\Sigma\rho}$ is the `sound speed', taken in its most general sense. The sound waves can be connected to compression or shear, whether the stress $\Sigma$ is taken to be a shear stress or a pressure. Some materials, typically gaseous can only sustain compression waves. For air, with $\rho\simeq 1.2$~kg/m$^3$, and $\Sigma\simeq 1.4~10^5$~Pa, the sound speed $u_{\Sigma\rho}$ would be about 340 m/s (the stress $\Sigma$ is the bulk modulus of the air, which is given by the product between the atmospheric pressure and the `adiabatic index' around $1.4$). The order of magnitude of the sound speed in various materials can be computed from values of densities and elasticity/pressure/shear modulus, etc. In general, there can be different elastic moduli depending on the directions of deformation. Nevertheless, for isotropic and homogeneous materials, only two moduli are enough to characterize the material~\cite{Landau1986}. Many pairs are possible. For instance, inside the Earth, the sound waves are `seismic waves', called `P-waves' (compression) and `S-waves' (shear), with speeds obtained with the formula of Eq.~\ref{Sigmarho}, by choosing $\Sigma$ to be respectively the P-wave modulus and shear modulus. For granite, the P-wave speed is typically around 5000~m/s, whereas the S-wave speed is 3000~m/s. In contrast, for medical ultrasounds, the relevant stress is the shear modulus of tissues, around $\Sigma\simeq 10^4$~Pa, with a density $\rho\simeq 10^3$~kg/m$^3$, giving a sound speed around 3~m/s. 

The sound speed $u_{\Sigma\rho}$ can also appear in disguise, for instance in astrophysics, and magnetohydrodynamics, where it is sometimes called the `Alfv{\'e}n speed', when the stress is built from a magnetic field strength $B$ as $\Sigma=B^2/\mu_0$, where $\mu_0$ is the permeability of the vaccum~\cite{Chandrasekhar2013}. Note that in the same way that the Boltzmann's constant was used to translate a temperature into an energy ($[k_B]=[E]/[\Theta]$), and the permittivity $\epsilon_0$ was used to translate charges into a strength ($[\epsilon_0]=[e]^2/[S]$), here the permeability is used to translate a magnetic field into a stress ($[\mu_0]=[B]^2/[\Sigma]$). These translation constants allow one to remain within the $\mathcal{M}$-$\mathcal{L}$-$\mathcal{T}$ system. 

\subsubsection{Stiffness and viscosity: visco-capillary speed}
\begin{equation}
u_{\Gamma\eta} \equiv \frac{\Gamma}{\eta}  \label{Gammaeta}
\end{equation}
This speed is crucial to the dynamics of capillary driven flows. In this context the stiffness is interpreted as a surface-tension, and the speed may be called the `visco-capillary speed'~\cite{Middleman1995,McKinley2005,PGG2013}. At the interface between \Maf{pure} water and air, the surface-tension is typically $\Gamma\simeq 7~10^{-2}$~N/m, and the viscosity of water is $\eta\simeq 10^{-3}$~Pa.s, such that the capillary speed is $\Gamma/\eta\simeq 10^2$~m/s. In contrast, the sound speed in water would be around $1400$~m/s, and the molecular speed would be $600$~m/s. \Maf{Note that such molecular speed would be expressed as $u_{Em}$ (Eq.~\ref{Em}), using the thermal energy and the mass of the molecules as factors}. For glycerol, since the surface-tension is similar and the viscosity is a thousand times bigger, the visco-capillary speed would be around $10$~cm/s. When there is no other significant mechanical quantity beyond $\Gamma$ and $\eta$, the visco-capillary speed is the natural speed scale. For instance, $u_{\Gamma\eta}$ is the speed at which viscous filaments gets thinner~\cite{Papageorgiou1995,Eggers1997,McKinley2000}, as depicted in Fig.~\ref{illu}a. This speed also occurs during the dynamics of spreading and coalescing droplets~\cite{Fardin2022}. Note that in the context of dilute surfactant solutions, a difference of surface-tension is used and the speed can be called the Marangoni speed~\cite{PGG1985,Edwards1991,Manikantan2020,Nikolov2021}. 

\subsubsection{Force and friction: terminal or active speed}
\begin{equation}
u_{F\zeta} \equiv \frac{F}{\zeta} \label{Fzeta}
\end{equation}
This pair is more often found in the form $F\simeq \zeta u$, where $u$ is the drift speed or terminal speed. Usually, $\zeta$ is understood as a constant, whereas $F$ and $u$ are variable. This viewpoint corresponds well to passive fluids, where the force is usually applied by the experimenter or set by gravity~\cite{Stokes1850,Leal2007}. One famous example of this speed is in the context of an object falling inside a viscous fluid, like a steel ball in corn syrup. In this context, $F=m g_0$, where $m$ is the mass of the ball. The friction or drag coefficient $\zeta$ can be estimated if we know the weight and the speed. For instance, for a ball of steel with a diameter of 1~cm, the weight is around $F\simeq 0.04$~N and the speed in syrup around $u\simeq 2.6$~cm/s, so $\zeta\simeq F/u\simeq 1.5$~kg/s. \Maf{Note that when the density of the falling object is comparable to that of the fluid, the driving force $F$ must take into account buoyancy. }

In Eq.~\ref{Fzeta} the force $F$ needs not be the weight. For instance, the standard acceleration of gravity can be superseded by a centrifugal acceleration, $g\simeq r\omega^2$, which can be orders of magnitude larger than the standard $g_0$. Then Eq.~\ref{Fzeta} can be written as $u \simeq mr\omega^2/\zeta$, a formula very useful in biology, chemistry, and physics, to separate objects based on their different sedimentation speeds. The equation can be rearranged as $u/g \simeq m/\zeta$. On the left, the sedimentation speed is divided by the effective acceleration, and is sometimes called the sedimentation coefficient, measured in Svedberg, after Theodor Svedberg, the Swedish chemist who got a Nobel prize for his study of colloids and proteins and the development of the \Maf{ultracentrifuge}~\cite{Svedberg1927,Claesson1972,Sharma2009}. By definition, one Svedberg is equal to $10^{-13}$~s, and indeed a speed divided by an acceleration is a time. The right hand side of the equation  reveals that this kinematic ratio of speed and acceleration can also be understood mechanically as a ratio of mass and friction. So the sedimentation coefficient is the simple time built from the mass and the friction, a pair on the same column of Table~\ref{masscary}, which we can add to the list of simple times we started in the previous section. 

Note that the most general formulation of this simple speed does not require the force to be connected to any mass. For sedimentation and centrifugation the force is known, but  in recent years, this simple formula has also been used the other way around, to estimate the magnitude of an unknown driving force $F$ from a known friction, as in the case of motile cells or organisms. For instance, considering a swimming bacteria, between turning points the bacteria \Maf{moves} at a roughly constant speed, $u\simeq 30~\mu$m/s. From Eq.~\ref{Fzeta}, we can obtain an estimate of the driving force $F$ from the speed $u$, if we also know the friction $\zeta$. In viscous fluids, as we saw in Eq.~\ref{zetaeta}, the friction can be related to the size $\ell$ of the moving object and to the viscosity of the fluid, as $\zeta \simeq \ell \eta$. The driving force can then be expressed as $F\simeq \eta \ell u$. For a swimming E-coli with $u\simeq 30~\mu$m/s and $\ell\simeq 2~\mu$m, the surrounding medium is around 10 times more viscous than water, so $\eta\simeq 10^{-2}$~Pa.s. Overall, $F\simeq 10^{-12}$~N, i.e. one piconewton, which is indeed the right order of magnitude, although the numerical prefactors we ignored can increase this force to a few tens of piconewtons~\cite{Marchetti2013,Schwarz2013}.

\subsubsection{Strength and action: the speed of light}
\begin{equation}
u_{SH} \equiv \frac{S}{H}  \label{SH}
\end{equation}
This last example gives a simple speed as a ratio between a strength and an action. We have already seen an example of such speed with the semi-classical speed of the electron, $u=\alpha_0 c$, with $\alpha_0$ the fine structure constant, which can be expressed as $\alpha_0=S_0/\hbar c$, where we recall that $S_0=k_C e^2$ is the electromagnetic strength between two elementary charges. Thus the electron speed is $u=S_0/\hbar$, an important example of simple speed built from strength and action. 

Note that the speed of light itself could be expressed from Eq.~\ref{SH}. Since $c=S_0/\hbar \alpha_0$, we could define an action $H=\hbar \alpha_0$, or a strength $S=S_0/\alpha_0$, which would provide slightly different ways to think about the speed of light.

\subsection{Non-uniform motion\label{sprexa}}
\href{https://youtu.be/jv6IKH7aO5s?si=55dEy2nCWsS_51GT}{Mechanics 5: Struggle in Motion}\\

\begin{figure*}
\centering
\includegraphics[width=18cm,clip]{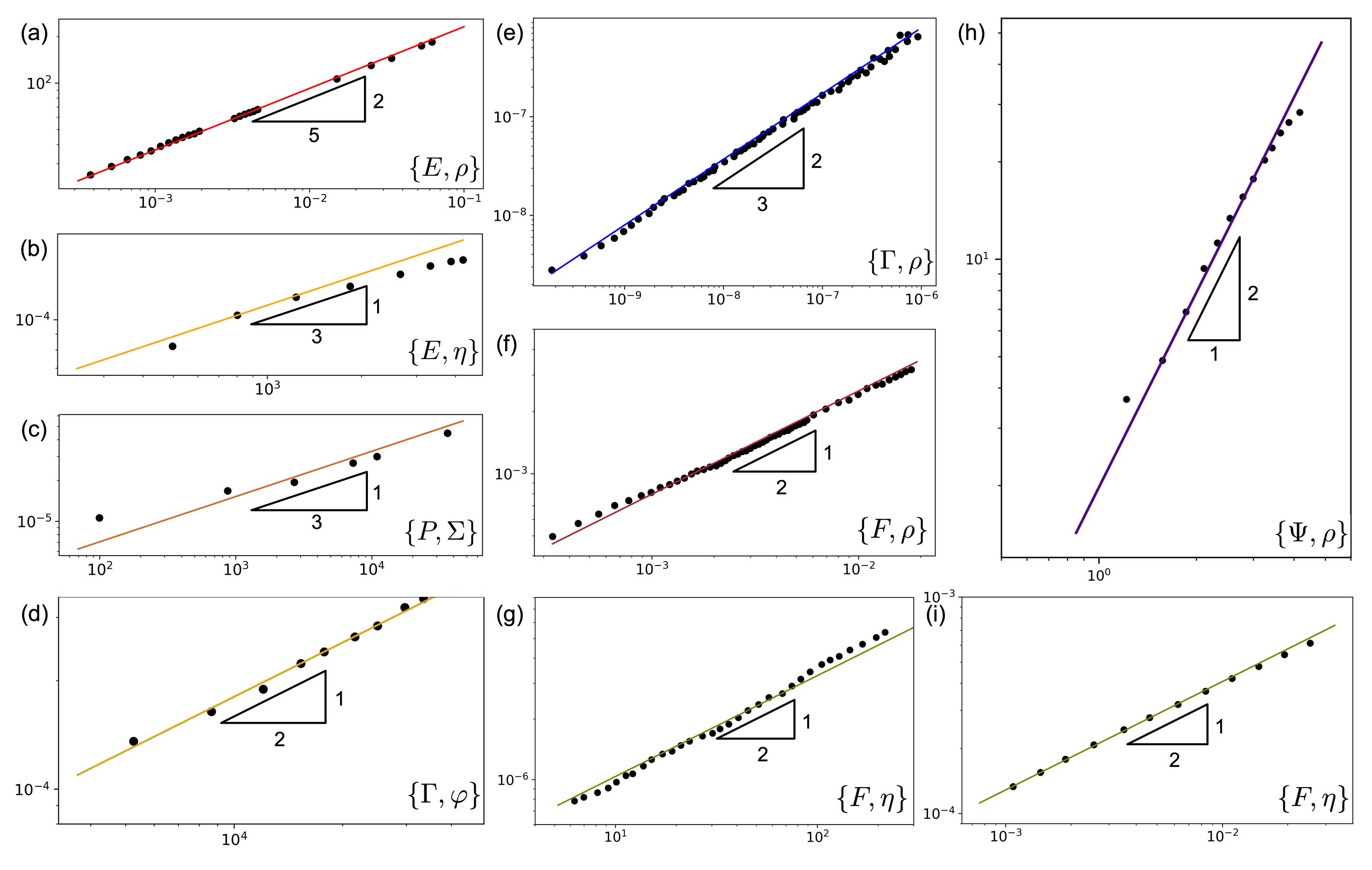}
\caption{Growth of a length $d$ over time $t$ for regimes observed in dynamics from atomic explosions to living cells. Lengths and times are respectively measured by the vertical and horizontal axes, in meters and seconds for all plots. All plots show power laws of the form $d=Kt^\alpha$. On each plot the value of $\alpha$ is represented by a triangle giving the slope of the line. The pair of mechanical quantities invoked to rationalize the slopes are given in the bottom-right corner. Note that in some cases the data have been truncated to isolate the range of validity of the regime in focus, undisturbed by the effect of additional mechanical quantities (see the conclusion of the review for a discussion of this point). (a) Blast radius of the Trinity explosion~\cite{Taylor1950b}. (b) Initial spreading of a spherical aggregate of cells in `partial wetting' conditions~\cite{Douezan2011}. (c) Grain size growth for the `Ostwald ripening' regime of `sintering'~\cite{German2009}. (d) Spreading of the `precursor film' of motile cells for the `complete wetting' of a spherical aggregate of cells~\cite{Douezan2011}. (e) Inertio-capillary pinching of a bridge of liquid mercury~\cite{Burton2004}. In this case, the time $t$ is the duration before pinch-off, so the ``actual time'' runs from right to left. (f) Spreading of a water droplet~\cite{Biance2004}. (g) Spreading of a single cell~\cite{Cuvelier2007}. (h) Distance traveled by a debris flow down an incline~\cite{Iverson2011}. (i) Radius of the neck for the coalescence of two air bubbles in silicone oil~\cite{Paulsen2014}.  
\label{figsprlaws}}
\end{figure*} 

From Antiquity to the Middle Ages, motion was practically synonymous with uniform motion, where distances grow linearly with time, as $d=u t$. The great leap made by Galileo, Kepler and Newton was in no small part driven by their departure from this narrow focus on motion at constant speed. Generations of thinkers had been obsessed by speed, with dimensions $\mathcal{L}\mathcal{T}^{-1}$, but the Renaissance shifted the attention toward acceleration, $\mathcal{L}\mathcal{T}^{-2}$, in particular with the study of free-fall, where the fallen distance grows quadratically, as $d=\Maf{\frac{1}{2}} g t^2$. Unfortunately, this revolutionary takeover turned into a new dogma, and for centuries acceleration became the imposed kinematic metric of motion. It is only toward the end of the 19th century that the existence of other types of motion resurfaced with the study of diffusion~\cite{Bird2006}. For diffusive processes a distance grows like a square root of time, $d=K t^\frac{1}{2}$, which is usually written as $d= (\mathcal{D} t)^\frac{1}{2}$, introducing the `diffusivity' or `diffusion coefficient' $\mathcal{D}= K^2$, with $[\mathcal{D}]=\mathcal{L}^2\mathcal{T}^{-1}$~\cite{Einstein1905b,Perrin1926,Sharma2009}. For reasons beyond our scope, kinematic quantities just like standard mechanical quantities are usually defined in such a way as to have integer exponents. 

Motions at constant speed, at constant acceleration, or ``diffusive motions'' are the three most historically significant examples of motion, but they are in no way more fundamental than other types of motions discovered since. For instance, following Taylor~\cite{Taylor1950,Taylor1950b}, we have seen that the blast of an explosion may advance according to a `power law' $d= K t^\frac{2}{5}$.  In this case, the kinematic parameter $K$ is neither a speed, nor an acceleration, nor given by a diffusivity. The kinematic quantity $[K]=\mathcal{L}\mathcal{T}^{-\frac{2}{5}}$ (or $[K^5]=\mathcal{L}^5\mathcal{T}^{-2}$ if we prefer integer exponents) is a more unusual combination of time and space. Like most kinematic quantities $K^5$ does not have a standard name, but it has every right to be named. In our \href{https://www.youtube.com/playlist?list=PLbMiQs7eX-bbNTc-7HwdWzohUs8yPw300}{video lectures}, we took the liberty of calling it the \textit{explosivity}. An explosion as the one studied by Taylor corresponds to a motion at constant explosivity. Just like a speed, an acceleration, or a diffusivity, an explosivity can be understood as a ratio of a pair of mechanical quantities. In Taylor's analysis, $K^5\simeq E/\rho$, where $E$ is the energy of the bomb and $\rho$ the density of the ambient medium. This relationship is a direct consequence of Eq.~\ref{General}. Indeed, since energy and density are five columns and two lines apart in Table~\ref{masscary}, we have: 
 \begin{equation}
 \Big[\frac{E}{\rho}\Big] = \mathcal{L}^{5} \mathcal{T}^{2}
 \end{equation}
We can understand Taylor's relation, $d\simeq (E/\rho)^\frac{1}{5} t^\frac{2}{5}$, as being the natural expression of the ratio $E/\rho$ when time is measured by $t$ and space by $d$. We shall use these two symbols, $t$ and $d$, instead of $\tau$ and $\ell$, in order to underline the fact that the length and time are here variable. 

This way to represent kinematics as some evolution law for a size $d(t)$ is pretty visual, so we will adopt it in this entire subsection, but as we shall see in section~\ref{perspec} it is by no mean the only perspective on kinematics. From this length versus time perspective the general formula in Eq.~\ref{General} can be expressed as a `scaling law' or \textit{regime}: 
\begin{align}
& Q_1/Q_2 \simeq d^{x_1-x_2} t^{y_1-y_2}\label{Generalregime2}\\
\Leftrightarrow ~ &d\simeq \Big(\frac{Q_1}{Q_2}\Big)^\frac{1}{x_1-x_2} t^{\frac{y_2-y_1}{x_1-x_2}}
\label{Generalregime}
\end{align}
This formula includes simple lengths, simple times and simple speeds as special cases, and it also includes all sorts of non-uniform motions. Any choice of two mechanical quantities immediately yields a regime. The mechanical quantity in the numerator drives motion, while the quantity in the denominator slows things down. We will say that the numerator is the \textit{impelling} factor, while the denominator is the \textit{impeding} factor, and we shall return to the subtleties of this duality in subsection~\ref{II}. 

Note the use of the approximate equality `$\simeq$' in Eq.~\ref{Generalregime}, which underlines the fact that this relationship may not be exact, depending on the precise definitions of the mechanical parameters ($Q_1$ and $Q_2$) and kinematic variables ($d$ and $t$). For now, we will consider that the mechanical quantities $Q_1$ and $Q_2$ provide a satisfying model of the dynamics if the two sides of Eq.~\ref{Generalregime} only differ by numerical factors `of order 1'. We will return to this point in section~\ref{scon}.

Note also that Eq.~\ref{Generalregime} includes \textit{growing regimes}, where $(y_2-y_1)/(x_1-x_2)>0$, and  \textit{shrinking regimes}, where $(y_2-y_1)/(x_1-x_2)<0$. These shrinking regimes have a divergent length $d$ at initial time, and $d$ only converges to zero for $t\rightarrow \infty$. In this review, we will focus on growing regimes. We differ a discussion of shrinking regimes to a future publication.   

In Table~\ref{masscary}, we defined 25 widely used mechanical quantities. Considering all pairs, would generate more than 300 regimes. If we remove the simple lengths, times, and speeds, and if we only focus on dynamics where the size grows over time ($d\sim t^\alpha$ with $\alpha>0$), there are still more than 100 possible regimes. This large number reflects the great diversity of ``physics'' that can be at play in different situations. In the following sub-sections, we will evidently not discuss all possibilities, but we will show that regimes of all sorts have already been used to describe dynamics across scales and disciplines. 

\subsubsection{Dynamics impelled by energy}
Let us first present a few regimes impelled by energy. If we put aside the simple lengths and times, we have seen two cases so far: the uniform regime given by energy and mass, i.e. $d\simeq (E/m)^\frac{1}{2} t$, and Taylor's regime of explosions, $d\simeq (E/\rho)^\frac{1}{5} t^\frac{2}{5}$. Taylor's regime is depicted in Fig.~\ref{illu}b, and the scaling is plotted in Fig.~\ref{figsprlaws}a. 

If we seek additional regimes impelled by energy, quantities on the same line or column as $E$ cannot be included since they yield simple lengths and times. Quantities on the line of index $y=-3$ cannot be included because they produce shrinking ($d\sim t^\alpha$, with $\alpha<0$) rather than growing ($\alpha >0$). Of the quantities that are left, we have chosen to highlight the 2D density $\Lambda\equiv Q(-2,0)$, the mass flux  $\varphi\equiv Q(-2,-1)$ \Maf{(understood as an interfacial friction $Q(-2,-1)=Q(0-2,-1)$)}, and the viscosity $\eta\equiv Q(-1,-1)$:
\begin{align}
& d\simeq (E/\Lambda)^\frac{1}{4} t^\frac{1}{2} \label{E2dens}\\
& d\simeq (E/\varphi)^\frac{1}{4} t^\frac{1}{4} \label{Eflux}\\
& d\simeq (E/\eta)^{\frac{1}{3}} t^{\frac{1}{3}}  \label{Evis} 
\end{align}
The first regime in Eq.~\ref{E2dens} can be understood as the equivalent of Taylor's regime in cylindrical geometry~\cite{Sedov1993}. In the case of explosions confined inside a cylinder of radius $\ell$, one can define a 2D density $\Lambda\simeq \rho \ell$, leading to a regime with $d\sim t^\frac{1}{2}$. This regime can be used to describe exploding-bridgewire detonators~\cite{Murphy2010}. 

The second example can be used to describe the dynamics of the radius of contact of `spin-coated' drops, as illustrated in Fig.~\ref{illu}c~\cite{Melo1989}. In that case, the spinning is associated with a centrifugal energy $E=\rho \ell^5 \omega^2$, where $\ell^3$ is the volume of a drop and $\omega$ is the rotational frequency. \Maf{One can then define a form of `interfacial friction' from the fluid viscosity as $\varphi=\eta/\ell$.  With these definitions, the spreading of the spun drop follows Eq.~\ref{Eflux}. Note that quantities with the same dimensions have been used to study the ``friction'' of fluids, polymers and elastomers on solid boundaries~\cite{Bocquet2007,Henot2018}. In this context, one can also define the so-called Navier slip length, $\ell\simeq \eta/\varphi$, a form of simple length measured by extrapolating the velocity profile beyond the boundary~\cite{Navier1823}. We will see another use of such kind of `interfacial friction' $\varphi$ in Eq.~\ref{capflux}.} 

The third example in Eq.~\ref{Evis} describes a regime where viscosity prevents the spreading of a source of energy. This regime could apply for point-like inputs of energy in very viscous fluids. This input could for instance come from explosions, lasers~\cite{Campanella2019}, or ultrasounds~\cite{Lauterborn1997,Gibaud2020}. Interestingly, this regime has also been used in a context far from explosions, to describe the spreading of aggregates of cells, as illustrated in Fig.~\ref{illu}e~\cite{Douezan2011}. In this context, a ball of cells comes in contact with a substrate, on which it starts to spread by cell migration. The cell-substrate adhesion $\Gamma$ and the size $\ell$ of the ball can be used to define an adhesion energy $E\simeq \Gamma \ell^2$, such that the spreading abides Eq.~\ref{Evis}, where $\eta$ is the aggregate viscosity. A comparison between this mechanical model and the data is shown in Fig.~\ref{figsprlaws}b. 

\subsubsection{Dynamics impelled by power}
We now turn our attention to dynamics driven by a constant power rather than a constant energy. Many choices of resisting quantities could be useful. We here choose to highlight four possibilities: 
\begin{align}
&d\simeq (P/\rho)^\frac{1}{5}  t^\frac{3}{5}\label{Newton}\\
&d\simeq (P/\zeta)^{\frac{1}{2}}  t \label{braking}\\
&d\simeq (P/\eta)^{\frac{1}{3}}  t^\frac{2}{3} \label{Pvis}\\
&d\simeq (P/\Sigma)^{\frac{1}{3}}  t^{\frac{1}{3}}\label{PSig}
\end{align}
Eq.~\ref{Newton} can be found in slightly different form in the context of \Maf{turbulent mixing}. In the design of stirrers for mixing of liquids inside vessels, it has been found that the mechanical input power $P$ required for mixing is given by $P\simeq \rho d^5 t^{-3}$, where $d$ is the agitator diameter, $t$ its period of rotation and $\rho$ is the density of the fluid~\cite{Seinfeld1992}. \Maf{Note that in this case the scaling does not relate a variable length and time, the approach adopted in the examples discussed so far. We will see in section~\ref{perspec} that any pair of mechanical quantities can be expressed from multiple perspectives. The mixing scaling, $P\simeq \rho d^5 t^{-3}$, is but a first example of what we will generically call a ``mechanical perspective'' in section~\ref{mechpersp}. }

Eq.~\ref{braking} gives another example of regime based on power. This equation defines the simple speed $(P/\zeta)^{\frac{1}{2}}$, so we could have put it in section~\ref{speeds}. This speed is relevant to dynamics characterized by a constant friction $\zeta$. In many situations, the friction $\zeta$ is not constant and depends on speed. In general one can define $\zeta$ from the friction force $F$ as $\zeta=F/u$, where $u$ is a speed. In the `inertial regime', the friction force is proportional to the square of the speed, such that $\zeta\simeq \rho u d^2$. Using $u \simeq d/t$ this definition of the friction would lead back to Eq.~\ref{Newton}. In contrast, in the `viscous regime', the friction force is proportional to speed and given by $\zeta\simeq \eta d$~\cite{Stokes1850}. If $d$ is the variable distance, this leads to the regime given in Eq.~\ref{Pvis}. However, in some cases $d$ is a constant length, for instance connected to the size of a vehicle. The quantity $\zeta$ would then be a constant parameter and Eq.~\ref{braking} may apply. 

Eq.~\ref{PSig} gives yet another example of regime driven by power, where the impeding quantity is a stress. This equation may be applicable to `sintering'~\cite{German2009}. In this process, the typical size of grains grows as $d\simeq (Kt)^\frac{1}{3}$, where in the so-called `Ostwald ripening' regime, the grain growth rate can be written as $K\simeq C \mathcal{D}\Omega \Gamma /RT$. The parameter $R$ is the ideal gas constant, and $\Omega$ is the molar volume, such that $\Sigma\equiv RT/\Omega$ is a characteristic thermal stress. The constant $C$ is a dimensionless solubility, $\Gamma$ is the solid-liquid surface energy and $\mathcal{D}$ is the solid diffusivity in the liquid. Thus, one can define the power associated with an increase in the size of the grains as $P\equiv C\mathcal{D}\Gamma$, such that the sintering equation becomes an example of Eq.~\ref{PSig}. An example of this regime is given in Fig.~\ref{figsprlaws}c~\cite{German2009}. 

\subsubsection{Dynamics impelled by force-density}
\begin{equation}
d\simeq (\Psi/\rho)  t^{2} \label{psirho}
\end{equation}
Because of its position in the table of mechanical quantities, force-density only allows a few regimes where it acts as the motor. In Eq.~\ref{psirho}, force-density is balanced by density. Since force-density is often taken to be $\Psi=\rho g$, the equation just states $d\simeq gt^2$, which is the `free-fall' equation. This regime applies to the early dynamics of materials driven by gravity before dissipation can set in. It applies for instance to the early dam-break flow~\cite{Janosi2004}, and to the debris flow down an incline~\cite{Iverson2011}, as depicted in Fig.~\ref{figsprlaws}h. In the first case, one of the walls of a reservoir of fluid is removed and one records the dynamics of the surge on a horizontal plane. In the second case, a mixture of fragmented rock and muddy water is similarly released, but down a steep incline plane. 

Note that Eq.~\ref{psirho} may also describe rises rather than falls, in the context of buoyancy. In this case, the force-density takes into account a difference in density between two materials $\Psi\simeq \mid\rho_f -\rho\mid g$. This version of Eq.~\ref{psirho} would for instance be useful to understand the initial rise of a mushroom cloud after a nuclear explosion such as Trinity~\cite{Taylor1950b}. Indeed, the blast generates a zone of very low density, which acts as a bubble inside the comparatively denser air.

\subsubsection{Dynamics impelled by stress}
Of the possible regimes driven by stress, we  have already seen the uniform regime associated with the sound speed, $d \simeq (\Sigma/\rho)^{\frac{1}{2}} t$. Let us also mention the following `diffusive' regime:  
\begin{align}
d & \simeq (\Sigma/\chi)^{\frac{1}{2}}  t^\frac{1}{2}\label{Stressspreadings2}
\end{align}
Like the regime at constant sound speed, this additional regime is most commonly found in aerodynamics~\cite{Landau1959}. Eq.~\ref{Stressspreadings2} is appropriate in situations where the density of the medium is uniformly changing at a rate $\chi$ (either compressing if $\chi >0$ or expanding if $\chi<0$). This situation is particularly relevant to some scenarios of star formation~\cite{Mac2004}. 

\subsubsection{Dynamics impelled by stiffness}
For the dynamics of drops and bubbles, surface-tension is often understood as a driving force. We here highlight three possible regimes impelled by surface-tension/stiffness:   
\begin{align}
&d\simeq (\Gamma/\rho)^{\frac{1}{3}}  t^{\frac{2}{3}} \label{inerticap}\\
&d\simeq (\Gamma/\varphi)^\frac{1}{2}  t^\frac{1}{2}  \label{capflux}\\
&d\simeq (\Gamma/\Phi)  t^{2}  \label{N1cap}
\end{align}
The first equation has been studied in great detail in the context of spreading, pinching and coalescence of simple liquids like water~\cite{Leger1992,Eggers1997,Basaran2002,Villermaux2007,Eggers2008,Bonn2009,Snoeijer2013,Kavehpour2015,Fardin2022}. We have seen with Eq.~\ref{Gammaeta} that combining a surface-tension and a viscosity yields a simple speed, which underlines what is sometimes called the `visco-capillary' regime, where viscosity is the principal impeding force. 
In contrast, Eq.~\ref{inerticap} describes the `inertio-capillary' regime. In the context of the spreading, pinching and coalescence of drops, Eq.~\ref{inerticap} dictates the dynamics unencumbered by viscosity, where the main impeding factor is `inertia', represented by the density $\rho$. This regime is for instance seen in the pinching of liquids with low viscosity~\cite{Keller1983}, like water, or mercury as in the example in Fig.~\ref{figsprlaws}e~\cite{Burton2004}. This regime has also been observed for spreading~\cite{Courbin2009,Eddi2013} and coalescence of low viscosity fluids~\cite{Eddi2013b}. Note that although the neck of a pinching drop decreases over time, its dynamics can be represented by a growing regime, $d\sim t^\alpha$, with $\alpha>0$, when the variable time $t$ is understood as the duration before pinch-off~\cite{Fardin2022}. 

The regime given in Eq.~\ref{capflux} describes dynamics driven by a surface energy $\Gamma$, but impaired by a mass flux or momentum density $\varphi$. An example of this regime is shown in Fig.~\ref{figsprlaws}d~\cite{Douezan2011}. In this example, the dynamics describe the spreading of the `precursor film' composed of the motile cells moving away in 2D after the contact of a spherical aggregate. In this context, the surface energy comes from the adhesion of cells with the substrate, and the parameter $\varphi$ is understood as a form of ``friction''~\cite{Douezan2011}.    

In Eq.~\ref{N1cap} the `1D density' $\Phi$ can be interpreted as a `normal stress coefficient', as mentioned in section~\ref{times}. The ratio $\Gamma/\Phi$ provides an acceleration, which has some relevance to the free-surface flows of visco-elastic liquids. In particular, Eq.~\ref{N1cap} has been discussed in the context of the pinching of so-called `second order fluids'~\cite{McKinley2005}. 

\subsubsection{Dynamics impelled by force}
Of the possible regimes driven by force, we choose to highlight three cases: 
\begin{align}
&d\simeq (F/m)  t^{2}\label{PFD}\\
&d\simeq (F/\rho)^{\frac{1}{4}}  t^{\frac{1}{2}}\label{Frho}\\
&d\simeq (F/\eta)^{\frac{1}{2}}  t^{\frac{1}{2}}\label{Feta}
\end{align}
In Eq.~\ref{PFD}, Newton's second law is barely disguised. Like Eq.~\ref{psirho}, this spreading law just describes dynamics at a constant acceleration $a=F/m$. We recall that under the assumption that $d\sim t^\alpha$, $F=ma$ just translates to $F\simeq md/t^2$. As mentioned in the introduction, this pair is of great historical significance, since it records the first time a kinematic--and not just geometric--quantity was understood mechanically. The force is the prototype for all impelling factors, and the mass is the prototype for all impeding factors. 

In Eq.~\ref{Frho}, the force is balanced by density rather than mass. This regime applies to spreading, coalescence and pinching of drops when the whole size of the drop has a substantial impact~\cite{Eggers1997,Basaran2002,Villermaux2007,Eggers2008,Bonn2009,Fardin2022}. In this context, the force can be expressed as $F\simeq \Gamma \ell$, where $\Gamma$ is the surface-tension and $\ell$ is the size of the drop. An example of this regime is given in Fig.~\ref{figsprlaws}f for the spreading of a water drop~\cite{Biance2004}.    

In Eq.~\ref{Feta}, the force is balanced by viscosity. This regime has been observed in a few different contexts. For instance, this regime describes the coalescence of drops in a viscous outer fluid, where $\eta$ is the viscosity of the outer fluid, as shown in Fig.~\ref{figsprlaws}i~\cite{Paulsen2014}. A similar regime can be used to describe the spreading of thin films, where $F/\eta$ is understood as an effective diffusivity, with $F\simeq E/\ell$, where $E$ is the Hamaker constant and $\ell$ the film thickness~\cite{Feslot1989,Popescu2012}. Since the Hamaker constant is usually on the order of the thermal energy, the effective diffusivity matches with Stokes diffusivity. The same equation was also used to describe the early spreading of single cells, as illustrated in Fig.~\ref{illu}d~\cite{Fardin2010}. In that case, the viscosity is that of the cell and the force can be understood either as coming from the stiffness of the cytoskeleton, or as a `protrusion force' originating from the polymerization of `actin'~\cite{Cuvelier2007,Fardin2010}. An example of such scaling of early cell spreading is given in Fig.~\ref{figsprlaws}g~\cite{Cuvelier2007}. 

\subsection{Impelling and impeding\label{II}}
\href{https://youtu.be/J5tm6IXt_fU?si=bPbBS5hD6SkaO3wm}{Mechanics 6: Holy Motors}\\

If two mechanical quantities are given, there is a unique regime associated with them. For this reason, we may use the notation $\lbrace Q_1,Q_2 \rbrace$ to stand for the regime associated with the pair of mechanical quantities $Q_1$ and $Q_2$. For instance, Taylor's regime is $\lbrace E,\rho \rbrace$, or $\lbrace \rho,E \rbrace$. The order of the quantities between brackets does not matter. In the pair $\lbrace E,\rho \rbrace$, the energy will always be the motor, or impelling factor, in accordance with Eq.~\ref{Generalregime}. Let us recall this equation here so there is no need to turn the pages: 
\begin{equation}
d\simeq \Big(\frac{Q_1}{Q_2}\Big)^\frac{1}{x_1-x_2} t^{\frac{y_2-y_1}{x_1-x_2}}
\end{equation}
In this equation, the mechanical quantity $Q_1$ is really in the numerator if $x_1-x_2>0$, that is if $x_1>x_2$. Graphically, this means that the impelling factor of a pair is always the quantity on the rightmost part of Table~\ref{masscary}. For instance, if the pair is $\lbrace E,\rho \rbrace$, since $E$ is on the right of $\rho$, then energy will be the impelling factor. 

Although given a pair, one is always driving, what is driving in one situation can be resisting in another and vice-versa. This fact was clearly not understood when the first few mechanical quantities were defined, and to this day it remains the source of a lot of confusion. We only need to look at the historical names of the quantities on Table~\ref{masscary} to see that some of them are quite heavily connoted. To the right of the table, where they are more likely to be motors, the names are markedly positive, like `action', `energy', `force', or `power'. All these ``macho'' terms in the English language are here to remind us that these quantities were thought as movers. For a lot of early scientists they were nothing short of the hand of God in the physical world. In contrast, terms like `friction', `mass', or `viscosity', were initially thought of as sticky, gooey, resisting or at most inert rather than active. In fact, these quantities can be active, they can be impelling motion rather than impeding it, under the right circumstances. We shall illustrate this versatility on two historically significant examples.  

\subsubsection{Boundary layers\label{BLsec}}
We have seen a few examples of dynamical regimes where viscosity was involved: $\lbrace \Gamma,\eta \rbrace$, $\lbrace E,\eta \rbrace$, and $\lbrace F,\eta \rbrace$. In all these cases the viscosity appeared as an impeding factor, living up to its name (``viscous'' comes from Latin ``viscosus'', meaning ``sticky''). However, in some situations the viscosity can actually drive motion, it can be the impelling factor. We just have to pair viscosity with a mechanical quantity on its left in Table~\ref{masscary}. One such pairing with great historical significance is the following: 
\begin{equation}
\lbrace \eta , \rho \rbrace\rightarrow d\simeq \Big(\frac{\eta}{\rho}\Big)^\frac{1}{2} t^\frac{1}{2}\label{BL}
\end{equation}
Here, viscosity is paired with density, and the resulting regime is central to the understanding of `boundary layers'~\cite{Landau1959}. In this context, the viscosity and density are that of a fluid and Eq.~\ref{BL} describes the thickness of the sheared layer of that fluid near a boundary. For instance, if a plate starts moving at $t=0$ in a quiescent fluid, the fluid in the immediate vicinity of the plate will start moving too (in the absence of slip), but the fluid far away will remain immobile. As time goes by, the size of the moving layer of fluid near the boundary will grow, according to Eq.~\ref{BL}. Since the power law has an exponent of $\frac{1}{2}$, this motion can be said to be `diffusive', and the mechanical ratio $\nu \equiv \eta/\rho$ can be interpreted as a (`momentum') diffusivity. It is also called the `kinematic viscosity'. The greater the viscosity the greater the diffusivity, so indeed, viscosity is here impelling motion! Note that as surprising as it may be Eq.~\ref{BL} does not involve the speed of the moving boundary. The dynamics of the boundary layer are actually independent of this speed only up to a point, where the flow becomes turbulent. So Eq.~\ref{BL}  only refers to `laminar boundary layers'~\cite{Landau1959}.  

\subsubsection{Kepler's law and levity\label{Klev}}
\begin{figure}
\centering
\includegraphics[width=8.5cm,clip]{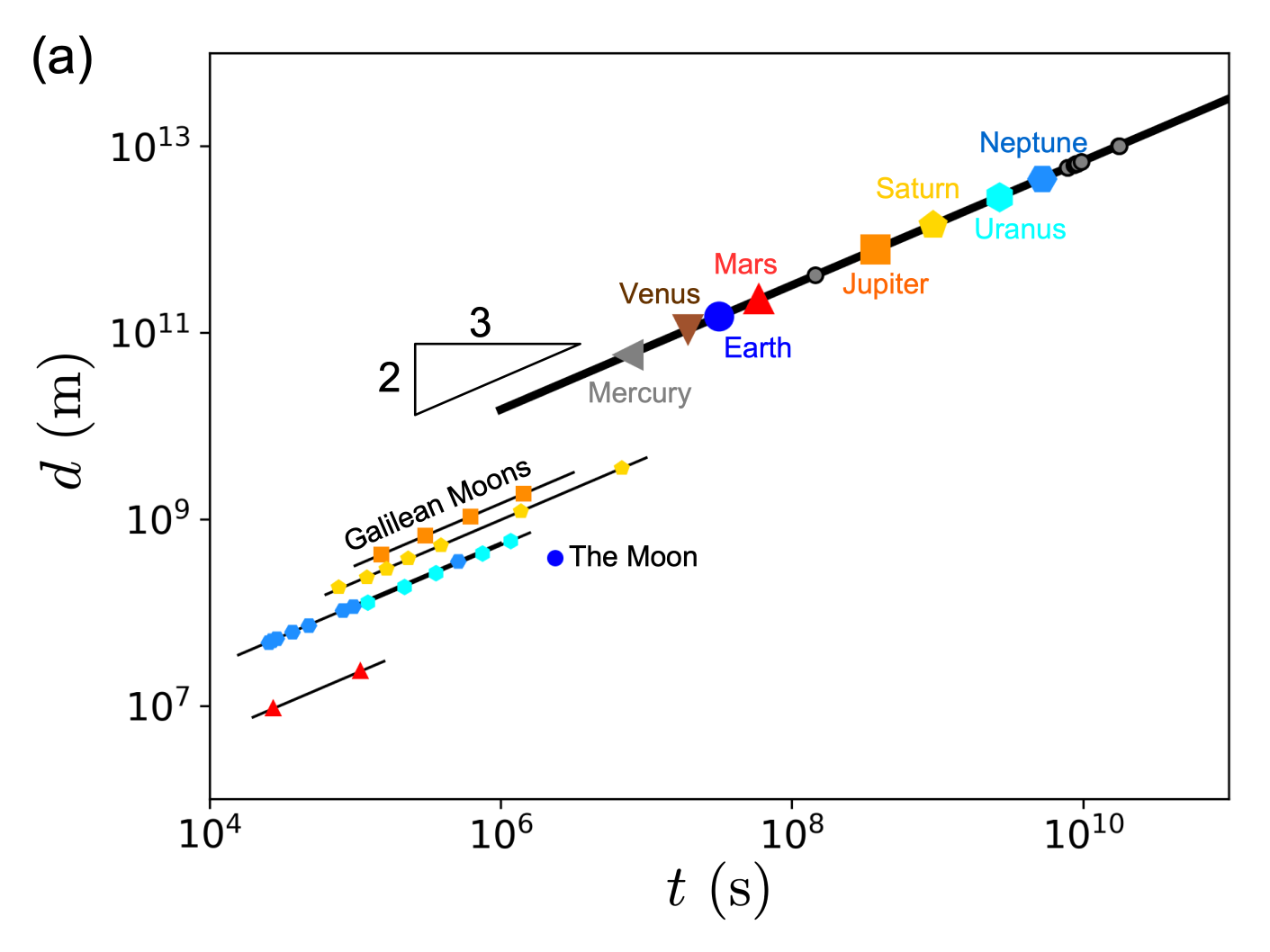}
\includegraphics[width=8.5cm,clip]{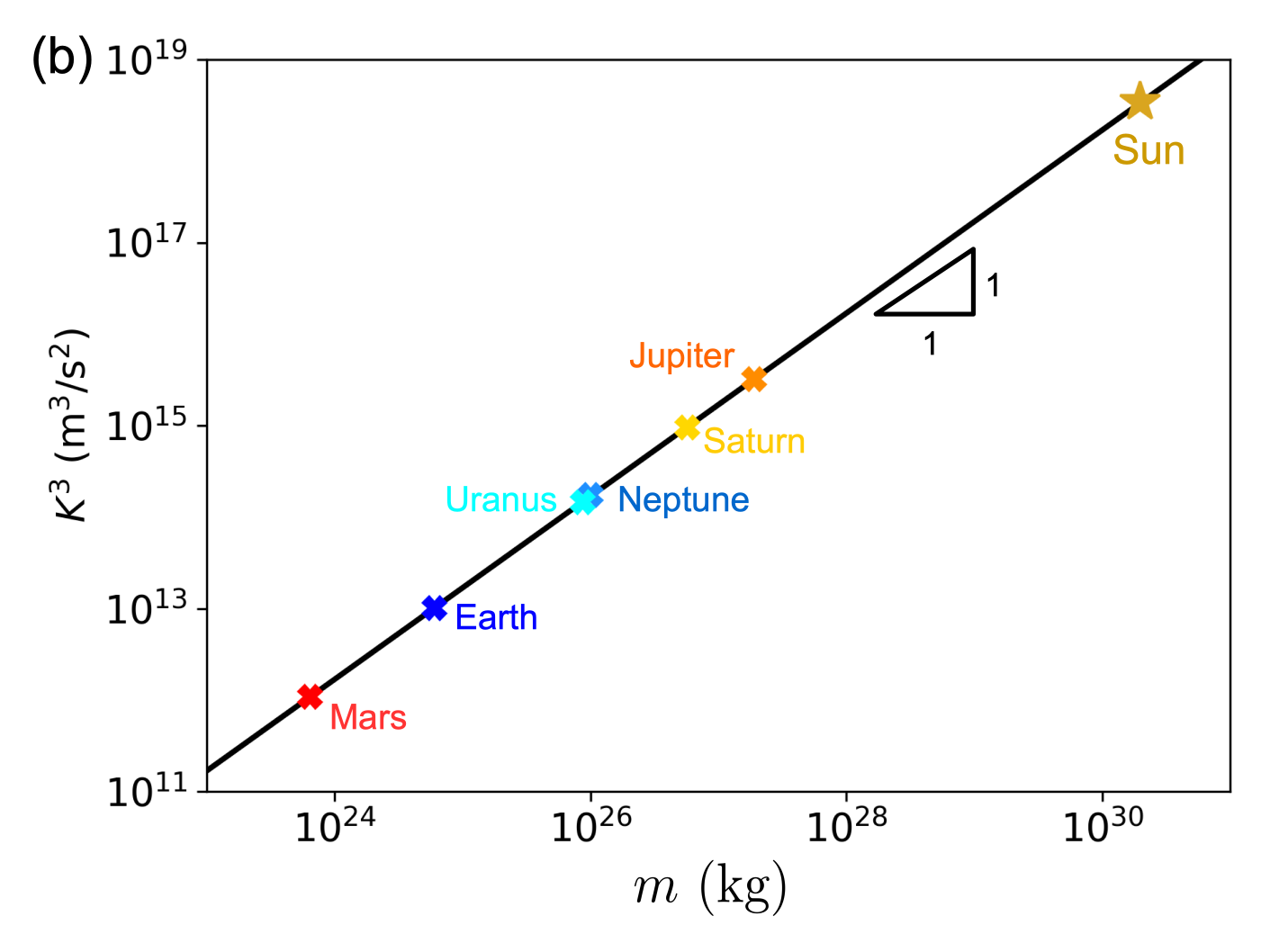}
\caption{(a) Kepler's law $d=Kt^\frac{2}{3}$, i.e. the square of the orbital period, $t$, is proportional to the cube of the length of the semi-major axis of its orbit, $d$. The law is verified for the `solar system', where the represented orbiting bodies are the planets and the dwarf planets (small grey disks). The law is also verified for the orbits of the satellites of the planets, represented by stars of a color corresponding to the associated planet. These satellites include the Galilean moons of Jupiter (orange). The values of the prefactor $K$ differ depending on the orbital system. (b) The values of $K^3$ for the solar system and for the satellite systems are plotted against the mass $m$ of the central object, which is the planet for its satellites, and the Sun for the orbiting planets.  One finds that $K^3\sim m$. \Maf{For instance, the data point for Jupiter corresponds to the value of the prefactor $K$ of the power law for the Galilean moons in (a) (small orange squares).}  
\label{NKG}}
\end{figure} 
The mass was originally understood as `inertial', a word from the beginning of the 18th century based on the adjective ``inert'', in-art, without art, without skill, from Latin ``inertem'', meaning ``unskilled, incompetent, inactive, helpless, weak, sluggish, worthless'', a long list of pretty negative attributes given by the etymological dictionary. That was how mass was initially perceived, as the paradigm of the impeding factor, as the exemplar of what is resisting motion. This view on the mass changed during the Renaissance, when it was realized that the mass could also be the source of motion, in the context of gravity. 

At the end of the 16th century, building on the observations of his master Tycho Brahe, Kepler had established his now famous `third law of planetary motion': $t^2 \sim d^3$. The square of a planet's orbital period, $t$, is proportional to the cube of the length of the semi-major axis of its orbit, $d$~\cite{Kepler1609}. If we want we can of course express this law in the following fashion: $d=Kt^\frac{2}{3}$, where the unknown kinematic prefactor $K$ must have dimensions $\mathcal{L}\mathcal{T}^{-\frac{2}{3}}$, so $[K^3]=\mathcal{L}^3\mathcal{T}^{-2}$, a rather odd kinematic quantity that is nowadays called the `gravitational parameter'. In the case of the solar system the data available to Kepler could establish that $K^3\simeq 3~10^{18}$~m$^3$/s$^2$. The mechanical interpretation of this kinematic quantity had to wait one more generation, with the work of Newton. 

Fig.~\ref{NKG}a demonstrates the validity of Kepler's law for the orbits of the planets of the solar system. Also included are the characteristics (periods and semi-major axes) of the orbits of the satellites of the various planets of the solar system. Concomitantly with Kepler's investigations, Galileo had discovered the four largest moons of Jupiter~\cite{Galilei1632}, now called the `Galilean moons'. As evident in Fig.~\ref{NKG}a, the \Maf{Galilean} moons, as well as the moons of other planets also follow power laws of the form $d=Kt^\frac{2}{3}$, but the value of the kinematic prefactor $K$ changes from one system to another. 

It was Newton who first realized that the prefactor $K$ is not a ``universal'' constant, \Maf{but that} it depends on the ``world'' under study~\cite{Newton1687}. Newton then set out to find a decomposition of the kinematic quantity $K$ into mechanical quantities. Using the notation introduced in this review, we could say that Newton sought a pair of mechanical quantities $Q_1(x_1,y_1)$ and $Q_2(x_2,y_2)$ such that $K=(Q_1/Q_2)^\frac{1}{x_1-x_2}$. As shown in Fig.~\ref{NKG}b, Newton realized that $K^3 \sim m$, the impelling mechanical quantity was the central mass of the orbit, $Q_1=m$, where $m$ is for instance the mass of the Sun for the solar system~\cite{Newton1687}. 

Since $[K^3]=\mathcal{L}^3\mathcal{T}^{-2}=[Q_1/Q_2]$, and $[Q_1]=[m]=\mathcal{M}$, Newton's second mechanical quantity had to have the following dimensions: $[Q_2]=\mathcal{M}\mathcal{L}^{-3}\mathcal{T}^2$. Thus, $Q_2=Q(-3,2)$. Taking a look back at Table~\ref{masscary}, we see that this mechanical quantity is at an odd position in comparison to all the others, which probably explains its peculiar fate. A quantity $Q(-3,2)$ is what we called a \textit{levity}, denoted by the symbol \myG. Since this mysterious quantity is on the left of the mass (in the table), it is impeding motion, whereas mass is impelling it: 
\begin{equation}
\lbrace m , \myG \rbrace\rightarrow d\simeq \Big(\frac{m}{\myG}\Big)^\frac{1}{3} t^\frac{2}{3}\label{Kepler}
\end{equation}
Kepler's law is here understood as a regime based on pairing mass and levity. Using a notation introduced at the end of the 19th century and still in use today~\cite{Boys1894}, we can identify the levity $\myG$ with the inverse of the so-called `universal constant of gravity' G. More precisely, $\myG\simeq 4\pi^2/G$ (assuming the central mass is much more massive than the orbiting objects). 

The purpose of defining such ``levity'' is to stress that Kepler's law--like any other regime--can be expressed from a pair of standard mechanical quantities with dimensions of the form $\mathcal{M}\mathcal{L}^{x}\mathcal{T}^y$. In contrast, the universal constant of gravity has dimensions $[G]=\mathcal{M}^{-1}\mathcal{L}^{3}\mathcal{T}^{-2}$. Other mechanical quantities with similar negative power of the mass have been used on occasion, like the `fluidity' with dimensions $\mathcal{M}^{-1}\mathcal{L}^{1}\mathcal{T}^{1}$~\cite{Bocquet2009}. These quantities are absolutely valid, but they are unnecessary, since they can be reduced to standard quantities by inversion. We do believe that the ``levity'' deserves its place in Table~\ref{masscary} and should be free to interact with all the other quantities. As we saw, pairing levity and mass leads to Kepler's law for orbital motion, but we invite the reader to try different pairings. For instance, $\{\myG, \rho\}$ gives rise to a simple time $(\myG/\rho)^\frac{1}{2}\simeq 1/(G\rho)^\frac{1}{2}$, which is called the `free-fall time', with important applications in astrophysics~\cite{Kippenhahn1990}. 


\section{Perspectives on motion\label{perspec}}
\href{https://youtu.be/lVLjflQI41w?si=m3h0d3Kvm6hI_2e6}{Mechanics 7: Perspectives}\\

Since Eq.~\ref{Generalregime} we have presented mechanics from one particular kinematic perspective. The assumption was that we observed some kind of motion, like an explosion, or the pinching of a water droplet, or the fall of an object, or the orbit of a planet. With these examples, we represented motion by plotting a length $d$ versus a time $t$, as in Fig~\ref{figsprlaws}. That length could be a size or a distance, and the time could be an indefinite duration, or a recurring period. We have seen that the interplay of a pair of mechanical quantities $Q_1$ and $Q_2$ produces a single kinematic power law, $d=K t^\alpha$, where the kinematic prefactor $K$ and the dimensionless exponent $\alpha$ are obtained from the underlying mechanical quantities. As soon as we know the relative dimensions of the two mechanical quantities $Q_1$ and $Q_2$, we know the scaling connecting kinematics and mechanics. 

Describing motion as the time series of a distance $d(t)$ is quite visual, however not all dynamics can easily be understood in this fashion, and it is sometimes much more practical to use different pairs of kinematic variables. To each pair of kinematic variables corresponds a different \textit{perspective} on the dynamics. In some situations it may even be useful to manipulate mechanical rather than kinematic variables. These different kinematic and mechanical perspectives provide complementary approaches on the same physics. The ``physics'' are set by the quantities $Q_1$ and $Q_2$, and the perspectives by the choice of variables. The exact forms of the scalings differ from one perspective to another, but they are always the direct consequence of dimensional analysis. 

\subsection{Kinematic perspectives}
For a given pair of mechanical quantities, we know since Eq.~\ref{General} that the dimensions of their ratio can be expressed as some combination of space and time: $[Q_1/Q_2] = \mathcal{L}^{x_1-x_2} \mathcal{T}^{y_1-y_2}$. So far, we have considered cases where the spatial dimension could be associated to a variable size $d$, and the time dimension to a variable time $t$. However, in some situations the more obvious variables are different kinematic combinations. 

\subsubsection{Velocity profiles\label{velprof}}
Let us first come back to the example of an explosion studied by Taylor. When the explosion is initially supersonic, Taylor showed that the radius of the blast will extend for some time according to $d\simeq (E/\rho)^\frac{1}{5} t^\frac{2}{5}$. This perspective on the dynamics is convenient, because the validity of this regime can simply be checked by measuring the radius of the blast on pictures captured at different instants~\cite{Taylor1950b,Mack1946}, as those reproduced in Fig.~\ref{illu}b. However, one may rather choose to record this explosion by measuring the speed of the blast $v$ as it passes by detectors placed at different distances $d$ from ground zero. What should we expect for $v(d)$? If $v$ is defined as the instantaneous speed of the front, then $v\equiv \partial d/\partial t =  2d/5t$, which can be expressed solely from the variable length $d$, as $v\simeq (E/\rho)^\frac{1}{2} d^{-\frac{3}{2}}$ (where numerical factors are absorbed by the approximate equality). This equation still displays the same physics combining energy and density, but it is expressed from the perspective of a speed versus a distance, rather than a distance versus time. We may write this shift in perspective by adding indices to the pair of mechanical quantities underlining this type of motion:
\begin{align}
\{E,\rho\}_{dt} \rightarrow d&\simeq \Big(\frac{E}{\rho}\Big)^\frac{1}{5} t^\frac{2}{5}\\
\{E,\rho\}_{vd} \rightarrow v&\simeq \Big(\frac{E}{\rho}\Big)^\frac{1}{2} d^{-\frac{3}{2}}
\end{align} 
In the second equation, one of the variables is a speed, and the other is a distance. One important application of this perspective is in hydrodynamics, where one is interested in representing so-called `velocity profiles'~\cite{Landau1959,Bird1987}. The scaling $\{E,\rho\}_{vd}$ provides an example of such approach, although it is rather exotic. Let us now consider two more famous cases. 

We have seen in Eq.~\ref{etaSigma} that combining a viscosity and a stress yields a simple time $\tau_{\eta\Sigma} \equiv \frac{\eta}{\Sigma}$. In the case of a Newtonian fluid like water in normal conditions, this simple time can be understood as the inverse of the deformation rate on a fluid with viscosity $\eta$, upon applying a shear stress $\Sigma$ at the boundary. The deformation rate can in turn be interpreted as a `velocity gradient', since $\mathcal{T}^{-1}=(\mathcal{L}\mathcal{T}^{-1})/\mathcal{L}$. So the interplay between viscosity and stress can also be written as $v/d \simeq \Sigma/\eta$, or as:
\begin{equation}
\{\Sigma,\eta\}_{vd} \rightarrow v\simeq \frac{\Sigma}{\eta} d
\end{equation}
In other words if a flow is given by the interplay between a viscosity and a stress, the `velocity profile' in this flow is linear. This scaling goes back all the way to Newton's definition of viscosity, but when this duo of stress and viscosity is expressed as a speed over distance, one speaks of the `Couette profile' or `Couette flow'~\cite{Landau1959}. This simple shear-flow is what naturally comes out of the interplay of stress and viscosity. Speed varies linearly with distance. 

Another very useful flow is the pipe flow. Here, the profile is not linear but parabolic. Why? In this situation, the viscosity of the fluid is still relevant, but the flow is not driven by a stress at the boundary, but by a difference in pressure along the conduit. The impelling factor of these dynamics is the gradient of pressure, which is also a force-density $\Psi$, since $[\Psi]=[\Sigma]/\mathcal{L}$. Then, indeed, the combination of a force-density and a viscosity leads to a parabolic profile: 
\begin{equation}
\{\Psi,\eta\}_{vd} \rightarrow v \simeq \frac{\Psi}{\eta}  d^2
\end{equation} 
When the interplay of viscosity and force-density is represented from this perspective one speaks of the `Hagen-Poiseuille flow'~\cite{Landau1959}. Note that in this equation, the distance $d$ is defined from the center of the pipe, and $v$ is the speed in the co-moving frame, that is the maximum speed in the center minus the speed in the lab frame. 

The scalings $\{\Psi,\eta\}_{vd}$, $\{\Sigma,\eta\}_{vd}$ and $\{E,\rho\}_{vd}$ give but three examples of velocity profiles, but they illustrate a general principle that can be applied to any pair of mechanical quantities. 

\subsubsection{Dispersion relations}
\begin{figure}
\centering
\includegraphics[width=8.5cm,clip]{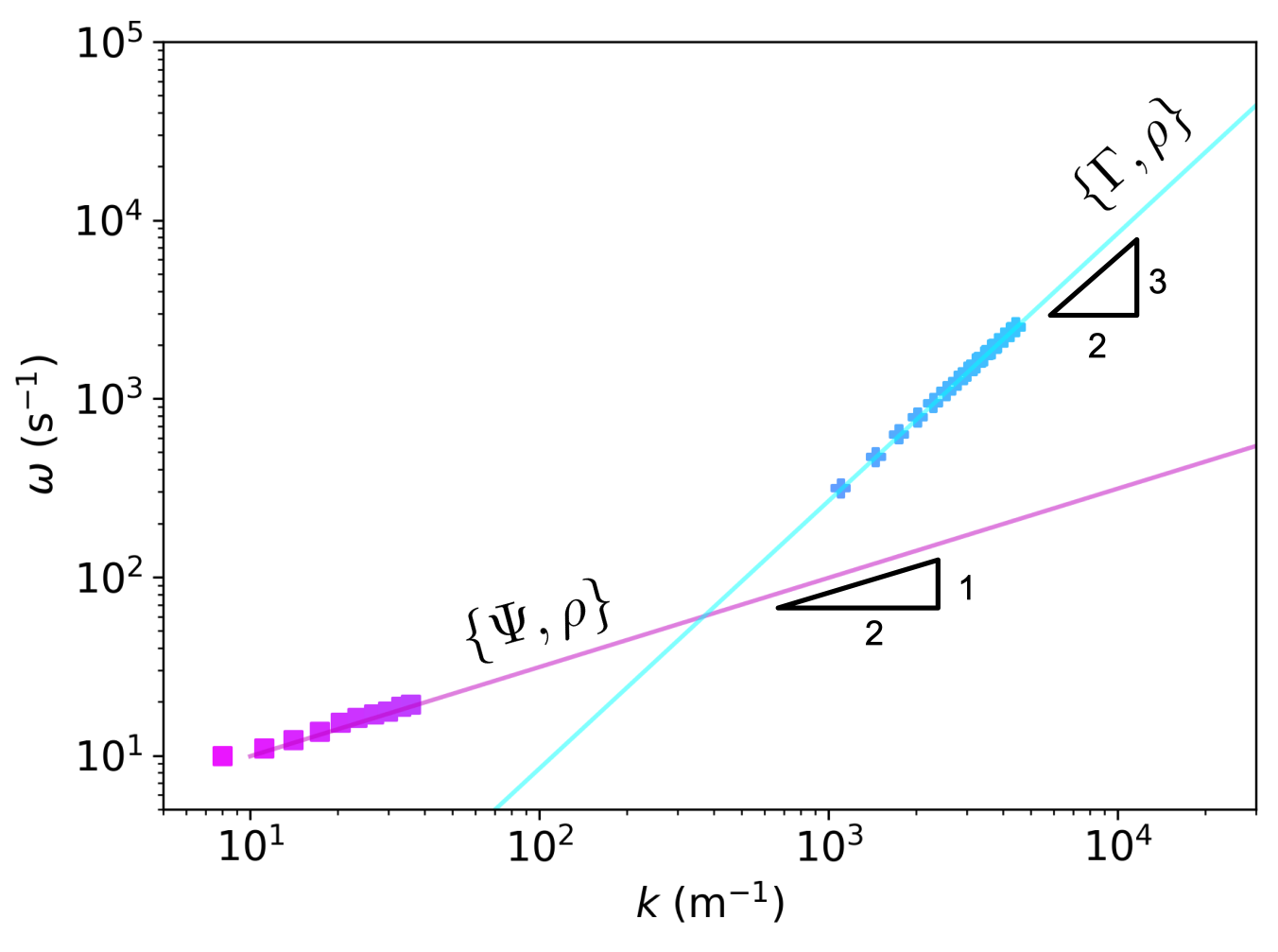}
\caption{Examples of dispersion relations of `capillary ripples'~\cite{Nikolic2012} and `gravity waves'~\cite{Wang2004} on the surface of water. The capillary ripples correspond to the pair $\{\Gamma,\rho\} \rightarrow \omega \simeq (\Gamma/\rho)^\frac{1}{2} k^{\frac{3}{2}}$. The gravity waves correspond to $\{\Psi,\rho\} \rightarrow \omega \simeq (\Psi/\rho)^\frac{1}{2} k^{\frac{1}{2}}$. 
\label{disprel}}
\end{figure} 
Another important kinematic perspective is that of `dispersion relations'~\cite{PGG2013}. This viewpoint on kinematics is rooted in the study of waves, where one is interested in relating the frequency of waves and their wavenumber, that is the inverse of their wavelength. We will use the symbol $\omega$ for the frequency and $k$ for the wavenumber. So this perspective connects the inverse of a time $\mathcal{T}^{-1}$ to the inverse of a length $\mathcal{L}^{-1}$. We will give two famous examples related to the waves on the surface of liquids like water.

For small waves, also called `ripples'~\cite{Rayleigh1890}, the relevant mechanical quantities are the surface-tension $\Gamma$, and the density $\rho$. We have seen this pair already, in Eq.~\ref{inerticap}, in the context of pinching, coalescence and spreading of low viscosity fluids. In these cases, we saw that the interplay of the pair of mechanical quantities could be expressed as a power law of the form $d\simeq Kt^\frac{2}{3}$, where the length $d$ was the radius of the neck for pinching and coalescence, or the radius of contact for spreading. The time $t$ was either the time elapsed since contact for spreading and coalescence, or the time remaining before pinch-off. For ripples, the size $d\simeq k^{-1}$ and the time $t\simeq \omega^{-1}$ are understood as the wavelength and period of the waves, so the perspective we started with, $\{\Gamma,\rho\}_{dt}$, can be translated to a dispersion relation $\{\Gamma,\rho\}_{\omega k}$: 
\begin{align}
\{\Gamma,\rho\}_{dt} \rightarrow d&\simeq \Big(\frac{\Gamma}{\rho}\Big)^\frac{1}{3} t^\frac{2}{3}\\
\{\Gamma,\rho\}_{\omega k} \rightarrow \omega&\simeq \Big(\frac{\Gamma}{\rho}\Big)^\frac{1}{2} k^{\frac{3}{2}}
\end{align} 
The regime $\{\Gamma,\rho\}_{\omega k}$ was first studied theoretically by Lord Kelvin~\cite{Thomson1871}, \Maf{but} it is Rayleigh who first tested this regime experimentally~\cite{Rayleigh1890}. An example of modern measurements of such scaling is given in Fig.~\ref{disprel}~\cite{Nikolic2012}. The dynamics of these small ripples have exactly the same underlying physics as the dynamics of droplets, they are all due to the interplay of surface-tension and density, but this mechanical struggle is seen from different angles. 

For Newtonian liquids like water, the dynamics of waves change when their wavelength becomes substantially larger than the capillary length $(\Gamma/\Psi)^\frac{1}{2}$, which we introduced in Eq.~\ref{GammaPsi}. In this regime, the dominant impelling factor becomes the weight density $\Psi\simeq \rho g$, and the impeding factor remains the `inertia' of the fluid encapsulated in its density $\rho$. Thus, the struggle is between $\Psi$ and $\rho$, and we have seen this balance before in Eq.~\ref{psirho}. Just as we did for ripples, we can compare the initial scaling with how the dynamics look when given as a dispersion relation: 
\begin{align}
\{\Psi,\rho\}_{dt} \rightarrow d&\simeq  \frac{\Psi}{\rho}  t^2\\
\{\Psi,\rho\}_{\omega k} \rightarrow \omega&\simeq \Big(\frac{\Psi}{\rho}\Big)^\frac{1}{2} k^{\frac{1}{2}}
\end{align} 
The kind of waves described by this regime are famously seen on the surface of oceans, seas, lakes and other bodies of water, but also in the sky~\cite{PGG2013}. We could not find a clear reference for the first use of this regime, but it was already used as a matter of fact by Kelvin at the end of the 19th century. Fig.~\ref{disprel} gives an example of this scaling in the case of oceanic surface waves~\cite{Wang2004}. 

The cases discussed here provide two historically important examples of dispersion relations. Nevertheless, here again the procedure is general, and any regime can be expressed from such perspective if need be. We invite the reader to select any pair from Table~\ref{masscary}, to derive the associated dispersion relation and to investigate if such relation has been observed.  

\subsubsection{Power spectra}
\begin{figure}
\centering
\includegraphics[width=8.5cm,clip]{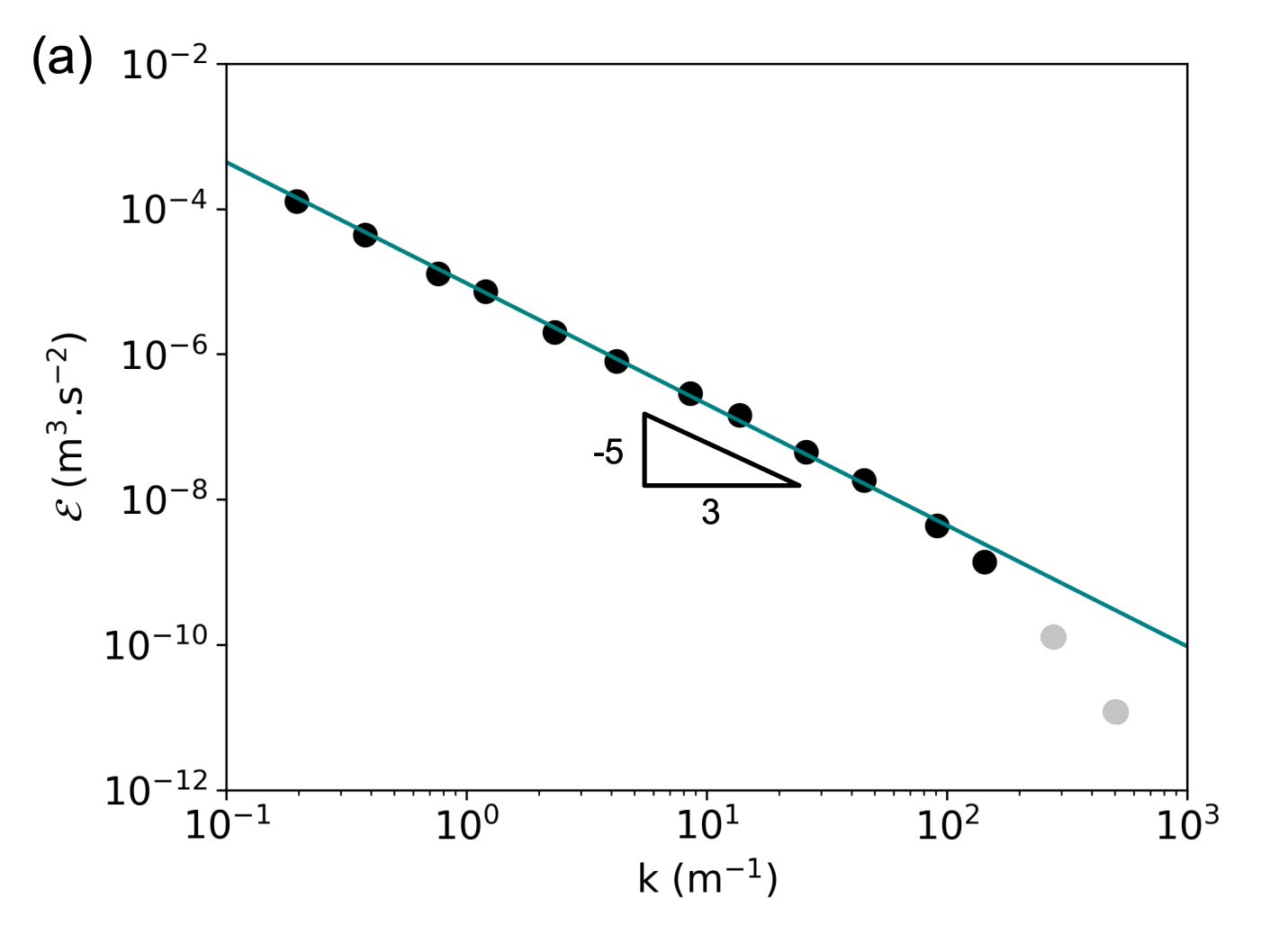}
\includegraphics[width=8.5cm,clip]{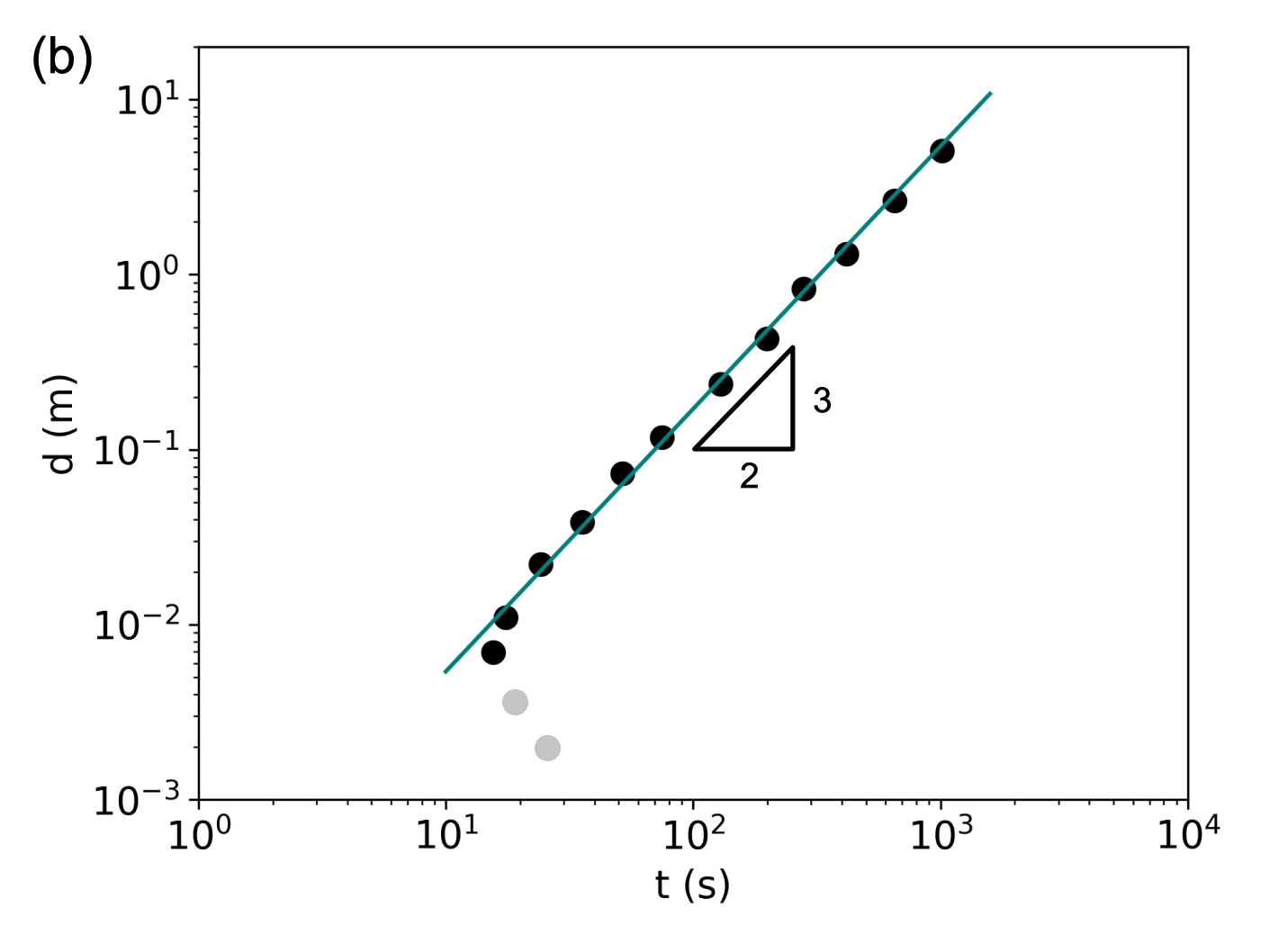}
\caption{(a) Kolmogorov's spectrum of `inertial turbulence' $\{\Pi,\rho\}_{\mathcal{E}k}$ (Eq.~\ref{Kolmo}), observed in a tidal channel by~\citet{Grant1962}. (b) Translation of the spectrum into the evolution law for the size of strained ``blobs'' $\{\Pi,\rho\}_{dt}$, using $d\equiv k^{-1}$ and $t\equiv (k^3 \mathcal{E})^{-\frac{1}{2}}$. Note that since the translation formulas neglect numerical factors the resulting plot is only qualitative. In both (a) and (b) the grey data points deviate from Kolmogorov's scaling due to the effect of a third mechanical quantity: the viscosity of the fluid. We refer the reader to the \href{https://youtu.be/1WP7k1VUf3Q?si=Muusfb3EDpNHB2I2}{9th episode} of our Mechanics series for a discussion of this crossover. It is beyond the scope of this review. 
\label{kolmof}}
\end{figure} 
In the case of velocity profiles or dispersion relations, the variables used ($v(d)$ or $\omega(k)$) are not too distant from the initial variables ($d(t)$), so it is not too difficult to identify the kinship between the resulting regimes. However, for practical or historical reasons, some perspectives may use comparatively complicated kinematic variables. This is for instance the case when considering so-called ``energy'' or ``power'' spectra, which are very useful in turbulence studies~\cite{Frisch1995}. 

In the context of turbulence one often measures what is routinely called an `energy spectrum' $\mathcal{E}(k)$. The first variable, $k$, is a wavenumber, as in the case of dispersion relations, so with $[k]=\mathcal{L}^{-1}$. The misleadingly called ``energy'' $\mathcal{E}$ is actually a kinematic quantity, with $[\mathcal{E}]=\mathcal{L}^3\mathcal{T}^{-2}$~\cite{Frisch1995}, so which could be thought of as a \textit{specific strength} (strength over mass). Fig.~\ref{kolmof}a gives a famous example of such energy spectrum measured in the case of turbulence in a tidal channel~\cite{Grant1962}. These measurements were among the first to validate a prediction from Kolmogorov~\cite{Frisch1995}, about the spectrum to be expected in the case where the turbulence is impelled by a power per unit volume $\Pi$ and impaired by the density of the fluid $\rho$: 
\begin{equation}
\{\Pi,\rho\}_{\mathcal{E}k} \rightarrow \mathcal{E} \simeq  \Big(\frac{\Pi}{\rho}\Big)^\frac{2}{3}  k^{-\frac{5}{3}}\label{Kolmo}
\end{equation}
Note that one usually uses a so called `dissipation rate' or `transfer rate' $\epsilon\equiv \Pi/\rho$, which is a `specific power', i.e. a power per unit mass, such that $[\epsilon]=[P]/\mathcal{M}$~\cite{Frisch1995}. This kinematic notation unfortunately obscures the underlying mechanical quantities, so we shall not adopt it. As we will see now, the spectrum given in Eq.~\ref{Kolmo} provides a great exercise of translation between different perspectives. 

The variable $\mathcal{E}$ is called the ``energy'' because its integral over all wavenumbers is defined from the variance of the flow speed, also understood as `specific energy': $\int_{0}^{\infty} \mathcal{E}(k) \,dk \equiv \frac{1}{2} \overline{v^2}$. If we maintain our neglect for numerical factors all the subtleties of this integration should not concern us, and we can simply write $\mathcal{E} k \simeq v^2$. The variable $\mathcal{E}$ is connected to the magnitude $v$ of the velocity fluctuations for each wavenumber $k$. Larger wavenumbers are in turn related to smaller distances, with $k\simeq 1/d$. Thus, we can use Kolmogorov's spectrum given in Eq.~\ref{Kolmo}, and $v \simeq (\mathcal{E}/d)^\frac{1}{2}$ to express the interplay of power-density and mass-density from a relationship between the amplitude of velocity fluctuations and the size of these fluctuations: 
\begin{equation}
\{\Pi,\rho\}_{vd} \rightarrow v \simeq  \Big(\frac{\Pi}{\rho}\Big)^\frac{1}{3}  d^{\frac{1}{3}}\label{Kolmo2}
\end{equation}
Velocity fluctuations over longer distances have larger amplitude. This equation encompasses the same physics as the initial Kolmogorov spectrum $\mathcal{E}(k)$, but it provides a slightly different perspective on the pair $\{\Pi,\rho\}$. From this alternate vantage point, we can more easily translate the dynamics to the perspective of a length versus a time, which we adopted for most of this review. Let us do this with an intermediate step in between: 
\begin{align}
\{\Pi,\rho\}_{\omega d} \rightarrow \omega &\simeq  \Big(\frac{\Pi}{\rho}\Big)^\frac{1}{3}  d^{-\frac{2}{3}}\label{Kolmo3}\\
\{\Pi,\rho\}_{dt} \rightarrow d &\simeq  \Big(\frac{\Pi}{\rho}\Big)^\frac{1}{2}  t^{\frac{3}{2}}\label{Kolmo4}
\end{align} 
These two perspectives are quite useful to the study of so-called `coherent structures', like the vortices, whirls, or eddies, which populate turbulent flows~\cite{Metais2013}. These structures have attracted a lot of attention in the last 30 years, and they provide an alternative way to think about turbulence, complementary to the statistical style of the spectral perspectives~\cite{Davidson2011}. These coherent structures are the face of velocity fluctuations in turbulent flows. The vortices can rotate at different rates $\omega$, which is called the `vorticity', with $\omega\simeq v/d$, so $[\omega]=\mathcal{T}^{-1}$~\cite{Landau1959}. This variable $\omega$ is also connected to the zones between vortices. There $\omega$ is the extension rate, when the vortices rotate in different directions, or the shear rate where locally the vortices rotate in the same direction. Through shear, extension, or rotation, $\omega$ is connected to the strain rate. So Eq.~\ref{Kolmo3} can be interpreted as relating the strain rate to the characteristic size of the strained region, whether it is a `vortex filament', or a `shear layer', or some more complicated creature~\cite{Davidson2011,Metais2013,Dubrulle2019}. 

To conclude, we can mention the perspective given in Eq.~\ref{Kolmo4}, a size $d$ versus a time $t$, which is the initial approach of this review. In Table~\ref{masscary} the mass-density $\rho$ and power-density $\Pi$ are two columns apart and three lines apart so we get a size growing like time with a power $\frac{3}{2}$. We know that one way to think about the variable $d$ is as the size of a strained region. We know that we can think of $t$ as the period of velocity fluctuations at this length scale, but also as the inverse of the strain rate $\omega$ on a fluid region of size $d$. Yet another way to think about it is as following the evolution of a strained ``blob'' of varying size $d$. We do not know a priori if the blobs are getting bigger or smaller, pinching like droplets. Some blobs might grow while other deflate, but anyway dimensional analysis gives us the power law relating size and time. Although tracking this process for a single blob entangled in a sea of countless others may be challenging, we can actually estimate what we would get by translating the measured energy spectrum, as shown in Fig.~\ref{kolmof}. For each measured values of $\mathcal{E}$ and $k$ we can compute the associated values of $d\simeq k^{-1}$ and $t\simeq (k^3 \mathcal{E})^{-\frac{1}{2}}$. Of course, since these translation formulas neglect all numerical factors, the resulting plot in Fig.~\ref{kolmof}b can only be regarded as qualitative, but the scaling is ensured. 

The four perspectives on turbulence given in Eqs.~\ref{Kolmo}-\ref{Kolmo4}--and any other we may like to adopt--are formally equivalent, but they collectively contribute to a richer and finer appreciation of the interplay between power-density and mass-density, the mechanical pair behind `inertial turbulence'. What we see is a combination of what \textit{is} and of the perspective we have taken, and greater insight is reached by comparing what is perceived from different perspectives. We have seen this in some detail for the case of `inertial turbulence' but the lesson is again general. In particular, we invite the reader to investigate the scalings associated with the growing number of `non-inertial' types of turbulence, from visco-elastic fluids~\cite{Steinberg2021} to active matter~\cite{Alert2022}. Although these more exotic turbulent flows might be represented as energy spectra, $\mathcal{E}\sim k^\alpha$, in these cases one finds that $\alpha\neq -\frac{5}{3}$, because mechanical pairs beyond $\{\Pi,\rho\}$ are involved--finding which pair is the subject of current research. 

\subsection{Mechanical perspectives\label{mechpersp}}
We have seen in the previous sub-section that the effects of any one pair of mechanical quantities can be expressed from different perspectives. So far, all the perspectives we considered were kinematic, which is to say that the pair of chosen variables always had dimensions of the form $\mathcal{L}^x\mathcal{T}^y$ (allowing $x$ or $y$ to be zero). For instance, we saw that Taylor's regime of explosions, $d\simeq (E/\rho)^\frac{1}{5} t^\frac{2}{5}$, could also be expressed as a velocity profile, $v\simeq (E/\rho)^\frac{1}{2} d^{-\frac{3}{2}}$. In the first perspective the variables are $d$ and $t$, in the second $v$ and $d$. In both cases they are kinematic quantities in a broad sense (i.e. including geometric and chronometric quantities). In some circumstances one may rather prefer to focus on mechanical variables. 

A proper discussion of mechanical perspectives would lead us astray from our purpose, but let us say a few words about it here, keeping explosions as our example. To illustrate our purpose, we shall consider the following equations, all equivalent formulations of Taylor's regime: 
 \begin{align}
E &\simeq   (\rho d^3) (d/t)^2 \label{Tay1}\\
E/d &\simeq   (\rho d^3) (d/t^2) \label{Tay2}\\
E/d^3 &\simeq   \rho  (d/t)^2 \label{Tay3}\\
E/t &\simeq    (\rho d^3) (d/t^2) (d/t) \label{Tay4}
\end{align} 
The terms of a regime can be reshuffled and grouped in countless ways. Some outcomes from this reshuffling can be interpreted in terms of mechanical variables, like the four cases above. 

Let us first consider Eq.~\ref{Tay1}. There, the dynamics of the explosion are presented as a balance of energy. On the left-hand side is the constant input of energy $E$ coming from the bomb. On the right-hand side the variables are grouped in such a way as to compensate each other and yield a constant result. The combination $[\rho d^3]=\mathcal{M}$ can be understood as a mechanical variable, the variable mass $\tilde{m}$ of ambient air swept-away by the blast (we use tildes for mechanical variables to differentiate them from the constant mechanical parameters of the regimes)~\cite{Bethe1947}. The entirety of the right-hand side can then be understood as the kinetic energy of the swept-up air, $\tilde{E}\simeq \tilde{m} v^2$, and Taylor's regime as $E\simeq \tilde{E}$, which can be read as saying that the energy of the bomb is converted into the kinetic energy of the blast front. 

If we now consider Eq.~\ref{Tay2} we can recover the Newtonian perspective on mechanics, which is dear to so many textbooks. The terms of this equation are now forces, since $[E/d]=[F]$. We can introduce the variable ``explosion force'' $\tilde{F}\simeq E/d$. The term $\rho d^3\simeq \tilde{m}$ is still the variable swept-up mass, and $d/t^2\simeq a$ is the acceleration of the explosion front. Overall Eq.~\ref{Tay2} can be written as a force balance, in the style of Newton's second law: $\tilde{F}\simeq \tilde{m} a$. 

In Eq.~\ref{Tay3} the dynamics are now given as a balance of pressures. The left-hand side, $\tilde{\Sigma}_1\simeq E/d^3$, is the energy density of the explosion. The right-hand side, $\tilde{\Sigma}_2 \simeq \rho v^2$, is the dynamic pressure emerging from the speed of the front, $v\simeq d/t$ . Taylor's regime corresponds to a front moving in such a way that the varying energy density equals the dynamic pressure. 

In Eq.~\ref{Tay4} the dynamics are given as a balance of power. The power of the explosion $\tilde{P}\simeq E/t$ is faced by $\tilde{m} a v$. 

In these four examples we illustrated how a given regime can be interpreted in terms of mechanical variables. Here these mechanical variables are just built from algebraic manipulations, but in some cases they can be measured by instruments designed for this purpose. For instance, in the context of explosions one may perform pressure measurements at various distances from ground zero. Many such measurements were for instance performed for the Trinity test, enough to confirm the validity of Eq.~\ref{Tay3} from this perspective (c.f. Barschall (1945) LA-352, Graves (1945) LA-354, Manley (1945) LA-360, Bright (1945) LA-366, Marley (1945) LA-431; all these reports are available on the Los Alamos National Laboratory \href{https://sgp.fas.org/othergov/doe/lanl/}{website}).

For more information on this topic, we invite the reader to watch our lecture series on explosions, and more particularly the episode dedicated to kinematic and mechanical perspectives (\href{https://youtu.be/bvVCvdB5Uzk?si=7BpPAD6HsvfYjpUe}{Explosions 8: Kinematic and Mechanical Variables})\\

\section{Simple dimensionless numbers\label{DimNum}}
\href{https://youtu.be/aun0CHanq94?si=w6CXSKyvzqmlqrqk}{Mechanics 8: The Right Angle}\\

We have seen in the previous section that different perspectives provide complementary approaches on the same ``physics''. In the framework of this review these ``physics'' are conveniently expressed by a pair of mechanical quantities $\{Q_1,Q_2\}$, taken from Table~\ref{masscary}, and these mechanical parameters manifest themselves as regimes. 
The exact forms of the regimes or `scalings' differ from one perspective to another, but they are always the direct consequence of dimensional analysis. They can be re-derived with a few lines of algebra. When we briefly considered mechanical variables we saw that the kinematic parts of a scaling could be combined in such a particular way as to provide a constant outcome. For instance, in the case of Taylor's regime, we saw that $E  \simeq   (\rho d^3) (d/t)^2$, and since the energy $E$ and density $\rho$ were constant, this meant that $d^3 (d/t)^2$ was also constant, despite being constructed from variables, $d$ and $t$. Even if we restrict ourselves to kinematic variables, for any pair of mechanical parameters $\{Q_1,Q_2\}$ there is always a perspective that can be built that exploit this property. Kinematic variables and mechanical parameters can combine to form mechanical variables, but they can also combine in such a way as to provide a so-called `dimensionless number', which provides a useful angle on the dynamics. 

\subsection{Constant variables}
\begin{figure}
\centering
\includegraphics[width=8.5cm,clip]{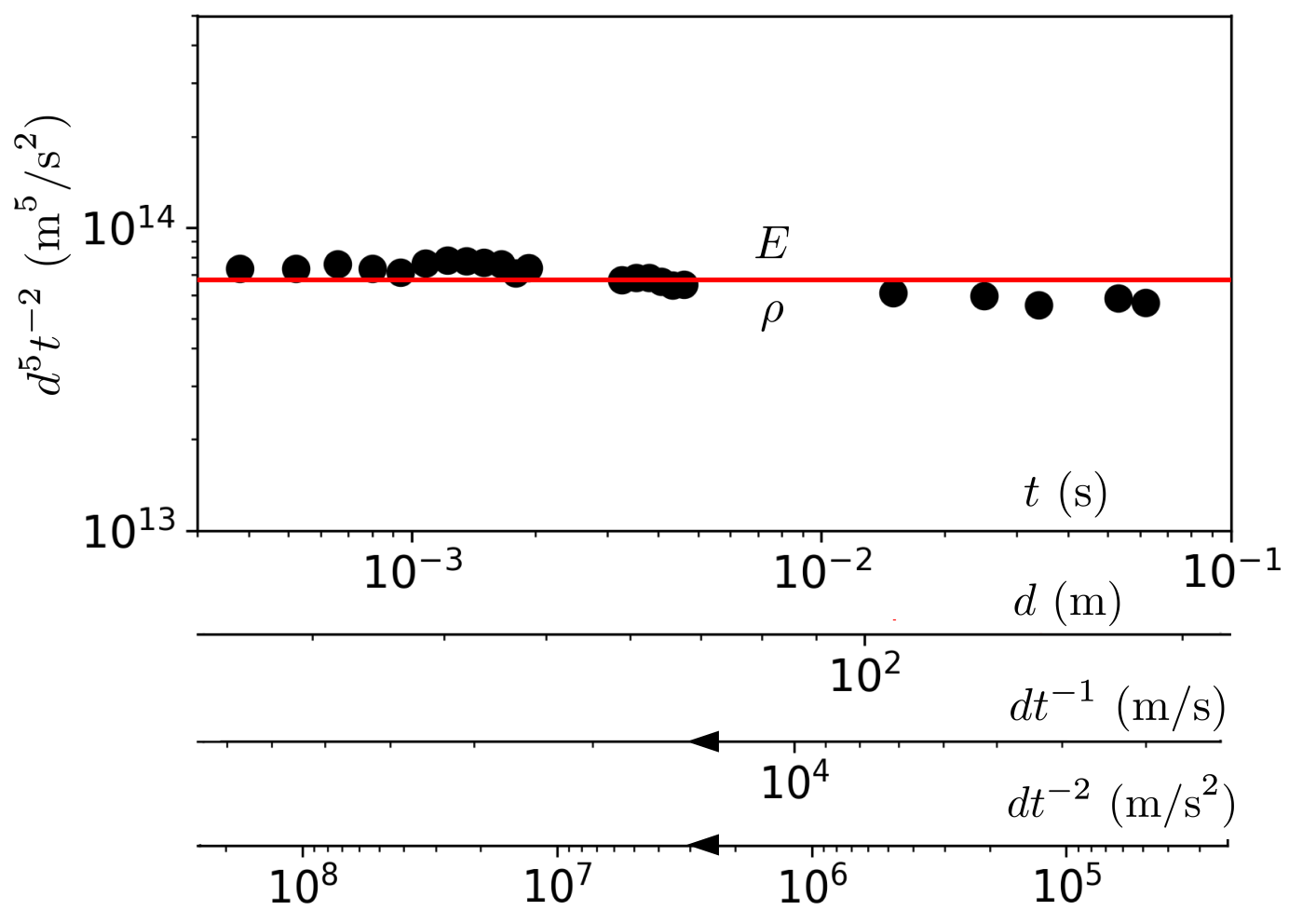}
\caption{The ``constant variable'' of the Trinity explosion, i.e. the ``explosivity'' $\kappa\equiv d^{5} t^{-2}$ plotted against the time since detonation $t$, the blast size $d$, the average front speed $d/t$, or the average front acceleration $d/t^2$. Note that the axes for the speed and acceleration run from right to left, since the speed and acceleration are initially greater. The horizontal red line is the ``unit of explosivity'' $\kappa_0\equiv E/\rho$. The data correspond to the set used by~\citet{Taylor1950b}, and shown as $d(t)$ in Fig.~\ref{figsprlaws}a. 
\label{cstvar}}
\end{figure} 
Let us consider the free fall as an example. In this context an object--like an apple--falls down to the ground. \Maf{Galileo had discovered that before any significant friction can set in the trajectory is characterized by a constant acceleration.} Because this fact is now taught in pretty much all schools throughout the world it is easy to overlook that a ``constant acceleration'' would have sounded like a complete oxymoron back then. Indeed ``acceleration'' is built from Latin ``accelerare'', meaning ``to hasten, quicken'', whereas ``constant'' comes from ``constantem'', meaning ``standing firm''. What did Galileo mean? The distance traveled since the object was dropped increases like the square of the elapsed time, $d=Kt^2$. If instead of following the distance over time we track the speed over time, we would find that it increases linearly, $v\simeq d/t \simeq K t$. Now, if we track the acceleration over time, we find that it is indeed independent of time, since $a\simeq v/t \simeq K$. All these considerations may sound trivial if we solely consider this one example of the free fall. However, the remarks made in this particular case can be generalized to any sort of regime. 

We just saw the famous example of a constant acceleration, but there are ``constant variables'' of all kinds. If we agree to say ``constant acceleration'', we should welcome this apparent contradiction as well. For any pair of mechanical quantities, for any regime, there is always a way to combine the initial variables to define a new variable, which will remain constant in the range of validity of the regime. The recipe to build such combination is simple. If we have two mechanical quantities $Q_1(x_1,y_1)$ and $Q_2(x_2,y_2)$, then assuming that the kinematics can be described by a variable length $d$ and time $t$, we know since Eq.~\ref{Generalregime2} that $Q_1/Q_2 \simeq d^{x_1-x_2} t^{y_1-y_2}$. Since both $Q_1$ and $Q_2$ are constant parameters, the combination of variables $\kappa\equiv d^{x_1-x_2} t^{y_1-y_2}$ is also constant. The new kinematic variable $\kappa$ is the \textit{constant variable} of this regime.  

In the case of Taylor's regime, the constant variable is thus $\kappa\equiv d^{5} t^{-2}$, which we called the ``explosivity''. As shown in Fig.~\ref{cstvar}, if we plot the values of this variable for the Trinity test we find that it remains roughly constant over time. In fact, this ``constant variable'' is not only constant over time, but also for all sizes $d$, all speeds $v\simeq d/t$, etc. In Fig.~\ref{cstvar}, changing the horizontal axis does not change the shape of the plot. For the x-axis, we could actually choose any kinematic quantity as long as it is not a power of the explosivity, which would just replicate the y-axis. The horizontal axis is largely arbitrary. The dimensions of the selected variable are inconsequential, as are its units. In contrast, for the vertical axis, the mechanics underlying this motion are suggesting a particular unit for the explosivity: $\kappa_0\equiv E/\rho$, with $[\kappa_0]=[\kappa]$, the ratio of energy and density is an explosivity. So we can use the ratio of energy and density as our ``natural'' or ``objective'' unit of explosivity (a term we shall specify in section~\ref{numun}). Another way to say this is that $N_{\scaleto{E\rho}{4pt}}\equiv\kappa/\kappa_0 \equiv \rho d^5 /E t^2$ provides a `dimensionless number' for the pair $\{E,\rho\}$. Indeed, $[N_{\scaleto{E\rho}{4pt}}]=\mathcal{M}^0\mathcal{L}^0\mathcal{T}^0=1$, which is to say that the quantity $N_{\scaleto{E\rho}{4pt}}$ has no dimensions. As we will see now these numbers are quite useful. 

\subsection{Dimensionless numbers}
In their simplest expression, dimensionless numbers provide a privileged perspective on each particular regime. Nevertheless, the term ``dimensionless number'' is also used to describe more complex combinations of kinematic and mechanical quantities, with no overall dimensions of space, time or mass. In this review, we will focus on \textit{simple dimensionless numbers}. For any pair of mechanical quantities $\{Q_1,Q_2\}$ ,we assume an initial perspective with two variables, $d$ and $t$, and we define the simple dimensionless number from Eq.~\ref{Generalregime2} as follows: 
\begin{equation}
N_{\scaleto{Q_1Q_2}{4pt}}\equiv \frac{Q_2}{Q_1} d^{x_1-x_2} t^{y_1-y_2}\label{GenN}
\end{equation}
By construction we have $N_{\scaleto{Q_1Q_2}{4pt}}\simeq 1$ in the range of validity of the regime.

\subsubsection{The Reynolds number}
To illustrate the concept of simple dimensionless numbers, let us come back to the boundary layer regime, $d\simeq (\eta/\rho)^{\frac{1}{2}}  t^{\frac{1}{2}}$~\cite{Landau1959}. The dimensionless number associated with this regime is the `Reynolds number' Re, which is probably one of the most famous of its kind~\cite{Rott1990}. 

According to Eq.~\ref{GenN} the dimensionless number associated with the pair $\{\eta,\rho\}$ is: 
\begin{equation}
N_{\scaleto{\eta\rho}{4pt}}\equiv \frac{\rho d^{2}}{\eta t}  \simeq \frac{\rho d v}{\eta}
\end{equation}
The dynamics of a laminar boundary layer are such that $N_{\scaleto{\eta\rho}{4pt}}\simeq 1$. Here, the dimensionless number is initially expressed from the size $d$ of the boundary layer, and the time $t$ since deformation started. However, as we saw in section~\ref{velprof}, in hydrodynamics the scalings are more often expressed from `velocity profiles', so from a speed $v\simeq d/t$ and a distance $d$. With these alternative variables $N_{\scaleto{\eta\rho}{4pt}}\simeq \rho d v/\eta$, which is the better known formula for the Reynolds number. The reader is invited to check that the following expressions are also satisfying formulations of the Reynolds number: $N_{\scaleto{\eta\rho}{4pt}}\simeq  \rho v^2 t/\eta \simeq \rho \omega d^2/\eta\simeq \rho \omega/\eta k^2\simeq \rho \mathcal{E}/\eta v \simeq \rho \mathcal{E}^\frac{1}{2}/\eta k^\frac{1}{2}$, etc. The form used depend on the choice of kinematic variables, and so on the context. 

All too often, the Reynolds number is narrowly defined from one particular perspective. For instance, Wikipedia states that ``The Reynolds number is the ratio of inertial forces to viscous forces''. In fact, because dimensionless numbers are dimensionless, they can be understood as ratios of any pair of quantities with the same dimensions. So the Reynolds number can indeed be defined as a ratio of (variable) forces, but also as a ratio of a number of other quantities. Here are some of the different ways to interpret the Reynolds number: 
\begin{align}
&\text{Re} = \frac{\text{force 1}}{\text{force 2}} = \frac{\rho d^2 v^2}{\eta d v}\\
&\text{Re} = \frac{\text{time 1}}{\text{time 2}} = \frac{\rho d^2/\eta}{d/v}\\
&\text{Re} = \frac{\text{length 1}}{\text{length 2}} = \frac{d}{\eta /\rho v}\\
&\text{Re} = \frac{\text{speed 1}}{\text{speed 2}} = \frac{v}{\eta /\rho d}\\
&\text{Re} = \frac{\text{stress 1}}{\text{stress 2}} = \frac{\rho v^2}{\eta v/d}\\
&\text{Re} = \frac{\text{density 1}}{\text{density 2}} = \frac{\rho}{\eta /d v}\\
&\text{Re} = \frac{\text{viscosity 1}}{\text{viscosity 2}} = \frac{\rho d v}{\eta}
\label{ReynoldsNumbers}
\end{align}
You can check that all these quantities have the correct dimensions by using Table~\ref{masscary}. Each particular formulation of the Reynolds number underscores a different way to think about the interplay of viscosity and density. 

Of the various ways to interpret the Reynolds number, one has a stronger standing than the others. As we saw in section~\ref{BLsec}, the ratio of viscosity and density gives rise to a `diffusivity' ($\nu\Maf{\equiv} \eta/\rho$, with $[\nu]=\mathcal{L}^2\mathcal{T}^{-1}$). The most natural way to express the Reynolds number is then as a ratio of diffusivities:  
\begin{equation}
\text{Re} = \frac{\text{diffusivity 1}}{\text{diffusivity 2}} = \frac{\nu_k}{\nu}
\label{Peclet}
\end{equation}
The diffusivity $\nu_k$ is traditionally called an `advection' or `convection' and written as $\nu_k \Maf{\equiv} dv$, or with alternate variables as $\nu_k\Maf{\equiv} d^2/t$ or $v^2 t$, depending on context. This combination of variables provides the ``constant variable'' of the pair $\{\eta,\rho\}$, and $\nu\Maf{\equiv}\eta/\rho$ is its ``objective unit'' \Maf{(a term we shall specify in section~\ref{numun}).}

\subsubsection{The Taylor-Sedov number}
The procedure we followed to construct the Reynolds number can be carried out for any regime. As we will see in the next sub-section some of the ensuing dimensionless numbers also have their own name, but most do not. Surprisingly, to the best of our knowledge the dimensionless number associated with Taylor's regime of explosions does not have a name. Since the `Taylor number' refers to something else already~\cite{Fardin2014}, let us call it the \textit{Taylor-Sedov number}, from the name of a Soviet physicist whom also contributed substantially to the understanding of explosion blasts~\cite{Sedov1993,Deakin2011}. 

According to Eq.~\ref{GenN} the dimensionless number associated with the pair $\{E,\rho\}$ is: 
\begin{equation}
N_{\scaleto{E\rho}{4pt}}\equiv \frac{\rho d^5}{E t^2}  \simeq \frac{\rho d^3 v^2}{E}
\end{equation}
The right-most expression gives the Taylor-Sedov number from the traditional ``hydrodynamic perspective'', using a speed and a length as variables. Dynamics following Taylor's $\frac{2}{5}$ regime correspond to a constant Taylor-Sedov number, $N_{\scaleto{E\rho}{4pt}}\simeq 1$. Contrary to the Reynolds number, this dimensionless number is not naturally expressed as a ratio of diffusivities, but as a ratio of explosivities.

\subsubsection{Multitudes of dimensionless numbers\label{dimfam}}
\begin{table}
\centering
\includegraphics[width=8cm,clip]{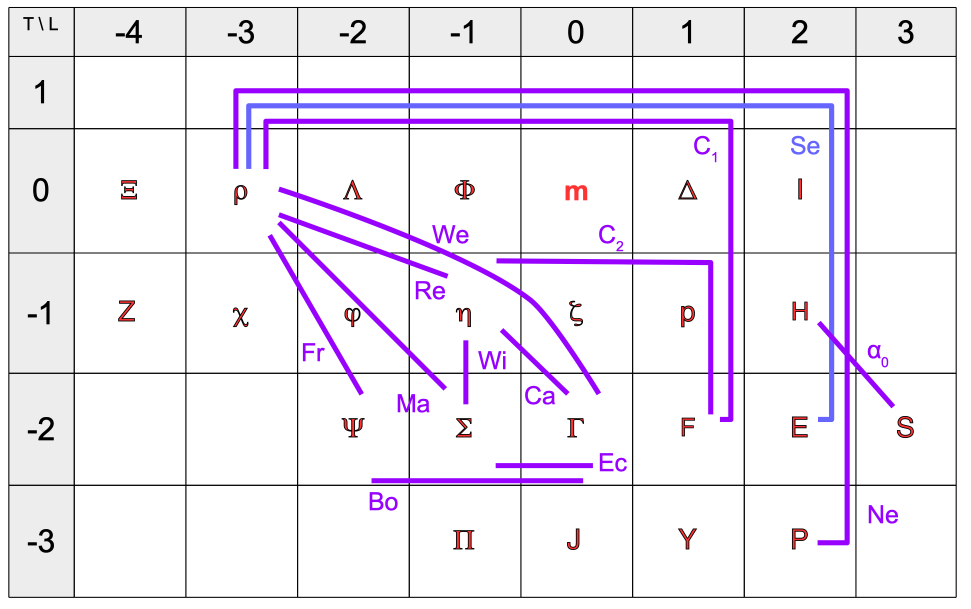}
\caption{Simple dimensionless numbers as relations between pairs of mechanical quantities. Some representative dimensionless numbers are shown to connect the mechanical quantities of Table~\ref{masscary}: Bo=$\Psi d^2/\Gamma$ (Bond number), Ec=$\Gamma /(\Sigma d)$ (Elasto-capillary number), Fr=$\rho^\frac{1}{2} v/(\Psi d)^\frac{1}{2}$ (Froude number), Ma=$v \rho^\frac{1}{2}/\Sigma^\frac{1}{2}$ (Mach number), Wi=$\eta v/(\Sigma d)$ (Weissenberg number), Ca=$\eta v/\Gamma$ (Capillary number), Re=$\rho d v/\eta$ (Reynolds number), We=$\rho v^2 d/\Gamma$ (Weber number), Ne=$P/(\rho v^3 d^2)$ (Newton or Power number), Se=$\rho d^3 v^2/E$ (Taylor-Sedov), $C_1=F/(\rho v^2 d^2)$ (\Maf{inertial drag coefficient}), $C_2=F/(\eta v d)$ (\Maf{viscous drag coefficient}). All numbers shown are standard, except for the Taylor-Sedov, which we defined in this review. The notation `Se' is here to imitate the traditional style, but as mentioned in the text we recommend the standard notation introduced in Eq.~\ref{GenN}. The structure constant $\alpha_0=S/Hv$ is usually defined in the special case where $H=\hbar$, $v=c$ and $S=S_0=k_C e^2$, in which case it is called the `fine structure constant'.  
\label{fig7}}
\end{table} 
To each pair of mechanical quantities corresponds a simple dimensionless number. Even if we restrict ourselves to the standard quantities in Table~\ref{masscary} we could build over three hundred different dimensionless numbers. Some of the most well-known are represented in Table~\ref{fig7}, with some listed here with a comparison between the traditional hydrodynamic formulations and the standard definitions following Eq.~\ref{GenN}: 
\begin{align}
\text{Bond/E{\"o}tv{\"o}s} \quad \text{Bo}&\equiv \frac{\Psi d^2}{\Gamma} \simeq N_{\scaleto{\Gamma\Psi}{4pt}}\\
\text{Elasto-capillary} \quad \text{Ec}&\equiv \frac{\Gamma}{\Sigma d} \simeq N_{\scaleto{\Gamma\Sigma}{4pt}}^{-1}\\
\text{Weissenberg/Deborah} \quad \text{Wi}&\equiv \frac{\eta v}{\Sigma d} \simeq N_{\scaleto{\eta\Sigma}{4pt}}^{-1}\\
\text{Mach} \quad \text{Ma}&\equiv \frac{\rho^\frac{1}{2} v}{\Sigma^\frac{1}{2}} \simeq N_{\scaleto{\Sigma\rho}{4pt}}^{\frac{1}{2}}\\
\text{Capillary} \quad \text{Ca}&\equiv \frac{\eta v}{\Gamma} \simeq N_{\scaleto{\Gamma\eta}{4pt}}\\
\text{Froude} \quad \text{Fr}&\equiv \frac{\rho^\frac{1}{2} v}{(\Psi d)^\frac{1}{2}} \simeq N_{\scaleto{\Psi\rho}{4pt}}^{\frac{1}{2}}\\
\text{Weber} \quad \text{We}&\equiv \frac{\rho d v^2}{\Gamma} \simeq N_{\scaleto{\Gamma\rho}{4pt}}
\end{align}
The first two examples correspond to simple lengths (Bo and Ec). The third example corresponds to a simple time (Wi). The forth and fifth examples correspond to simple speeds (Ma and Ca). The sixth example corresponds to a regime at constant acceleration (Fr). The last example correspond to a more ``exotic'' regime, where the associated constant variable is a quantity with no standard name, and dimensions $\mathcal{L}^3\mathcal{T}^{-2}$.  
 
As we first saw with the Reynolds number, Re, dimensionless numbers are often referred to using the first two letters of the person most often associated with that number. This is true for the Bond, Mach, Froude or Weber numbers. This is almost true for the Weissenberg number, which uses Wi, instead of We, to avoid confusion with the Weber number. Some numbers, like the elasto-capillary or capillary numbers imitate the style of one capitalized letter followed by a lower case letter, although no surnames are attached to it. This odd nomenclature can become quite unpractical when dealing with an increasing number of regimes, and can be the source of bitter priority disputes. It is one of the reasons why we advocate the more neutral and less reverent notation $N_{\scaleto{Q_1 Q_2}{4pt}}$, as defined in Eq.~\ref{GenN}. Note also that in some cases the name used for a particular dimensionless number depends on the choice of kinematic variables. For instance, one usually speaks of the `Weissenberg number' to describe Wi$\equiv \eta v/\Sigma d$, but of the `Deborah number' for De$\equiv \eta/\Sigma t$~\cite{Dealy2010,Poole2012}. 

As is apparent, for a given pair $\{Q_1,Q_2\}$, the traditional dimensionless numbers usually coincide with the definitions of $N_{\scaleto{Q_1Q_2}{4pt}}$. However, in certain cases they differ by some power, vestige of historical circumstances. Indeed, dimensionless numbers are defined \textit{modulo an overall power}, which is to say that if a combination of kinematic variables and mechanical parameters is dimensionless, then any power of that combination is necessarily dimensionless. So for a dimensionless number $N$, $N^\alpha$ will also be dimensionless for all values of $\alpha$. Underneath this trivial fact, if $\kappa$ is a constant variable for a regime, the $\kappa^\alpha$ will also be constant.  

The general definition provided in Eq.~\ref{GenN} ensures that all dimensionless numbers constructed in that way are linear in the underlying mechanical quantities, in contrast to Ma or Fr. In addition, since both $N$ and $N^{-1}$ could be equally valid definitions Eq.~\ref{GenN} ensures that the impelling and impeding factors always occupy the same place from one number to another. For instance, historically, the `elasto-capillary number' Ec$=\Gamma/(\Sigma d)$~\cite{McKinley2005} has been defined in contradiction to the Bond number, such that Ec$<1$ means that $d>\ell_{\scaleto{\Gamma\Sigma}{4pt}}$, whereas Bo$<1$ means that $d<\ell_{\scaleto{\Gamma\Psi}{4pt}}$. When we use traditional definitions, all these little discrepancies pile up and end up seriously obstructing the use of these dimensionless quantities. This is another reason to prefer the notation introduced in Eq.~\ref{GenN}.

\subsection{Numbers and units\label{numun}}
We have seen that given a pair of mechanical quantities we can derive its kinematic outcome and express it from a number of perspectives. One of these perspectives combines all variables and parameters into a dimensionless number. How is this last viewpoint special? How to use dimensionless numbers to further our understanding of the relationship between mechanics and kinematics?

\subsubsection{Rescaling}
\begin{figure*}
\centering
\includegraphics[width=8.5cm,clip]{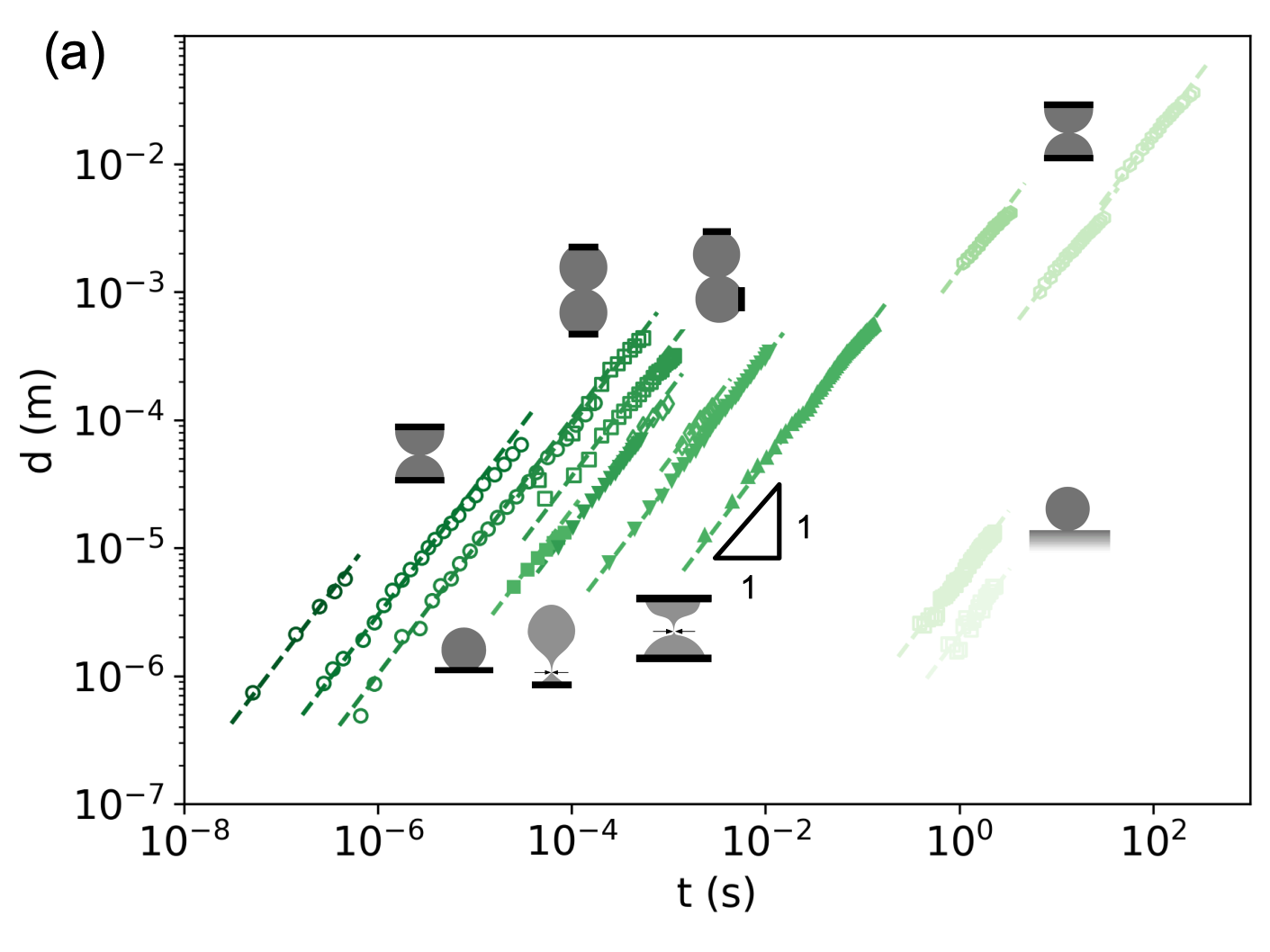}
\includegraphics[width=8.5cm,clip]{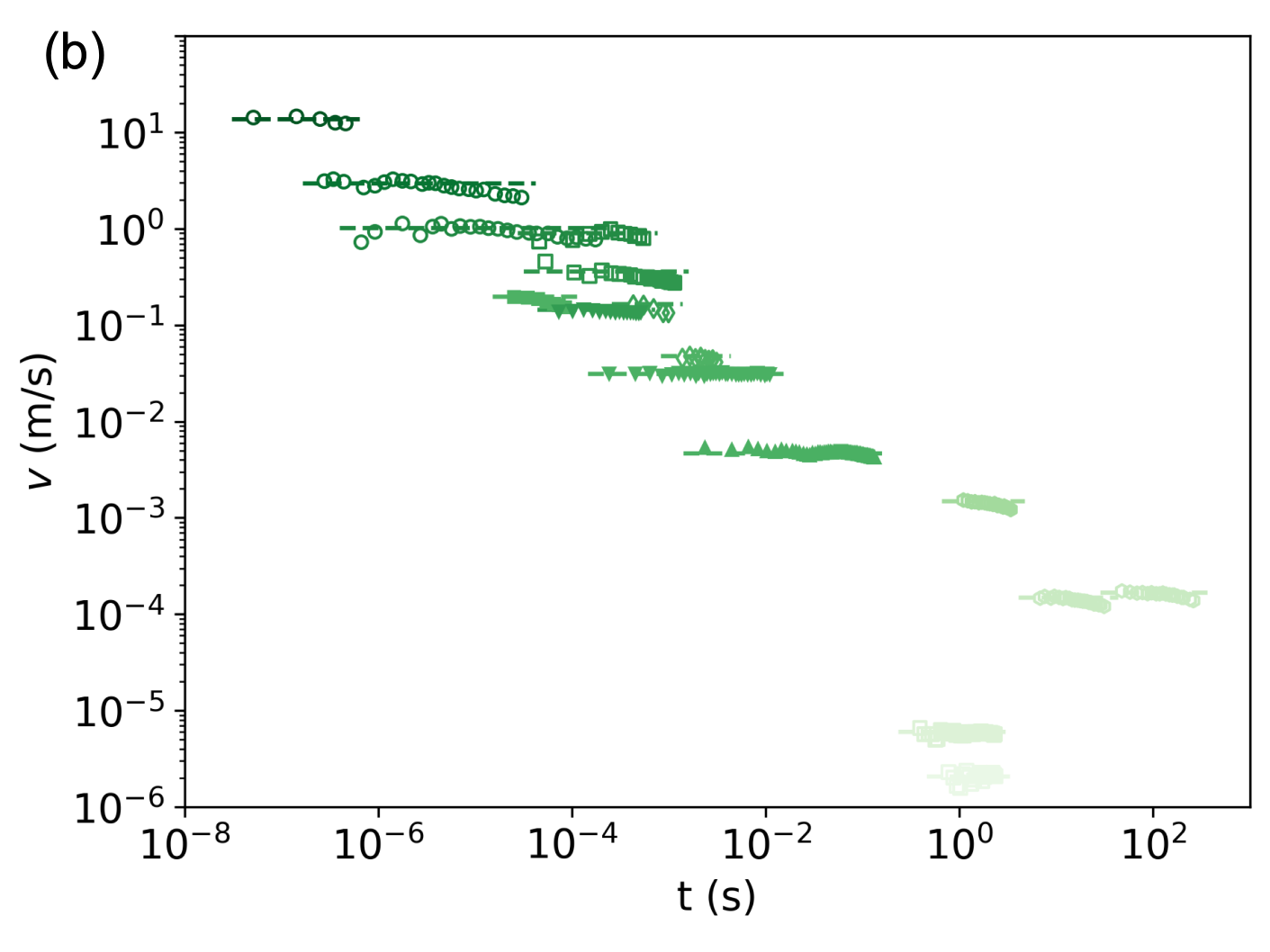}
\includegraphics[width=8.5cm,clip]{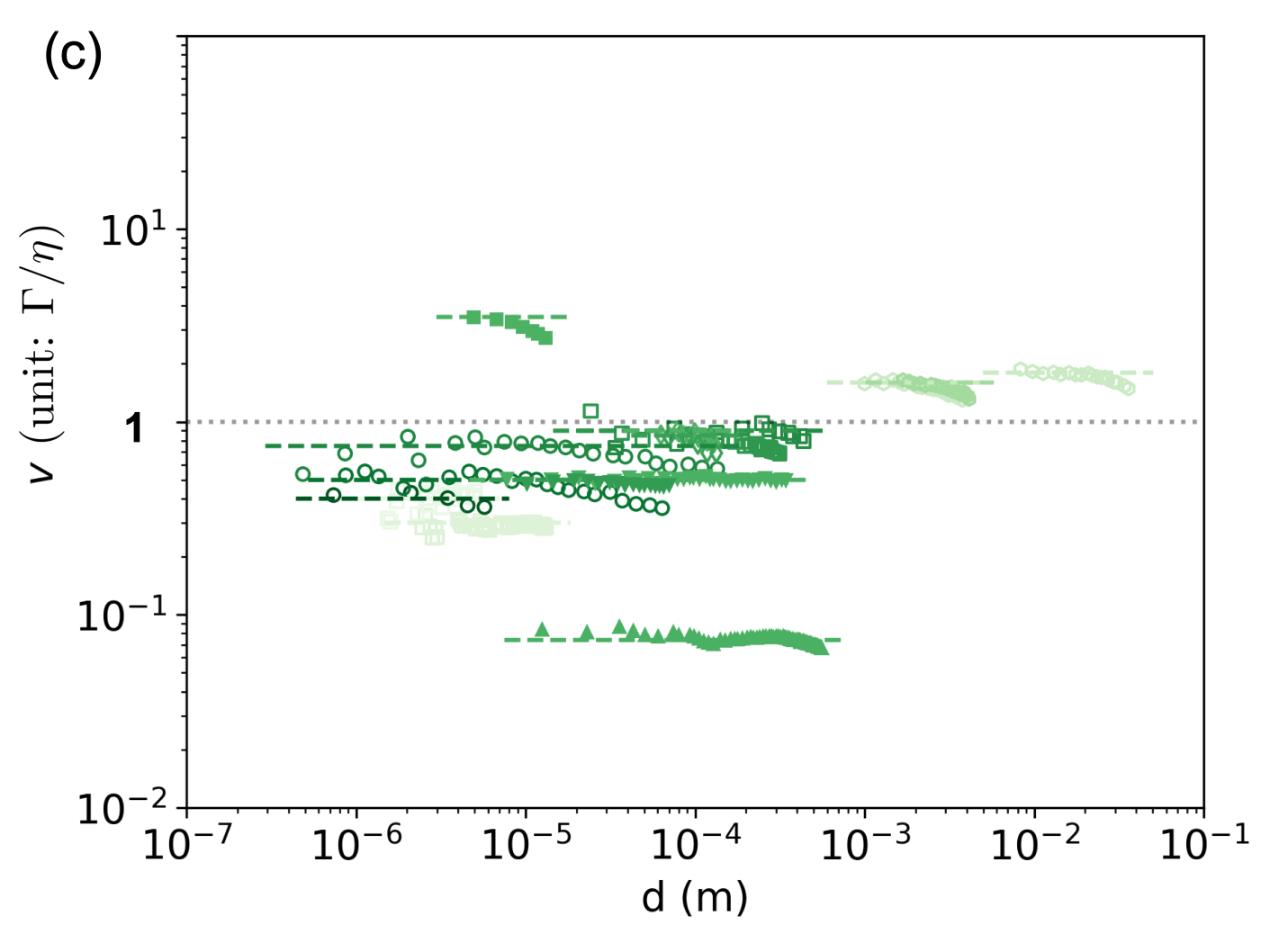}
\includegraphics[width=8.5cm,clip]{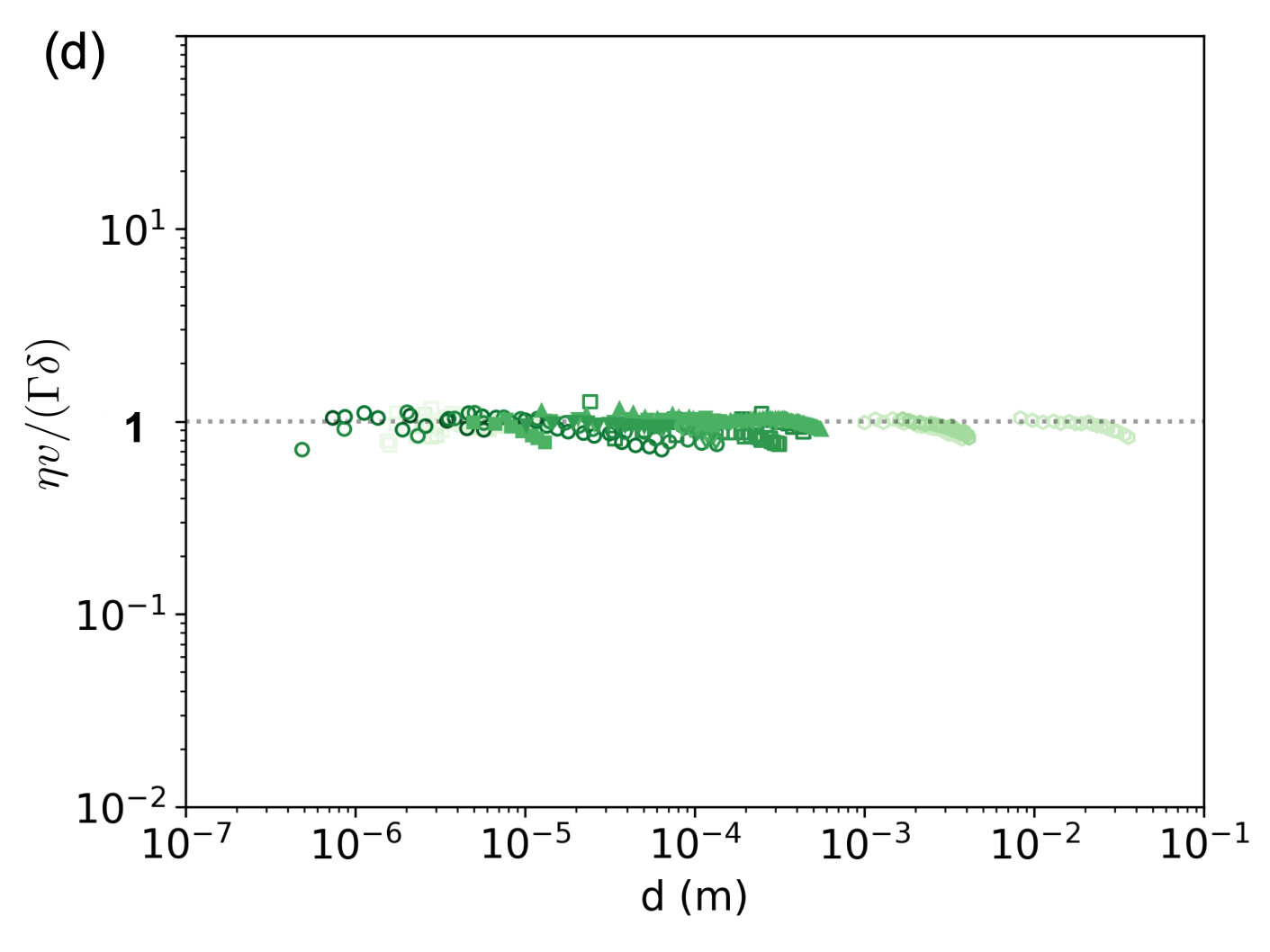}
\caption{Illustration of ``rescaling'' on a set of data exhibiting the visco-capillary regime $\{\Gamma,\eta\}$. The data include various configurations of pinching~\cite{McKinley2000,Burton2005,Bolanos2009}, coalescence~\cite{Yao2005,Aarts2005,Aarts2008,Paulsen2011,Rahman2019} and spreading~\cite{Eddi2013}. The data sets are available in the supplementary files of our recent review of this subject~\cite{Fardin2022}. (a) The visco-capillary regime seen from the ``canonical perspective'' of a length versus a time, $\{\Gamma,\eta\}_{dt}$. The length $d$ is the radius of the neck or contact area. The time $t$ is the duration since contact for spreading and coalescence, and the duration before pinch-off for pinching. (b) The speed of the neck or of the edge of the contact area is plotted against time, $\{\Gamma,\eta\}_{vt}$. Since the speed is the ``constant variable'' of this regime, the data sets fall on plateaus. (c) Upon using the mechanical ratio $\Gamma/\eta$ as ``objective unit'' of speed the data sets are all found to be close to 1, $\eta v/\Gamma\simeq 1$. The vertical axis can also be interpreted as giving the value of the simple dimensionless number of the regime, $N_{\scaleto{\Gamma\eta}{4pt}}=\text{Ca}$. The actual ordinates of each plateau give the value of the scaling constant $\delta_{\scaleto{\Gamma\eta}{4pt}}$ in each experiment. (d) When the scaling constant is included in the definition of the dimensionless number $N_{\scaleto{\Gamma\eta}{4pt}}^*\equiv \eta v/\Gamma \delta_{\scaleto{\Gamma\eta}{4pt}}$, all data sets naturally collapse on a single plateau given by $N_{\scaleto{\Gamma\eta}{4pt}}^*=1$. \Maf{Note that in (c) and (d) the size $d$ is used as horizontal axis, but any variable other than the speed $v$ or powers of it could have been used (c.f. Fig~\ref{cstvar}). }
\label{VCfig}}
\end{figure*} 
So far we have illustrated regimes associated with pairs of mechanical quantities, $\{Q_1,Q_2\}$, by giving single examples where the mechanical parameters took set values. For instance when we discussed Taylor's regime $\{E,\rho\}$, we focused--like him--on the Trinity explosion, where $E\simeq 10^{14}$~J and $\rho\simeq 1$~kg/m$^3$~\cite{Taylor1950b}. To gain greater insight on the use of dimensionless numbers we need to consider a collection of dynamics governed by the same mechanical pair, but with different values for the impelling and impeding factors. We will review the case of explosions in detail in an upcoming article, so let us here consider the `visco-capillary' regime as an example, i.e. the regime combining surface-tension and viscosity, $\{\Gamma,\eta\}$, which we mentioned in section~\ref{speeds} on simple speeds. 

As shown in Fig.~\ref{VCfig}a the visco-capillary regime has been observed in a number of situations where viscous fluids are driven by surface-tension~\cite{Fardin2022}. As mentioned in section~\ref{speeds} this regime is for instance observed for the pinching of viscous liquid bridges~\cite{McKinley2000}. In that case $t$ is the duration before pinch-off, and so the ``actual time'' runs from right to left. This pinching configuration is just one out of many possible set-ups exhibiting the regime $\{\Gamma,\eta\}$. Similar visco-capillary dynamics can also occur with rising bubbles pinching-off~\cite{Burton2005,Bolanos2009}. In that case, the viscosity is that of the outer fluid. This regime has also been found in a number of slightly different configurations of droplet coalescence~\cite{Yao2005,Aarts2005,Aarts2008,Paulsen2011,Rahman2019}. For these examples, the neck between the drop grows and so time is running from left to right. As we will see shortly, the differences in setups actually have a marginal impact. The visco-capillary regime is observed during pinching, coalescence, and spreading of drops onto substrates~\cite{Eddi2013}. For spreading the size $d$ is the radius of contact. 

All lines in Fig.~\ref{VCfig}a have the same slope but the intercepts are different, because the values of viscosity and surface-tension are different in each case, giving rise to visco-capillary speeds ranging from over 10~m/s for water with a bit of glycerol, in dark green~\cite{Paulsen2011}, to slightly over 1~$\mu$m/s for colloid-polymer mixtures, in faint green~\cite{Aarts2008}. What we are seeing in Fig.~\ref{VCfig}a is how all these experiments look like from the perspective of a length versus a time, the ``canonical perspective'' of this review. But we are free to use a different perspective, and in particular to adopt a perspective where one of the axis is the ``constant variable'', which in this case is the speed $v\simeq d/t$.  In Fig.~\ref{VCfig}b we kept the time $t$ as the other variable, but we could have chosen the size $d$, or anything we want other than the speed (or powers of the speed). As we saw already in Fig.~\ref{cstvar} the second variable is largely irrelevant. The data sets would have still looked the same: horizontal lines. Since $\{\Gamma,\eta\}\rightarrow v\simeq \Gamma/\eta$, in Fig.~\ref{VCfig}b, the different ordinates of the horizontal lines associated with each experiment reflect the different values of $\Gamma/\eta$. Experiments found higher on the plot correspond to higher values of surface-tension or lower values of viscosity. 

In Fig.~\ref{VCfig}b, regardless of the fluid and set-up, the speeds are measured in meters per second. Of course, we could have used any arbitrary unit we want like cm/min or feet/hour. The choice of unit is completely subjective. We can--if we want--measure the speed of these pinching, coalescing and spreading droplets in relation to the length of our feet (ft), and to a faction of the rotation period of our planet (hour). We are allowed to do this, but we should recognize how presumptuous we are to expect that the dynamics of droplets would be best described by such provincial choices. When for instance we say that the pinching of glycerol happens at a speed around 54 feet per hour, the number we get, 54, is due to two different things. First, and hopefully, it is related to some actual natural phenomenon, which was recorded sometime at the turn of the millennium~\cite{McKinley2000}. Second, the number 54 is connected to the choice of unit. If we choose different units, we get a different number, like $27.6$~cm/min. So these two concepts of units and numbers are obviously related. That is not really contentious. 54 or 27.6, these numbers are a bit random, as random as our choice of units. 

Can we instead find a way to define more ``objective'' or ``natural'' units, less bound to our preferences? Units that would be set by the mechanics at play? Yes we can, this unit of speed is given by the ratio of surface-tension and viscosity, $v_0\equiv\Gamma/\eta$. For each experiment we know the surface-tension, and we know the viscosity so we can compute their ratio and use it directly as our unit. This unit is more objective than any of our choices because it is given directly by the mechanical quantities dictating the dynamics. 

In contrast to the subjective units like meters per second, the value of the \textit{objective unit} changes from one experiment to another. Once we have our objective units we can then plot all curves together, as shown in Fig.~\ref{VCfig}c. The curves now start to overlap, revealing their inherent similarity. With these objective units, all speeds are reasonably ``close to 1''. Note in addition that plotting the speed $v$ ``in units of $\Gamma/\eta$'' is the same as plotting the dimensionless number of the regime, which in this case is the Capillary number $\text{Ca}= N_{\scaleto{\Gamma\eta}{4pt}}$. In Fig.~\ref{VCfig}c, it is said that the dynamics from Fig.~\ref{VCfig}a or b have been `rescaled'. 

Nevertheless the overlap between the different curves in Fig.~\ref{VCfig}c is not perfect. For all curves we may say that $N_{\scaleto{\Gamma\eta}{4pt}}\simeq 1$, but not that $N_{\scaleto{\Gamma\eta}{4pt}}=1$. Now that we are reaching the end of our exploration of scalings based on pairs of mechanical quantities, it is time to come back to the difference between the approximate equality, `$\simeq$', which we have been relying on, and a stricter kind of equality, `$=$'. 

\subsubsection{Scaling constants\label{scon}}
Throughout this review, except on rare occasions, we have used approximate equalities, `$\simeq$', which connect left and right-hand sides with the same `order of magnitude'. This sign has helped us absorb a number of numerical factors `of order 1', which tend to cloud the expressions of regimes. For instance, in a formula such as $\Omega=(4\pi/3) r^3$, giving the volume of a sphere of radius $r$, the symbols between brackets should not be given the same status as the term $r^3$. When focusing on scaling, we would just write $\Omega\simeq r^3$, neglecting the numerical factors. However, as we have just seen with Fig.~\ref{VCfig}, in the end if we want to neatly overlap all dynamics pertaining to the same underlying mechanics we need to pay more attention to these factors. 

A dimensional equation like $[E/\rho] = [d]^5 [t]^{-2}$ is unimpeachable and exact regardless of the values of energy, density, size and time. This is why we use the sign `$=$'. When the brackets are dropped, $E/\rho \simeq d^5 t^{-2}$, the approximate equality `$\simeq$' is not so much an expression of imprecision as a requirement of adequacy between the chosen kinematic variables $d$ and $t$, and the values of the mechanical quantities $E$ and $\rho$. One may say that if the dynamics of an explosion blast are due to the interplay of energy and density, then $E/\rho$ and $d^5 t^{-2}$ must indeed have the same order of magnitude. More generally, assuming for simplicity that we have identified a variable length $d$ and a variable time $t$ to describe the kinematics, if we have $Q_1/Q_2\simeq d^{x_1-x_2} t^{y_1-y_2}$, we can safely assume that the dynamics are impelled by $Q_1$ and impeded by $Q_2$. However, the agreement between the two sides may not be exact. We can write this from the ``canonical perspective'' of a length versus a time: 
\begin{equation}
d= \delta_{\scaleto{Q_1Q_2}{4pt}} \Big(\frac{Q_1}{Q_2}\Big)^\frac{1}{x_1-x_2} t^{\frac{y_2-y_1}{x_1-x_2}}\label{sccst}
\end{equation}
This equation is identical to Eq.~\ref{Generalregime}, but we have introduced a (dimensionless) numerical correction $\delta_{\scaleto{Q_1Q_2}{4pt}}$, in order to be able to use a strict equality. This kind of prefactor has been called by different names, like the `similarity constant', or the `scaling constant', or even the ``fudge factor'' by more facetious commentators. In contrast to the rest of the equation, this constant cannot be derived from dimensional analysis, but dimensional analysis imposes that its value remains `of order 1', roughly between 0.1 and 10. Unfortunately, making this constraint more precise would go beyond the scope of this review since it requires considering more than two mechanical quantities. Nevertheless, we can already clarify a few things about this scaling constant. 

The scaling constant $\delta_{\scaleto{Q_1Q_2}{4pt}}$ of a regime is a black box, containing all sorts of influences beyond that of the mechanical pair underlying the dynamics. For instance, consider Archimedes' simple length $\ell_{\scaleto{m\rho}{4pt}}\equiv (m/\rho)^\frac{1}{3}$. Suppose we are dealing with a sphere of diameter $d$, so its volume is $\Omega=(\pi/6)d^3$, and its mass is $m=\rho\Omega$, and so $d=\delta (m/\rho)^\frac{1}{3}$, with $\delta=(6/\pi)^\frac{1}{3}\simeq 1.2$. In this case, the scaling constant $\delta$ includes shape effects. Its value would be different if we were dealing with a cube or a pyramid, or some more complicated figure. The description of shapes may require more than a single length (height, width, etc.) and would then involve more than a pair of mechanical quantities. 

When we discussed the Hooke-Rayleigh time $\tau_{m\Gamma}$ we saw a similar effect in the time rather than space dimension. The period of oscillation of a spring and mass system is $\tau=\delta (m/\Gamma)^\frac{1}{2}$, with $\delta=2\pi\simeq 6.3$. Such correction can very well be absorbed by redefining the variables. For instance here, we can use the angular frequency $\omega\equiv 2\pi/\tau$ to reach $\omega= (\Gamma/m)^\frac{1}{2}$, where there is no more correction factor. 

In some cases, as in the previous two examples, the scaling constant can be disposed off by an appropriate redefinition of the variables. In other cases, a redefinition of the mechanical quantities may also be helpful. For instance, if the standard way to measure the energy (or `yield') of an explosion was to fit the dynamics of the blast by a power law $d=Kt^\frac{2}{5}$ and then to set $E\equiv \rho K^5$, then obviously we would have $d=\delta (E/\rho)^\frac{1}{5} t^\frac{2}{5}$, with $\delta=1$, as long as our measurement of the air density $\rho$ is correct. Whenever we face a regime with a scaling constant that conveniently reduces to one, it is probably because the mechanical quantities of this regime are actually defined in that context. However, this is rarely the case, since mechanical quantities are free to interact with so many partners, like all those given in Table~\ref{masscary}. The difficulties in understanding the value of the scaling constant $\delta$ then lie in the fact that $\delta$ usually connects a given regime to other manifestations of its mechanical factors, beyond the range of validity of the regime. For instance, for explosions, the energy may be rather defined from measurements of the final blast radius $\ell$, as $E\equiv \frac{4\pi}{3}\Sigma \ell^3$, where $\Sigma$ is the bulk modulus of the air (according to Eq.~\ref{ESigma}), or from the initial speed $u$, as $E\equiv \frac{1}{2} m u^2$, where $m$ is the ejected mass (according to Eq.~\ref{Em}). The numerical factors used in these definitions ($\frac{4\pi}{3}$, or $\frac{1}{2}$) get carried over from one equation to another and end up pilling up in the scaling constant $\delta$. 

In the case of the pinching, coalescing and spreading fluids in Fig.~\ref{VCfig}c, the ordinates of the plateaus in each data set give an average value of the scaling constant $\delta_{\scaleto{\Gamma\eta}{4pt}}$ for each experiment. We refer the reader to our recent meta-analysis of this subject, where values of the constant are listed for all experiments shown in Fig.~\ref{VCfig}c~\cite{Fardin2022}. The constants are all reasonably ``close to 1'', but they are influenced by the geometric details of each set-up. For instance, in the case of the pinching of a liquid thread between two plates, theoretical analyses proposed slightly different values of $\delta_{\scaleto{\Gamma\eta}{4pt}}$ depending on subtle differences: $\delta_{\scaleto{\Gamma\eta}{4pt}}=0.1666$~\cite{McKinley2000}, $\delta_{\scaleto{\Gamma\eta}{4pt}}=0.0709$~\cite{Papageorgiou1995}, $\delta_{\scaleto{\Gamma\eta}{4pt}}=0.0304$~\cite{Eggers1993,Brenner1996}, $\delta_{\scaleto{\Gamma\eta}{4pt}}=0.0108$~\cite{Brenner1996}. The experiment on the pinching of glycerol reproduced in Fig.~\ref{VCfig} ($\blacktriangle$) seemed to favor Papageorgiou's value~\cite{McKinley2000}. 

Ultimately, one may decide to include the scaling constants into the definition of the objective unit or constant variable, that is into the definition of the simple dimensionless number, in order to reach a more satisfying overlap of the data, as shown in Fig.~\ref{VCfig}d for the visco-capillary regime. If the scaling constant is defined from the canonical perspective in Eq.~\ref{sccst}, then amending Eq.~\ref{GenN} we may write: 
\begin{equation}
N_{\scaleto{Q_1Q_2}{4pt}}^* \equiv  \delta_{\scaleto{Q_1Q_2}{4pt}}^{x_2-x_1} \frac{Q_2}{Q_1} d^{x_1-x_2} t^{y_1-y_2}\equiv \delta_{\scaleto{Q_1Q_2}{4pt}}^{x_2-x_1} N_{\scaleto{Q_1Q_2}{4pt}}
\end{equation}
Then all data sets naturally collapse on a single plateau given by $N_{\scaleto{\Gamma\eta}{4pt}}^*=1$. Systematic deviations form this plateau can only be witnessing the growing effect of mechanical quantities beyond the initial pair, for instance the effect of the density of the fluid $\rho$, or of the `Laplace force' $F\simeq \Gamma \ell$, when approaching the size $\ell$ of the whole drop~\cite{Fardin2022}. Since no regime extends forever, such deviations are bound to happen, but this topic is however beyond the scope of this review.   

\section{Conclusion}
With Archimedes, Newton, Taylor, and all those who sought to explore the mechanical underpinning of space and time, this review has demonstrated the parsimonious efficiency of dimensional analysis. Whether we want to explain the magnitude of a particular volume, an acceleration, or the more exotic motion of an explosion blast, the mechanical philosophy suggests a bold idea: to invoke an extra dimension, the dimension of mass $\mathcal{M}$, beyond the visible dimensions of space, $\mathcal{L}$, and time, $\mathcal{T}$. The scope of this simple idea is immense. 

What we see as motion, size, or duration is understood as a form of shadow, cast on a plane by a much broader play. The players are the mechanical quantities, which generations of researchers have inferred from their effects. The table of mechanical quantities that we built for this review provides a first map, drawn from centuries of exploration of this mechanical pantheon. \Maf{We hope that this table will incite historians of science to trace back the steps of past thinkers from one spot on the map to another, to define pressure, energy, power, action, etc. Why did so many great minds sailed these waters? }How does this mechanical chart relate to the world we live in? We wish to do our part in answering this question, and this review is a first step. In this review, we have shown what emerges from the interplay of duos of mechanical quantities. This type of pair interaction is the elementary building block of a dimensional analysis of mechanics, to which we can refer to more succinctly as \textit{dimensional mechanics}. 

As we saw, numerous experiments have shown that a range of complex motions can be cast as single lines in logarithmic scale. These straight regimes are drawn on a plane of space and time, and yet we find that they can be more deeply understood as a ``reflection'', or ``projection'', or ``shadow'' of something going on at a higher dimension. What is going on is a ``struggle'', or ``balance'', between ``competing'' mechanical factors. Dimensional mechanics help us formalize all these colloquialisms, help us find the ``causes'' of these scalings. These causes are embodied by the mechanical quantities. Any pair of mechanical quantities is associated with a regime, a line in the kinematic plane. The slope of this line is given by the dimensions of the mechanical ratio. Because the standard mechanical quantities all share a dimension of mass $\mathcal{M}$, such mass disappears from their ratio: 
 \begin{equation}
 \Big[\frac{Q_1}{Q_2}\Big] = \frac{\cancel{\mathcal{M}}\mathcal{L}^{x_1}\mathcal{T}^{y_1}}{\cancel{\mathcal{M}}\mathcal{L}^{x_2}\mathcal{T}^{y_2}} =\mathcal{L}^{x_1-x_2} \mathcal{T}^{y_1-y_2}
 \end{equation}
Hence kinematics emerge from a confrontation of mechanical terms. Motion, change, comes from ratios of constant mechanical parameters. Mathematical division formalizes the age-old intuition that motion results from a ``tug of war'' between ``forces'' and ``masses'', between what we more generally called impelling and impeding factors. 

The type of scaling observed in a particular context depends on the dimensions of the underlying mechanical parameters. Some experiments may evidence characteristic lengths, or times, or speeds, but others may record more intricate relationships between space and time. For instance, from the ``canonical perspective'' of a variable length measured relative to a variable time one may see scalings of the form $d\sim t^\alpha$, with $\alpha=\frac{1}{2}$, or $\frac{2}{5}$, or $\frac{1}{4}$, or $\frac{2}{3}$, or $\frac{3}{2}$, etc. These exponents are not fundamentally weirder than the more traditional $\alpha=1$ of uniform motion, or $\alpha=2$ of uniformly accelerated motions. We just had less time to get used to them. In this review, we have tried to find a diverse array of examples from different fields, but the lists of scalings that we compiled is but a fraction of what could be gathered from a more thorough investigation. We invite readers to participate in this encyclopedic enterprise, and we will welcome any correspondence to that end. 

Throughout this review we have assumed that the ``relevant'' mechanical quantities were known in each particular context, and so from these parameters the regimes could be derived by dimensional analysis. We understood the ``play'' so we could make sense of its ``shadow'', as seen from different perspectives. For instance, in the case of the Trinity explosion, knowing the energy $E$ of the bomb and the density $\rho$ of the air to be the relevant parameters, Taylor could derive that $d\sim t^\frac{2}{5}$, or that $v\sim t^{-\frac{3}{5}}$. We have shown how a knowledge of mechanics implies the kinematics. But how is such knowledge gained in the first place? Mechanics implies kinematics, but the reverse is not so simple. Assuming a ``canonical perspective'', we can symbolize this as follows: 
\begin{align}
\text{Mechanics} &\mathrel{\substack{\textstyle\rightarrow\\[-0.6ex]
                      \textstyle\leftsquigarrow }} \text{Kinematics} \nonumber \\
      \{Q_1,Q_2\}  &\mathrel{\substack{\textstyle\rightarrow\\[-0.6ex]
                      \textstyle\leftsquigarrow }}  d\simeq K t^\alpha              
\end{align}
For a given pair of mechanical quantities $\{Q_1,Q_2\}$ there is a single associated kinematic regime. However, a single regime can be associated with a multitude of possible \textit{mechanical models} (what we symbolized with the squiggly arrow). We saw this in passing with the Bohr radius, which could be expressed from the electromagnetic strength and the kinetic energy of the electron, $\{S,E\}$ (Eq.~\ref{SE}), or from the Planck constant and the momentum of the electron, $\{H,p\}$ (Eq.~\ref{Hp}). Both models gave the same result. This ``redundancy'' is not at all unique to the Bohr radius, it is fundamentally entrenched in the asymmetric relationship between mechanics and kinematic. 

Kinematics have a lower dimension than mechanics. What we see as motion is only a projection of what goes on, and information is lost in such shadow play. We might see a dog or a duck on the screen, but it may actually be the hand of the puppeteer. For a given pair of mechanical quantities, $Q_1(x_1,y_1)$ and $Q_2(x_2,y_2)$, the resulting regime is $d\simeq K t^\alpha$, with $\alpha=(y_2-y_1)/(x_1-x_2)$. Only the \textit{relative dimensions} of the mechanical quantities matter, i.e. the differences $y_2-y_1$ and $x_1-x_2$. Thus, a given exponent $\alpha$ may actually come from an array of mechanical pairs. For instance, we saw diffusive regimes due to $\{E, \Lambda\}$, $\{\Sigma, \chi\}$, $\{\Gamma,\varphi\}$,  $\{F, \rho\}$, $\{F, \eta\}$, or $\{\eta, \rho\}$. Taking a look back at the table of mechanical quantities will confirm that these pairs have the same relative placement. So, if we do observe $d\sim t^\frac{1}{2}$, how can we know which mechanical pair is behind this motion? We cannot; that is if we only observe a single regime\ldots

On paper, power laws such as $d\simeq K t^\alpha$ are `self-similar'~\cite{Barenblatt2003}, they seemingly extend to arbitrarily small or large scales. In practice, no single regime extends indefinitely. For instance, Taylor showed that the Trinity explosion of July 1945 in New Mexico followed $d\simeq (E/\rho)^\frac{1}{5} t^\frac{2}{5}$. As shown in Fig.~\ref{figsprlaws}a, the data do support this model, but only for the selected time range, from a fraction of a millisecond to a fraction of a second after detonation. Taylor knew very well that this regime was only transient~\cite{Taylor1950,Taylor1950b}. If extended indefinitely in the future, the blast would have reached New York by now! And if extended to the instant of detonation, this regime would predict a diverging front speed, since $v\sim t^{-\frac{3}{5}}$. Evidently, the balance of energy and density cannot account for the whole dynamics of the explosion. We actually know what we should expect. At small time scales, the mass of the bomb will have an effect and we would get a constant initial speed from $\{E,m\}$, according to Eq.~\ref{Em}. At large scale, we would reach the final blast radius set by the ambient pressure, $\{E,\Sigma\}$, according to Eq.~\ref{ESigma}. Dynamics are never truly `self-similar', because the complete play always involves more than two players. 

We have learned much by focusing on pairs of mechanical quantities, and we would be right to expect that a deeper understanding of dimensional mechanics may come by progressively enlarging our set of mechanical parameters. We have already started exploring the effects of trios and quartets, and they greatly extend the reach of dimensional mechanics. Our \href{https://www.youtube.com/playlist?list=PLbMiQs7eX-bbNTc-7HwdWzohUs8yPw300}{online series of video lectures on explosions} documents this progression for this particular example, and starting with \href{https://youtu.be/1WP7k1VUf3Q?si=Muusfb3EDpNHB2I2}{episode 9} the series on mechanics generalizes our findings. \Maf{We are planning to summarize these investigations in upcoming publications, but we can already mention a few salient points, which may resonate with some of the questions that this review might have raised. }

\Maf{When considering three mechanical quantities, two cases should be distinguished: whether the three quantities are aligned or not. When the quantities are aligned they lead to three parallel regimes. For instance, in the special case where the quantities are on the same line of Table~\ref{masscary}, this configuration can be used to rationalize shapes with distinct dimensions along different directions, and more generally to discuss the mechanical underpinning of geometry. Readers wishing to get ahead may already investigate situations characterized by trios such as $\{E, \Sigma,\Psi\}$, which may be useful for cratering~\cite{Holsapple1993}, or $\{E, \Gamma,\Psi\}$, which may be useful for drop impact~\cite{Laan2014}. When considering trios of mechanical quantities that are not aligned, the three regimes associated with the three pairs of the trio intersect at a single point, a special event where dynamics take a turn, and which can serve as the locus of fully objective units. For instance, looking back at the dispersion relation in Fig.~\ref{disprel}, we can easily envision that the regimes of waves $\{\Psi,\rho\}$ and ripples $\{\Gamma,\rho\}$ will intersect at a wavenumber given by the inverse of the capillary length implied by the pair $\{\Gamma,\rho\}$. The coordinates of the point of intersection are fully characterized by the trio $\{\Psi,\Gamma,\rho\}$. Another example would consider the trio $\{\Gamma,\rho,\eta\}$, to capture the combined effects of inertia and viscosity on capillary flows. We discussed this case in a recent publication~\cite{Fardin2022}. }

\Maf{The key insight brought by considering a fourth quantity is even more intriguing. Quartets of mechanical quantities  include four connected trios associated with intersections. The six associated regimes connect these turning points to one another, allowing for instance to track the succession of events of an explosion from the initial dynamics driven by the energy of the bomb $E$ and impaired by the ejected mass $m$, to the gradual transformation of the blast into a simple sound wave, depending solely on the density $\rho$ and pressure $\Sigma$ of the ambient air. As shown in our  \href{https://www.youtube.com/playlist?list=PLbMiQs7eX-bbNTc-7HwdWzohUs8yPw300}{online series of video lectures on explosions} a quartet such as $\{E,m,\rho,\Sigma\}$ also provides a new kind of dimensionless number, depending solely on the constant mechanical parameters. In this particular instance, this number would be $\mathcal{N}\equiv E\rho/m\Sigma$ (modulo an overall power). Such kind of dimensionless number allows to distinguish two broad classes of dynamics, here detonations ($\mathcal{N}>1$) and deflagrations ($\mathcal{N}<1$). More profoundly, such kind of dimensionless number provides an ``objective number base'', to use instead of 10 in logarithmic representations. Base 10 is an international convention for numbers, just like kilograms, meters, and seconds are international conventions for mass, length and time. If dimensional analysis is practiced beyond the mechanical duos we have been focusing on here, these four conventions are replaced by standards set by the physics of the situation at play. }

\Maf{Although the role of the mass dimension and the associated mechanical quantities is initially to ``explain'' the kinematics, what dimensional analysis reveals is that the ``physics'' of a situation also inform the appropriate ways to represent what is seen. The kind of rescaling displayed in Fig.~\ref{VCfig} can be brought to a whole new level by enlarging the experimental range, including the effects of more than two mechanical quantities}. 

Dimensional analysis is so much more than a trick to circumvent the more ``rigorous'' use of differential equations. Dimensions provide the fundamental structure of physics, and the dimensions of mechanics are its most robust backbone. 


\subsection*{Acknowledgments} 
VS thanks UIC and UChicago students who enrolled in his many courses and provided student perspectives in discussions on dimensions and dimensional analysis, as well as feedback on the supplementary videos. VS also wishes to thank his Soft Matter ODES-lab students. VS acknowledges Matt Tirrell, Juan de Pablo, Paul Nealey, and the Pritzker School of Molecular Engineering at UChicago for financial support during his sabbatical. MH thanks DGAPA-UNAM for funding his own sabbatical, which got this review started. VS, MH and MAF acknowledge the joyful atmosphere of the Ladoux-M{\`e}ge lab, the stimulating haven that made this project possible. We owe a great debt to our predecessors and mentors, and many in the fluid mechanics, soft matter, and biophysics communities for inspiration and exemplary applications of dimensional analysis.

\normalem

\end{document}